\documentclass[aip,jcp,reprint]{revtex4-1}
\usepackage{graphicx}
\usepackage{appendix}
\usepackage{rotating}
\usepackage{simplewick}
\usepackage{amssymb}
\usepackage{amsmath}
\usepackage{txfonts}
\usepackage{mathrsfs}
\usepackage{bm}
\usepackage{dcolumn}
\usepackage{color}

\newcommand{\stirling}[2]{\genfrac{\{}{\}}{0pt}{}{#1}{#2}}

\begin{document}
\title{Finite-temperature many-body perturbation theory for electrons: Algebraic recursive definitions, second-quantized derivation, linked-diagram theorem, general-order algorithms, grand canonical and canonical ensembles}

\date{\today}
\author{So \surname{Hirata}}\email[Email:~]{sohirata@illinois.edu}
\affiliation{Department of Chemistry, University of Illinois at Urbana-Champaign, Urbana, Illinois 61801, United States}

\begin{abstract}
A comprehensive and detailed account is presented for the finite-temperature many-body perturbation theory for electrons 
that expands in power series all thermodynamic functions on an equal footing. Algebraic recursions  in the style of the Rayleigh--Schr\"{o}dinger 
perturbation theory are derived for the grand potential, chemical potential, internal energy, and entropy in the grand canonical ensemble
and for the Helmholtz energy, internal energy, and entropy in the canonical ensemble, leading to their sum-over-states analytical formulas at any
arbitrary order. For the grand canonical ensemble, these sum-over-states formulas are systematically 
transformed to sum-over-orbitals reduced analytical formulas by the quantum-field-theoretical techniques 
of normal-ordered second quantization and Feynman diagrams extended
to finite temperature. It is found that the perturbation corrections to energies entering the recursions have to be treated as a nondiagonal matrix, whose off-diagonal
elements are generally nonzero within a subspace spanned by degenerate Slater determinants. They give rise to
a unique set of linked diagrams---renormalization diagrams---whose resolvent lines are displaced upwards, which are distinct from the well-known anomalous diagrams
of which one or more resolvent lines are erased. A linked-diagram theorem is introduced that proves the size-consistency of the finite-temperature many-body perturbation theory at any order. General-order algorithms implementing the recursions establish
the convergence of the perturbation series towards the finite-temperature full-configuration-interaction limit unless the series
diverges. Normal-ordered Hamiltonian at finite temperature sheds light on the relationship between the finite-temperature Hartree--Fock and first-order many-body
perturbation theories. 
\end{abstract}

\maketitle        

\section{Introduction}

Strong electron correlation\cite{Strongcorrelation4,Zgid2011,Strongcorrelation5,Strongcorrelation1,Strongcorre2015,Tenno2015,Strongcorrelation3,Strongcorrelation6,SWZhang,ReiherDMRG,Strongcorrelation2,Cui2020,Piecuch2021} is said to occur in systems with many low-lying excited states, rendering their ground states 
quasidegenerate. Not only are they difficult to characterize theoretically and therefore a 
worthy computational challenge,\cite{Strongcorr2017,PhysRevX2020} but they are also technologically important, serving 
as a basis of useful materials whose structures and properties can change dramatically
upon external stimuli. Electronic excitations, phase transitions, and crossovers underlying these large, abrupt
changes may also occur thermally, giving rise to such fascinating phenomena as 
Mott transitions,\cite{Mott2,Mott1} Peierls distortion,\cite{Peierls1,Peierls2,highTc1} and high-$T_\text{c}$ superconductivity.\cite{Mott1,highTc1,Mott2,Kittel,Tinkham,Kresin}

It can be imagined that in such systems, thermally averaged properties are more accurately computable
than are the zero-temperature properties of a minimum-energy state, whose mean-field wave function tends to be unstable.\cite{Overhauser1,Overhauser2,Cizekstability,Yamada} 
It may even be argued that a finite-temperature treatment is more realistic and meaningful for systems whose very attractiveness derives from 
its large response to perturbations including temperature variations. For these and other reasons, there has been a surge of interest
in finite-temperature electron-correlation theories recently.\cite{sanyal,Mandal2,mandal,Shepherd,Zgid2,Zgid,WhiteChan2,Harsha1,Harsha2,Rubenstein2020,Harsha2020,Maestro2020,Chan2020,Malone2020} 

Despite being a cornerstone of finite-temperature electron-correlation  theories,\cite{march,thouless1972quantum,mattuck1992guide,Fetter}  many-body perturbation theory (MBPT) in the grand canonical ensemble\cite{matsubara,bloch,kohn,luttingerward,balian,blochbook,SANTRA} 
has not been fully developed, having gaps in its details: The grand potentials and chemical potentials are treated
differently\cite{JhaHirata} with no analytical formulas available for the lowest-order corrections to the chemical potential,
internal energy, or entropy for a long time.\cite{Hirata1storder,Hirata2ndorder} The time-dependent, diagrammatic formulation exploiting the isomorphism of the Schr\"{o}dinger and Bloch equations 
is elegant at the outset,\cite{matsubara,bloch,kohn,luttingerward,balian,blochbook,SANTRA}  but it becomes quickly inscrutable with the exhaustiveness of 
diagram enumeration being uncertain.  
Convergence of the finite-temperature MBPT to the well-established zero-temperature MBPT 
has been suspect,\cite{kohn,luttingerward,Hirata_PRA} compelling Kohn and Luttinger\cite{kohn} to conclude at one point that the theory
``is in general not correct.''\cite{kohn} Convergence towards the exact, i.e., full-configuration-interaction (FCI) limit at finite temperature\cite{Kou} 
is also unverifiable either numerically or analytically because of the lack of a Rayleigh--Schr\"{o}dinger-type recursion.
The relationship between the widely used finite-temperature Hartree--Fock (HF) theory\cite{Mermin}
and finite-temperature MBPT is tenuous, and as a result, the physical meaning\cite{Hirata2017} of their orbital energies and thus (quasiparticle) energy 
bands is still unknown.\cite{Jain} 

We recently introduced\cite{Hirata1storder,Hirata2ndorder} a new finite-temperature MBPT in the grand canonical ensemble, which expands in power series the grand potential, chemical potential, internal energy,
and entropy on an equal footing. We obtained sum-over-states and sum-over-orbitals analytical formulas for the first- and second-order
perturbation corrections to these thermodynamic functions in a transparent algebraic derivation 
involving only the combinatorial identities and the energy sum rules of the Hirschfelder--Certain degenerate perturbation theory (HCPT).\cite{Hirschfelder}
Sum-over-states formulas in the canonical ensemble were also reported up to the third order.\cite{JhaHirata_canonical}
These formulas were verified
by the exact numerical agreement with the benchmark data computed 
as the $\lambda$-derivatives of the finite-temperature FCI (Ref.\ \onlinecite{Kou}) (where $\lambda$ is the dimensionless perturbation strength). 
Using these analytical formulas, we elucidated\cite{Hirata_PRA} the root cause of
the nonconvergence at the correct zero-temperature limit,\cite{kohn} which has to do with the nonanalytic nature of 
the Boltzmann factor and thus plagues most any finite-temperature MBPT.  

This article is a comprehensive account of this new finite-temperature MBPT that 
expands all thermodynamic functions in uniform perturbation series. We introduce algebraic recursions, in the style of the Rayleigh--Schr\"{o}dinger perturbation theory,\cite{shavitt}
generating sum-over-states analytical formulas at any arbitrary order for both grand canonical and canonical ensembles.
They are  implemented into general-order algorithms that compute the perturbation corrections at any arbitrary order and demonstrate 
their convergence towards the exact (i.e., finite-temperature FCI) limits.\cite{Kou} 
For the grand canonical ensemble, we introduce systematic methods of converting them 
into sum-over-orbitals reduced analytical formulas, which will be useful
for actual condensed-matter applications. One of the conversion methods is the normal-ordered second quantization,\cite{march,sanyal}
which is fully developed in this study, and the other is the Feynman diagrams,\cite{bloch,balian,blochbook,mattuck1992guide} the rules of which are  
stipulated not just for the grand potential, but also for the chemical potential and entropy. 
The second-quantization rules shed light on the relationship between finite-temperature HF 
and first-order MBPT. 

To derive the correct reduced analytical formulas, it has been found that the perturbation energies entering the recursions have to be treated
as a nondiagonal matrix (not as scalars). 
Off-diagonal elements of this matrix, which are in general nonzero within each subspace spanned by degenerate Slater determinants, are shown to give rise 
to the renormalization terms\cite{shavitt} of the Rayleigh--Schr\"{o}dinger perturbation equations 
that are not necessarily unlinked. They are responsible for a unique set of linked diagrams
whose resolvent lines (i.e., factors of the energy denominators) are shifted, but not erased (the existence of `anomalous' diagrams\cite{kohn,SANTRA} with one or more resolvent lines erased is well known).
Since such `renormalization' diagrams originate from the degenerate perturbation theory and occur for the first time at the third order,
it is unclear whether the time-dependent, diagrammatic derivation\cite{matsubara,bloch,kohn,luttingerward,balian,blochbook}
or density-matrix formulation\cite{SANTRA} correctly take them into account. 

We furthermore prove the linked-diagram theorem of the finite-temperature MBPT, 
which asserts that all thermodynamic functions 
in the grand canonical ensemble are diagrammatically linked and therefore size-consistent (size-extensive)\cite{HirataTCA} at any perturbation order. 
The proof is based on the zero-temperature linked-diagram theorem\cite{GellmannLow,brueckner,goldstone,hugenholtz,Frantz,Manne} (see also Refs.\ \onlinecite{shavitt,Harris,Hirata2017}) and 
the systematic cancellation of unlinked anomalous diagrams. Finally, we document the numerical results of the perturbation corrections to the thermodynamic 
functions in both ensembles in a wide range of temperatures.  

\section{Algebraic recursive definitions\label{sec:recursions}}

In this section are presented the Rayleigh--Schr\"{o}dinger-like algebraic recursion relationships 
of the sum-over-states analytical formulas for the perturbation corrections to all thermodynamic functions for electrons in the grand canonical ensemble. 
Those for the canonical ensemble are readily inferred from 
these by restricting the summations to $\bar{N}$-electron
states and setting all chemical potentials to zero. They are relegated to Appendix \ref{app:canonical}.  
The sum-over-states analytical formulas have limited practical utility except to
produce benchmark data. They nevertheless serve as a mathematical basis of second-quantized and diagrammatic
derivations of reduced (sum-over-orbitals) analytical formulas expressed in terms of molecular integrals, which are the foci of 
Secs.\ \ref{sec:second} and \ref{sec:diagrams}. They also lead to 
the linked-diagram theorem proven in Sec.\ \ref{sec:linkeddiagramtheorem}.
Translational, rotational, and vibrational degrees of freedom are suppressed.

\subsection{Grand canonical ensemble} 

The grand partition function for electrons is given by\cite{Kou}
\begin{eqnarray}
\Xi &=& \sum_I e^{-\beta E_I + \beta \mu N_I},
\end{eqnarray}
where $I$ runs over all states, $E_I$ is the exact (FCI)
energy of the $I$th state, $N_I$ is the number of electrons of the same state, and $\beta = (k_\text{B}T)^{-1}$ ($k_\text{B}$ is the Boltzmann constant and $T$ is the temperature).
The exact chemical potential, $\mu$, is the one that keeps the average number of electrons at $\bar{N}$, the value ensuring the electroneutrality of the system. 
It is the root of the following equation:
\begin{eqnarray}
\bar{N} &=& \frac{1}{\beta} \frac{\partial}{\partial \mu} \ln \Xi.
\end{eqnarray}
The exact thermodynamic functions such as grand potential ($\Omega$), internal energy ($U$), and entropy ($S$) are derived from $\Xi$.
\begin{eqnarray}
\Omega &=& -\frac{1}{\beta} \ln \Xi, \\
U&=& -\frac{\partial}{\partial \beta} \ln \Xi  + \mu\bar{N}, \label{U} \\
S &=& k_\text{B}\beta^2 \frac{\partial \Omega}{\partial \beta}. \label{S}
\end{eqnarray}
They bear the following well-known relationship:
\begin{eqnarray}
\Omega &=& U -\mu \bar{N}- TS.
\end{eqnarray}

\subsection{Recursion for $\Omega^{(n)}$} 

The $n$th-order perturbation correction to any thermodynamic function $X$ ($X = \Xi$, $\mu$, $\Omega$, $U$, or $S$) is given by\cite{Hirata2017,JhaHirata}
\begin{eqnarray}
X^{(n)} = \frac{1}{n!} \left. \frac{\partial^n X(\lambda)}{\partial \lambda^n} \right|_{\lambda = 0} , \label{perturbationtheory}
\end{eqnarray}
where $X(\lambda)$ is defined exactly by the finite-temperature FCI (Ref.\ \onlinecite{Kou}) with a perturbation-scaled Hamiltonian, $\hat{H} = \hat{H}_0 + \lambda \hat{V}$, with 
$\lambda$ being the dimensionless perturbation strength. 

Using the Taylor expansions of exponential and logarithm,
\begin{eqnarray}
e^{a+b} &=& e^a + b e^a + \frac{b^2}{2!} e^a + \frac{b^3}{3!} e^a + \dots, \label{exp} \\
\ln (a+b) &=& \ln a + \frac{b}{a} - \frac{b^2}{2a^2} + \frac{b^3}{3a^3} - \dots, \label{log}
\end{eqnarray}
which are rapidly convergent when $a \gg b$, and defining
\begin{eqnarray}
D_I^{(n)} = E_I^{(n)} - \mu^{(n)} N_I
\end{eqnarray}
to minimize clutter, we obtain the recursion for $\Xi^{(n)}$ ($n \geq 1$) as
\begin{eqnarray}
\frac{\Xi^{(n)}}{\Xi^{(0)}} &=& ( -\beta) \langle D_I^{(n)} \rangle + \frac{(-\beta)^2}{2!} \sum_{i=1}^{n-1} \langle D_I^{(i)}D_I^{(n-i)} \rangle 
\nonumber\\&&
+ \frac{(-\beta)^3}{3!} \sum_{i=1}^{n-2}\sum_{j=1}^{n-i-1} \langle D_I^{(i)}D_I^{(j)} D_I^{(n-i-j)} \rangle 
\nonumber\\ &&
+ \dots + \frac{(-\beta)^n}{n!}  \langle (D_I^{(1)})^n\rangle, \label{recursionXi}
\end{eqnarray}
where the bracket denotes a zeroth-order thermal average:
\begin{eqnarray}
\langle X_I \rangle \equiv \frac{\sum_I X_I e^{-\beta E_I^{(0)}+\beta\mu^{(0)}N_I}}{\sum_I e^{-\beta E_I^{(0)}+\beta\mu^{(0)}N_I}}
=\frac{\sum_I X_I e^{-\beta D_I^{(0)}}}{\sum_I e^{-\beta D_I^{(0)}}} . \label{X}
\end{eqnarray}
State index $I$ is shown explicitly in the bracket to distinguish it from other state indexes that can coappear in
the summand, as in Eq.\ (\ref{projector}). A double bracket denotes a thermal average of a Dirac bracket; see, e.g., Eq.\ (\ref{double}). 
In Eq.\ (\ref{VRVRVRV}), a Brueckner bracket is represented by the same symbol, but this is appropriate since it too stands for a thermal average.

Postponing until Sec.\ \ref{sec:HCPT} the important discussion of $E^{(n)}_I$, we can use Eq.\ (\ref{log}) to arrive 
at the following recursion for $\Omega^{(n)}$, which in turn depends on the above recursion for $\Xi^{(n)}$:
\begin{eqnarray}
-\beta \Omega^{(n)} &=& \frac{\Xi^{(n)}}{\Xi^{(0)}} - \frac{1}{2} \sum_{i=1}^{n-1} \frac{\Xi^{(i)}}{\Xi^{(0)}}\frac{\Xi^{(n-i)}}{\Xi^{(0)}} 
\nonumber\\&&
+ \frac{1}{3} \sum_{i=1}^{n-2}\sum_{j=1}^{n-i-1} \frac{\Xi^{(i)}}{\Xi^{(0)}}\frac{\Xi^{(j)}}{\Xi^{(0)}}\frac{\Xi^{(n-i-j)}}{\Xi^{(0)}}
\nonumber\\&&
- \dots - \frac{(-1)^n}{n} \left( \frac{\Xi^{(1)}}{\Xi^{(0)}} \right)^n \label{recursionOmega0}
\end{eqnarray}
for $n \geq 1$.
Although nontrivial, we can rewrite the above into
\begin{eqnarray}
-\beta \Omega^{(n)} &=& \frac{\Xi^{(n)}}{\Xi^{(0)}} - \frac{(-\beta)^2}{2!} \sum_{i=1}^{n-1}  \Omega^{(i)} \Omega^{(n-i)}
\nonumber\\&&
- \frac{(-\beta)^3}{3!} \sum_{i=1}^{n-2}\sum_{j=1}^{n-i-1}  \Omega^{(i)} \Omega^{(j)} \Omega^{(n-i-j)} 
\nonumber\\&&
- \dots - \frac{(-\beta)^n}{n!} ( \Omega^{(1)})^n. \label{recursionOmega}
\end{eqnarray} 
A proof of the equivalence of Eqs.\ (\ref{recursionOmega0}) and (\ref{recursionOmega}) is given in Appendix \ref{app:StirlingS2}.
This is further simplified to a form that does not depend on $\Xi^{(n)}$:
\begin{eqnarray}
\Omega^{(n)} &=& \langle D_I^{(n)} \rangle + \frac{(-\beta)}{2!} \sum_{i=1}^{n-1} \left( \langle D_I^{(i)}D_I^{(n-i)} \rangle - \Omega^{(i)}\Omega^{(n-i)} \right) 
\nonumber\\&& 
+ \frac{(-\beta)^2}{3!} \sum_{i=1}^{n-2}\sum_{j=1}^{n-i-1} \left( \langle D_I^{(i)}D_I^{(j)} D_I^{(n-i-j)} \rangle - \Omega^{(i)}\Omega^{(j)} \Omega^{(n-i-j)}\right) \nonumber\\&&
+ \dots + \frac{(-\beta)^{n-1}}{n!}  \left\{ \langle (D_I^{(1)})^n\rangle - (\Omega^{(1)})^n  \right\}  \label{recursionOmega2}
\end{eqnarray}
for $n \geq 1$.

Next, we describe how $\langle E_I^{(n)} \rangle$, $\langle E_I^{(i)}E_I^{(n-i)} \rangle$, etc., can be evaluated so that 
the above recursion for $\Omega^{(n)}$ can be executed. See Sec.\ \ref{sec:mu} for the recursion for $\mu^{(n)}$, which is also needed. 
The above recursion is tantalizing with the linked-diagram theorem of $\Omega^{(n)}$, which will be discussed in Sec.\ \ref{sec:linkeddiagramtheorem}. 

\subsection{Recursion for $\langle E_I^{(n)}\rangle$ \label{sec:HCPT}} 

The perturbation corrections to energy, $E_I^{(i)}$ ($1 \leq i \leq n$), and their thermal averages, $\langle E_I^{(i)} \rangle$, are crucial quantities entering the recursion for $\Omega^{(n)}$. They are identified as 
those defined by the Hirschfelder--Certain degenerate perturbation theory (HCPT).\cite{Hirschfelder} The more familiar M\o ller--Plesset perturbation theory (MPPT)\cite{moller}
will not suffice here because a finite-temperature MBPT accesses all states, many of whose zeroth-order energies are exactly degenerate. 
Other degenerate or quasidegenerate perturbation theories may also fall short because HCPT is the proper Rayleigh--Schr\"{o}dinger perturbation theory 
for degenerate and nondegenerate reference states, complying with the canonical definition of perturbation theory, i.e., Eq.\ (\ref{perturbationtheory}),
although there are equivalent degenerate perturbation theories under different names.\cite{klein} 

For a nondegenerate reference state,\cite{shavitt} HCPT reduces to MPPT.
The Schr\"{o}dinger equation may be written with perturbatively expanded energy and wave function as
\begin{eqnarray}
&& ( \hat{H}_0 + \lambda \hat{V} ) (\Phi_I^{(0)} + \lambda \Phi_I^{(1)} + \lambda^2 \Phi_I^{(2)} + \dots) \nonumber\\
&& = (E_I^{(0)} + \lambda E_I^{(1)} + \lambda^2 E_I^{(2)} + \dots)(\Phi_I^{(0)} + \lambda \Phi_I^{(1)} + \lambda^2 \Phi_I^{(2)} + \dots) \nonumber\\ \label{PT}
\end{eqnarray}
with intermediate normalization $\langle \Phi_I^{(0)} | \Phi_I^{(n)} \rangle = \delta_{n0}$, where $\Phi_I^{(0)}$ is the nondegenerate reference 
(zeroth-order) wave function of the $I$th state, which is a single Slater determinant. Expanding and equating terms carrying the $n$th power of $\lambda$, 
we obtain the recursion for the $n$th-order correction to the wave function, $\Phi_I^{(n)}$, as
\begin{eqnarray}
(E_I^{(0)} - \hat{H}_0) \Phi_I^{(n)} &=& \hat{V} \Phi_I^{(n-1)} - \sum_{i=1}^{n} E_I^{(i)} \Phi_I^{(n-i)} \label{MPPTrec1}
\end{eqnarray}
or
\begin{eqnarray}
\Phi_I^{(n)} &=& \hat{R}_I \left( \hat{V} \Phi_I^{(n-1)} - \sum_{i=1}^{n-1} E_I^{(i)} \Phi_I^{(n-i)} \right), \label{MPPTrec2}
\end{eqnarray}
where $\hat{R}_I$ is the resolvent operator:
\begin{eqnarray}
\hat{R}_I = \sum_{A}^{\text{denom.}\neq 0} \frac{| \Phi_A^{(0)} \rangle \langle \Phi_A^{(0)} |}{E_I^{(0)} - E_A^{(0)}}. \label{resolvent}
\end{eqnarray} 
Here, ``denom.$\neq0$'' limits the summation to just over the states $\Phi_A^{(0)}$ that are not degenerate with the $I$th state,
and therefore to the cases where $E_I^{(0)} - E_A^{(0)} \neq 0$. The second term in the right-hand side of Eq.\ (\ref{MPPTrec1}) 
is known as the renormalization term,\cite{shavitt} which is entirely diagrammatically unlinked (non-size-consistent) in MPPT. It vanishes exactly 
by cancelling out the unlinked contribution of the same magnitude in the first term, leaving only the linked (size-consistent) contribution as 
the perturbation correction to the wave function. Multiplying Eq.\ (\ref{MPPTrec1}) by $\Phi_I^{(0)*}$ from left and integrating, we have
\begin{eqnarray}
E_I^{(n)} &=& \langle \Phi_I^{(0)} | \hat{V} | \Phi_I^{(n-1)} \rangle,
\end{eqnarray} 
defining the $n$th-order energy correction according to MPPT.

For $M$-tuply degenerate reference states ($M \geq 2$),\cite{Hirschfelder} HCPT and MPPT differ from each other materially. 
The perturbation expansion of the Schr\"{o}dinger equation now becomes a $M$-by-$M$ matrix equation of the form,
\begin{widetext}
\begin{eqnarray}
&& ( \hat{H}_0 + \lambda \hat{V} ) \left\{ \left( \begin{array}{c} \Phi_1^{(0)} \\ \vdots \\ \Phi_M^{(0)} \end{array}\right) 
+ \lambda\left( \begin{array}{c} \Phi_1^{(1)} \\ \vdots \\ \Phi_M^{(1)} \end{array}\right)   
+ \lambda^2 \left( \begin{array}{c} \Phi_1^{(2)} \\ \vdots \\ \Phi_M^{(2)} \end{array}\right)  + \dots\right\}  
\nonumber\\&& 
= \left\{ \left( \begin{array}{ccc} E_{11}^{(0)} & \cdots & E_{M1}^{(0)} \\ \vdots && \vdots \\ E_{1M}^{(0)} & \cdots & E_{MM}^{(0)} \end{array}\right) 
+ \lambda\left( \begin{array}{ccc} E_{11}^{(1)} & \cdots & E_{M1}^{(1)} \\ \vdots && \vdots \\ E_{1M}^{(1)} & \cdots & E_{MM}^{(1)}\end{array}\right)   
+ \lambda^2 \left( \begin{array}{ccc} E_{11}^{(2)} & \cdots & E_{M1}^{(2)} \\ \vdots && \vdots \\ E_{1M}^{(2)} & \cdots & E_{MM}^{(2)} \end{array}\right)  + \dots\right\}
\left\{ \left( \begin{array}{c} \Phi_1^{(0)} \\ \vdots \\ \Phi_M^{(0)} \end{array}\right) 
+ \lambda\left( \begin{array}{c} \Phi_1^{(1)}  \\ \vdots \\ \Phi_M^{(1)} \end{array}\right)   
+ \lambda^2 \left( \begin{array}{c} \Phi_1^{(2)}  \\ \vdots \\ \Phi_M^{(2)} \end{array}\right)  + \dots\right\} , \nonumber \\ \label{DPT} 
\end{eqnarray}
where all $M$ zeroth-order states, $\{\Phi_I^{(0)}\}$ ($1 \leq I \leq M$), are degenerate, and therefore $E_{IJ}^{(0)} = E_I^{(0)}\delta_{IJ}$. 
Collecting terms that are proportional to $\lambda^n$, we obtain
\begin{eqnarray}
\left( \begin{array}{c} (E_I^{(0)}- \hat{H}_0 ) \Phi_1^{(n)} \\ \vdots \\ (E_I^{(0)}- \hat{H}_0 ) \Phi_M^{(n)} \end{array}\right) 
&=&  \left( \begin{array}{c} \hat{V} \Phi_1^{(n-1)} \\ \vdots \\ \hat{V}  \Phi_M^{(n-1)} \end{array}\right)   
- \sum_{i=1}^{n} \left( \begin{array}{ccc} E_{11}^{(i)} & \cdots & E_{M1}^{(i)} \\ \vdots && \vdots \\ E_{1M}^{(i)} & \cdots & E_{MM}^{(i)} \end{array}\right)  
\left( \begin{array}{c} \Phi_1^{(n-i)} \\ \vdots \\ \Phi_M^{(n-i)} \end{array}\right)  \label{HCPTrec1}
\end{eqnarray}
or
\begin{eqnarray}
\left( \begin{array}{c} \Phi_1^{(n)} \\ \vdots \\ \Phi_M^{(n)} \end{array}\right) 
&=&  \hat{R}_I \left\{ \left( \begin{array}{c} \hat{V} \Phi_1^{(n-1)} \\ \vdots \\ \hat{V}  \Phi_M^{(n-1)} \end{array}\right)   
- \sum_{i=1}^{n-1} \left( \begin{array}{ccc} E_{11}^{(i)} & \cdots & E_{M1}^{(i)} \\ \vdots && \vdots \\ E_{1M}^{(i)} & \cdots & E_{MM}^{(i)} \end{array}\right)  
\left( \begin{array}{c} \Phi_1^{(n-i)} \\ \vdots \\ \Phi_M^{(n-i)} \end{array}\right) \right\} , \label{HCPTrec2}
\end{eqnarray}
where the definition of $\hat{R}_I$ remains unchanged [Eq.\ (\ref{resolvent})]. 
Multiplying the $J$th row of Eq.\ (\ref{HCPTrec1}) by $\Phi_I^{(0)*}$ from the left and integrating for all $I$ and $J$ ($1 \leq I, J \leq M$), we have for $n \geq 1$,
\begin{eqnarray}
\bm{E}^{(n)} \equiv \left( \begin{array}{ccc} E_{11}^{(n)} & \cdots & E_{1M}^{(n)} \\ \vdots && \vdots \\ E_{M1}^{(n)} & \cdots & E_{MM}^{(n)} \end{array}\right)  
= \left( \begin{array}{ccc} \langle \Phi_1^{(0)} |\hat{V}| \Phi_1^{(n-1)}\rangle & \cdots & \langle \Phi_1^{(0)} |\hat{V}| \Phi_M^{(n-1)}\rangle
\\ \vdots && \vdots 
 \\ \langle \Phi_M^{(0)} |\hat{V}| \Phi_1^{(n-1)}\rangle & \cdots & \langle \Phi_M^{(0)} |\hat{V}| \Phi_M^{(n-1)}\rangle \end{array}\right), \label{HCPTrec3}
\end{eqnarray}
\end{widetext}
which is Hermitian.\cite{Hirschfelder}
Owing to the degeneracy, $\Phi_1^{(0)}$ through $\Phi_M^{(0)}$ can mix with one another and remain as valid zeroth-order references insofar as 
they are orthonormal. We  seek a unique (up to a phase) set of the zeroth-order wave functions that brings
$\bm{E}^{(1)}$ through $\bm{E}^{(n)}$ into a diagonal form, so that the above $M$-fold coupled Schr\"{o}dinger equation [Eq.\ (\ref{DPT})] 
is separated into $M$ independent Schr\"{o}dinger equations of the form of Eq.\ (\ref{PT}). 
The eigenvalues of $\bm{E}^{(n)}$ are then identified as the $n$th-order HCPT energy corrections,\cite{Hirschfelder}
although the foregoing formulation is considerably simpler than that of Ref.\ \onlinecite{Hirschfelder}. Given above 
is an abridged version that has just enough details to make the recursion for $\Omega^{(n)}$ self-contained.

At this point, it may appear hopeless to try to derive compact analytical formulas of $\langle E_I^{(n)} \rangle$ written in terms of molecular integrals because 
eigenvalues are only procedurally defined and cannot be written in a closed form in general (for $M \geq 5$
according to the Abel--Ruffini theorem). Fortunately, however, we can still derive analytical formulas without knowing the eigenvalues.\cite{Hirata1storder,Hirata2ndorder} 
Since these energy corrections are  thermally averaged [Eq.\ (\ref{X})] 
with the same Boltzmann weight within each degenerate subspace, we only need the sum of the eigenvalues
(not the individual eigenvalues) in order to calculate the thermal average correctly. This sum is equal to the sum of diagonal elements owing to the trace invariance
for a cyclic permutation of a matrix product,
\begin{eqnarray} 
\text{Tr}\, \bm{U}^{-1}\bm{E}^{(i)}\bm{U} &=& \text{Tr}\, \bm{E}^{(i)}, \\
\text{Tr}\, \bm{U}^{-1}\bm{E}^{(i)}\bm{U} \bm{U}^{-1}\bm{E}^{(j)}\bm{U}  &=& \text{Tr}\, \bm{E}^{(i)}\bm{E}^{(j)}, 
\end{eqnarray}
etc., where $\bm{U}$ is the unitary matrix that brings 
all of $\bm{E}^{(i)}$ ($1 \leq i \leq n$) into a diagonal form. 
Each matrix element (before diagonalization) can be written in
a closed form, lending the trace and thus thermal average to analytical expressions.
In our previous studies,\cite{Hirata1storder,Hirata2ndorder} we used the Slater--Condon rules\cite{szabo}
to evaluate these matrix elements and derived compact analytical formulas for the thermal averages 
with custom-made combinatorial identities.
Such an order-by-order approach reaches an impasse at higher orders, and hence, in this study, 
we switch to the second-quantized and diagrammatic formulations expounded on in the subsequent sections. 
For the sum-over-states recursion and general-order algorithm based on it, we can still diagonalize $\bm{E}^{(i)}$ and use 
its eigenvalues in principle, but we elect to leave $\bm{E}^{(i)}$ undiagonalized because the diagonalization step is simply superfluous.
 
The foregoing argument must not be misconstrued to mean that only diagonal elements of $\bm{E}^{(i)}$ matter. 
As we demonstrate in Sec.\ \ref{sec:second}, off-diagonal elements of $\bm{E}^{(i)}$ enter thermal averages via the renormalization term
[the second term of Eq.\ (\ref{HCPTrec1}) or (\ref{HCPTrec2})], giving rise to unique diagrams---the renormalization diagrams with displaced resolvent lines 
(distinguished from the anomalous diagrams with missing resolvents) appearing for the first time at the third order.

 To summarize, $\langle E^{(n)} \rangle$, $\langle E^{(i)}E^{(n-i)} \rangle$, etc.\ are evaluated by thermally averaging the traces of 
 $\bm{E}^{(i)}$, $\bm{E}^{(i)}\bm{E}^{(n-i)}$, etc.\ using the energy recursion [Eq.\ (\ref{HCPTrec3})], which in turn depends on the recursion
 for wave functions [Eq.\ (\ref{HCPTrec2})]. 
 In other words, for the purpose of executing the recursions, we may use the following substitutions,
 where $E_I^{(i)}$ is the $I$th eigenvalue, whereas $E_{IJ}^{(i)}$ is the $IJ$th element of the matrix $\bm{E}^{(i)}$:
 \begin{eqnarray}
 \langle E_I^{(i)} \rangle &=& \langle E_{II}^{(i)} \rangle, \\
  \langle E_I^{(i)} E_I^{(j)} \rangle &=& \langle E_{IJ}^{(i)} \hat{P}_I E_{JI}^{(j)} \rangle, \label{projector} 
 \end{eqnarray}
 etc., where the inner projector, $\hat{P}_I$, restricts the summation over $J$ (whose summation symbol is suppressed according to Einstein's convention) to determinants
 within the degenerate subspace of $I$. It is not only advantageous, but also necessary to treat the energy corrections
that enter the recursions as a nondiagonal matrix rather than scalars. 

\subsection{Recursion for $\mu^{(n)}$\label{sec:mu}} 

The recursion for $\mu^{(n)}$ comes from\cite{Hirata2ndorder}
\begin{eqnarray}
\left( \frac{\partial \Omega^{(n)}}{\partial \mu^{(0)}}\right)_{\mu^{(1)},\dots,\,\mu^{(n)}} = 0. \label{mu0deriv}
\end{eqnarray}
Substituting this to Eq.\ (\ref{recursionOmega}), we have
\begin{eqnarray}
 \frac{\partial}{\partial \mu^{(0)}}\frac{\Xi^{(n)}}{\Xi^{(0)}} = 0.
\end{eqnarray}
Using the recursion for $\Xi^{(n)}/\Xi^{(0)}$ [Eq.\ (\ref{recursionXi})] in the above, we arrive at the following recursion
for $\mu^{(n)}$ ($n \geq 1$): 
\begin{eqnarray}
 \mu^{(n)} \langle N_I^2 -\bar{N}^2 \rangle
&=& \langle E_I^{(n)} (N_I-\bar{N}) \rangle  
\nonumber\\ &&
+ \frac{(-\beta)}{2!} \sum_{i=1}^{n-1} \langle D_I^{(i)}D_I^{(n-i)} (N_I-\bar{N}) \rangle 
\nonumber\\&&
+ \frac{(-\beta)^2}{3!} \sum_{i=1}^{n-2}\sum_{j=1}^{n-i-1} \langle D_I^{(i)}D_I^{(j)} D_I^{(n-i-j)} (N_I-\bar{N}) \rangle 
\nonumber\\&&
+ \dots + \frac{(-\beta)^{n-1}}{n!}  \langle (D_I^{(1)})^n (N_I-\bar{N})\rangle,  \label{recursionmu}
\end{eqnarray}
where we used the identity,
\begin{eqnarray}
 \frac{\partial}{\partial \mu^{(0)}}\langle X_I \rangle = \langle X_IN_I \rangle - \langle X_I \rangle \langle N_I \rangle. \label{XN}
\end{eqnarray}
Its application to the first term of Eq.\ (\ref{recursionXi}) leads to
\begin{eqnarray} 
\frac{\partial}{\partial \mu^{(0)}}\langle D_I^{(n)} \rangle 
&=& \langle E_I^{(n)} (N_I-\bar{N}) \rangle   -\mu^{(n)} \langle N_I^2   - \bar{N}^2 \rangle  ,
\end{eqnarray}
which was also used in Eq.\ (\ref{recursionmu}). 

\subsection{Recursion for $U^{(n)}$} 

Differentiating Eq.\ (\ref{U}) $n$ times with $\lambda$, we obtain
\begin{eqnarray}
U^{(n)} - \mu^{(n)} \bar{N} =  \left( \frac{\partial(-\beta \Omega^{(n)})}{\partial (-\beta) }\right)_{\mu^{(0)},\dots,\,\mu^{(n)}}.
\end{eqnarray}
Substituting Eq.\ (\ref{recursionOmega}) into this, we arrive at
\begin{eqnarray}
U^{(n)}-\mu^{(n)} \bar{N}  &=& \frac{\partial}{\partial(- \beta)} \frac{\Xi^{(n)}}{\Xi^{(0)}}- (-\beta) \sum_{i=1}^{n-1} (U^{(i)}-\mu^{(i)}\bar{N}) \Omega^{(n-i)}
\nonumber\\&&
- \frac{(-\beta)^2}{2!} \sum_{i=1}^{n-2}\sum_{j=1}^{n-i-1} (U^{(i)} -\mu^{(i)}\bar{N}) \Omega^{(j)} \Omega^{(n-i-j)} 
\nonumber\\&& 
- \dots - \frac{(-\beta)^{n-1}}{(n-1)!} (U^{(1)} -\mu^{(1)}\bar{N}) (\Omega^{(1)})^{n-1}. \label{recursionUgc}
\end{eqnarray}
The first term can be expanded as
\begin{eqnarray}
\frac{\partial}{\partial (-\beta)} \frac{\Xi^{(n)}}{\Xi^{(0)}} &=& 
 \langle  D_I^{(n)} \rangle - ( -\beta) \langle  D_I^{(n)} \rangle \langle  D_I^{(0)} \rangle  
+ {(-\beta)}\sum_{i=1}^{n} \langle D_I^{(i)}D_I^{(n-i)} \rangle 
\nonumber\\&&
- \frac{(-\beta)^2}{2!} \sum_{i=1}^{n-1} \langle D_I^{(i)}D_I^{(n-i)} \rangle \langle  D_I^{(0)} \rangle
\nonumber\\&&
+ \frac{(-\beta)^2}{2!} \sum_{i=1}^{n-1}\sum_{j=1}^{n-i} \langle D_I^{(i)}D_I^{(j)} D_I^{(n-i-j)} \rangle \nonumber\\&&
\nonumber\\&&
- \frac{(-\beta)^3}{3!} \sum_{i=1}^{n-2}\sum_{j=1}^{n-i-1} \langle D_I^{(i)}D_I^{(j)} D_I^{(n-i-j)} \rangle  \langle  D_I^{(0)} \rangle 
\nonumber\\&&
+ \frac{(-\beta)^3}{3!} \sum_{i=1}^{n-2}\sum_{j=1}^{n-i-1}\sum_{k=1}^{n-i-j} \langle D_I^{(i)}D_I^{(j)} D_I^{(k)}    D_I^{(n-i-j-k)} \rangle 
\nonumber\\&&
- \dots - \frac{(-\beta)^n}{n!}  \langle (D_I^{(1)})^n\rangle \langle D_I^{(0)} \rangle 
\nonumber\\&&
+ \frac{(-\beta)^n}{n!}  \langle (D_I^{(1)})^n D_I^{(0)} \rangle, \label{recursionXibeta}
\end{eqnarray}
where we used Eq.\ (\ref{recursionXi}) and 
\begin{eqnarray}
\frac{\partial}{\partial (-\beta)} \langle X_I \rangle = \langle X_I D_I^{(0)} \rangle - \langle X_I \rangle \langle D_I^{(0)} \rangle .
\end{eqnarray}
Therefore, the recursion for $U^{(n)}$ ($n \geq 1$) is given by
\begin{eqnarray}
U^{(n)}  &=&
 \langle  E_I^{(n)} \rangle + {(-\beta)}\sum_{i=1}^{n} \langle D_I^{(i)}D_I^{(n-i)} \rangle 
 \nonumber\\&&
- (-\beta) \sum_{i=1}^{n} \Omega^{(i)} (U^{(n-i)}-\mu^{(n-i)}\bar{N})  
 \nonumber\\&&
+ \frac{(-\beta)^2}{2!} \sum_{i=1}^{n-1}\sum_{j=1}^{n-i} \langle D_I^{(i)}D_I^{(j)} D_I^{(n-i-j)} \rangle 
\nonumber\\&&
- \frac{(-\beta)^2}{2!} \sum_{i=1}^{n-1}\sum_{j=1}^{n-i} \Omega^{(i)} \Omega^{(j)} (U^{(n-i-j)} -\mu^{(n-i-j)}\bar{N}) 
\nonumber\\&&
+ \dots 
+ \frac{(-\beta)^n}{n!}  \langle (D_I^{(1)})^n D_I^{(0)} \rangle 
\nonumber\\&&
- \frac{(-\beta)^n}{n!}   (\Omega^{(1)})^n (U^{(0)}-\mu^{(0)}\bar{N}) , \label{recursionUgc2}
\end{eqnarray}
where we used $\langle D_I^{(0)} \rangle = U^{(0)}-\mu^{(0)}\bar{N}$.

\subsection{Recursion for $S^{(n)}$} 

The recursion for $S^{(n)}$ is the concatenation of the recursions for $\Omega^{(n)}$, $\mu^{(n)}$, and $U^{(n)}$ given above
because
\begin{eqnarray}
S^{(n)} = k_\text{B}\beta \left( U^{(n)} - \mu^{(n)} \bar{N} - \Omega^{(n)} \right).
\end{eqnarray}

\section{Second-quantized derivation\label{sec:second}}

In this section, we derive the reduced analytical formulas of $\Omega^{(n)}$ and $\mu^{(n)}$ for the few lowest $n$'s,\cite{Hirata1storder,Hirata2ndorder} starting from 
the sum-over-states analytical formulas obtained from the recursions. We rely on the rules of normal-ordered second quantization
at finite temperature,\cite{march,sanyal} expounded on in Appendix \ref{app:secondquantization}.
Second-quantized derivations of $U^{(n)}$ and $S^{(n)}$ follow essentially the same procedure and will not be repeated.

\subsection{Zeroth order}

The zeroth-order formulas initiate the recursions.
They correspond to the Fermi--Dirac theory discussed in many textbooks.\cite{march,thouless1972quantum,mattuck1992guide,Fetter} 
We therefore only document the results.\cite{Kou,Hirata1storder,Hirata2ndorder}
\begin{eqnarray}
\Omega^{(0)} &=& E_\text{nuc.} + \frac{1}{\beta} \sum_p \ln f_p^+, \\
U^{(0)} &=& E_\text{nuc.} + \sum_p \epsilon_p f_p^-, \\
S^{(0)} &=& -k_\text{B} \sum_p \left( f_p^- \ln f_p^- + f_p^+ \ln f_p^+ \right),
\end{eqnarray}
where $E_\text{nuc.}$ is the nucleus-nucleus repulsion energy, and 
$f_p^-$ ($f_p^+$) is the Fermi--Dirac occupancy (vacancy) function for the $p$th spinorbital given by
\begin{eqnarray}
f_p^- = \frac{1}{1+e^{\beta(\epsilon_p - \mu^{(0)})}}, \label{occupancy} \\
f_p^+ = \frac{e^{\beta(\epsilon_p - \mu^{(0)})}}{1+e^{\beta(\epsilon_p - \mu^{(0)})}}. \label{vacancy}
\end{eqnarray}
Lastly, $\mu^{(0)}$ is the root of the following equation embodying the zeroth-order electroneutrality condition:
\begin{eqnarray}
\bar{N} = \sum_p f_p^-. \label{Nbar}
\end{eqnarray} 

\subsection{First order}

According to the recursion for $\Omega^{(n)}$ [Eq.\ (\ref{recursionOmega2})], using Einstein's convention
inside thermal averages $\langle \dots \rangle$, we can write $\Omega^{(1)}$ as
\begin{eqnarray}
\Omega^{(1)} &=& \langle D_I^{(1)} \rangle 
= \langle \langle I | \hat{V} |I \rangle \rangle -  \mu^{(1)} \langle \langle I | \hat{N} |I \rangle \rangle, \label{double}
\end{eqnarray}
where $\langle \langle I | \hat{V} |I \rangle \rangle$ is the thermal average of $\langle I | \hat{V} |I \rangle$.
Consulting with Appendix \ref{app:secondquantization}, we can evaluate this using the normal-ordered second quantization as
\begin{eqnarray}
\Omega^{(1)} &=& \langle \langle I | \langle E^{(1)}_I\rangle + F_{pq} \{ \hat{p}^\dagger \hat{q} \} 
+  (1/4) \langle pq || rs \rangle \{ \hat{p}^\dagger \hat{q}^\dagger \hat{s}\,\hat{r}\}  |I \rangle \rangle \nonumber\\
&&- \mu^{(1)} \langle\langle I |  \bar{N} + \{ \hat{p}^\dagger \hat{p} \} |I \rangle\rangle \label{survival} \\
&=& \langle E^{(1)}_I\rangle - \mu^{(1)} \bar{N} \label{Omega1constants} \\
&=& \sum_p F_{pp} f_p^- -\frac{1}{2} \sum_{p,q} \langle pq||pq \rangle f_p^- f_q^- - \mu^{(1)} \bar{N}  \label{Omega1reduced} \\
&\equiv& \langle D_I^{(1)} \rangle_{L}. \label{Omega1linked}
\end{eqnarray}
In Eq.\ (\ref{survival}), we substituted the normal-ordered perturbation operator [Eq.\ (\ref{V})] and number operator [Eq.\ (\ref{number})]
derived in Appendix \ref{app:Hamiltonian}. Going from Eq.\ (\ref{survival}) to (\ref{Omega1constants}), 
we invoked the rule that the thermal average of a normal-ordered product of operators is always zero, 
allowing only the constant terms in $\hat{V}$ and $\hat{N}$ to survive. 
In Eq.\ (\ref{Omega1reduced}), we referred to Eq.\ (\ref{e1}) for $\langle E^{(1)}_I\rangle$ and Eq.\ (\ref{Fock}) for $F_{pq}$, whereas 
$\langle pq || rs \rangle$ is an antisymmetrized two-electron integral, and $\mu^{(1)}$ will be considered below.
Therefore, as in the zero-temperature normal ordering,\cite{shavitt} an effort has been prepaid to derive the analytical formula for $\langle E^{(1)}_I\rangle$ 
during the derivation of the normal-ordered form of $\hat{V}$ (see Appendix \ref{app:Hamiltonian}). 
The end result [Eq.\ (\ref{Omega1reduced})] is recognized as diagrammatically linked and 
thus size-consistent,\cite{shavitt,HirataTCA} and can be symbolically written as Eq.\ (\ref{Omega1linked}) with subscript ``$L$'' standing for ``linked.'' 
Algebraically, a linked term is the one that is not a simple product of two or more extensive quantities ($\mu^{(n)}$, $f_p^\pm$, and $F_{pq}$
are intensive, while $\langle E_I^{(n)} \rangle$ and $\bar{N}$ are extensive). 

We start with the sum-over-states formula of $\mu^{(1)}$ [Eq.\ (\ref{recursionmu})], which reads 
\begin{eqnarray}
\mu^{(1)} \langle N_I^2 -\bar{N}^2 \rangle = \langle E_I^{(1)}(N_I - \bar{N}) \rangle . \label{mu1_sq}
\end{eqnarray}
The thermal average in the left-hand side is evaluated as
\begin{eqnarray}
\langle N_I^2 -\bar{N}^2 \rangle &=& \langle \langle I | \hat{N}^2 -\bar{N}^2 |I \rangle\rangle  \nonumber\\
&=& \langle\langle I | ( \bar{N} + \{ \hat{p}^\dagger \hat{p} \} ) (  \bar{N} + \{ \hat{q}^\dagger \hat{q} \} ) |I \rangle \rangle -\bar{N}^2 \nonumber  \\
&=& \bar{N}^2  +
 \contraction[0.5ex]{\langle\langle I| \{ \hat{p}^\dagger }{\hat{p}}{ \}  \{ }{\hat{q}}
\contraction[1.0ex]{ \langle\langle I| \{ }{\hat{p}}{^\dagger \hat{p} \}  \{ \hat{q}^\dagger }{\hat{q}}
 \langle\langle I| \{ \hat{p}^\dagger \hat{p} \}  \{ \hat{q}^\dagger \hat{q} \} |I\rangle \rangle 
 -\bar{N}^2 \nonumber \\
 &=& \sum_p f_p^- f_p^+ \equiv \langle N_I^2 -\bar{N}^2 \rangle_{L}, \label{NminusN}
 \end{eqnarray}
 which is linked owing to the cancellation of the unlinked terms, $\bar{N}^2$. 
 The right-hand side of Eq.\ (\ref{mu1_sq}) is evaluated as
\begin{eqnarray}
&& \langle E_I^{(1)} (N_I - \bar{N}) \rangle 
\nonumber\\&& 
= \langle E_{II}^{(1)} (N_I - \bar{N}) \rangle 
\nonumber \\&& 
= \langle \langle I | \hat{V} (\hat{N} - \bar{N}) |I \rangle \rangle \nonumber \\
&& =  \langle \langle I | (\langle E^{(1)}_I\rangle + F_{pq} \{ \hat{p}^\dagger \hat{q} \} 
+ (1/4) \langle pq || rs \rangle \{ \hat{p}^\dagger\hat{q}^\dagger \hat{s} \hat{r} \} )( \{ \hat{t}^\dagger \hat{t} \} ) | I\rangle \rangle \nonumber \\
&& =  \langle \langle I | F_{pq} 
\contraction[1.0ex]{\{ }{\hat{p}}{^\dagger \hat{q} \}  \{\hat{t}^\dagger }{\hat{t}} 
\contraction[0.5ex]{\{ \hat{p}^\dagger }{\hat{q}}{ \}  \{}{\hat{t}}
\{ \hat{p}^\dagger \hat{q} \} \{ \hat{t}^\dagger \hat{t} \} 
| I\rangle \rangle \nonumber \\
&&= \sum_{p} F_{pp} f_p^- f_p^+ \equiv \langle E_I^{(1)} (N_I - \bar{N}) \rangle_{L}, \label{E1N} 
\end{eqnarray}
which is also linked. 
Taken together, Eq.\ (\ref{mu1_sq}) can be symbolically rewritten as 
\begin{eqnarray}
\mu^{(1)} \langle N_I^2 -\bar{N}^2 \rangle_{L} = \langle E_I^{(1)}(N_I - \bar{N}) \rangle_{L}, \label{mu1_linked}
\end{eqnarray}
which reduces to 
\begin{eqnarray}
\mu^{(1)}  \sum_p f_p^- f_p^+ = \sum_{p} F_{pp} f_p^- f_p^+. \label{mu1reduced}
\end{eqnarray}

For both $\Omega^{(1)}$ and $\mu^{(1)}$, we recover the reduced analytical formulas reported
earlier.\cite{Hirata1storder,Hirata2ndorder}

\subsection{Second order} 

From the $\Omega^{(n)}$ recursion [Eq.\ (\ref{recursionOmega2})], we have
\begin{eqnarray}
\Omega^{(2)} &=& \langle D_I^{(2)} \rangle + \frac{(-\beta)}{2!} \left( \langle D_I^{(1)} D_I^{(1)} \rangle - \Omega^{(1)} \Omega^{(1)} \right),  \label{Omega2sq}
\end{eqnarray}
whose components are expanded by using the $E_I^{(n)}$ recursions [Eqs.\ (\ref{HCPTrec2}) and (\ref{HCPTrec3})] as
\begin{eqnarray}
\langle D_I^{(2)} \rangle &=& \langle E_I^{(2)} \rangle - \mu^{(2)} \langle N_I \rangle \nonumber\\
&=& \langle E_{II}^{(2)} \rangle - \mu^{(2)} \langle \langle I | \hat{N} | I \rangle \rangle \nonumber\\
&=& \langle \langle I | \hat{V} \hat{R}_I \hat{V} | I \rangle \rangle - \mu^{(2)} \bar{N} \label{D2sq} 
\end{eqnarray}
and
\begin{eqnarray}
\langle D_I^{(1)} D_I^{(1)} \rangle &=& \langle E_I^{(1)}E_I^{(1)} \rangle - 2 \mu^{(1)} \langle E_I^{(1)} N_I \rangle 
+ \left( \mu^{(1)} \right)^2 \langle N_I^2 \rangle \nonumber\\
&=& \langle E_{IJ}^{(1)}\hat{P}_I E_{JI}^{(1)} \rangle - 2 \mu^{(1)} \langle E_I^{(1)} (N_I - \bar{N}) \rangle  
\nonumber\\
&& - 2 \mu^{(1)} \langle E^{(1)}_I\rangle \bar{N} 
+ \left( \mu^{(1)} \right)^2 \langle N_I^2 - \bar{N}^2 \rangle +   \left( \mu^{(1)} \right)^2 \bar{N}^2 \nonumber \\
&=& \langle \langle I | \hat{V} \hat{P}_I \hat{V} |I \rangle \rangle - 2 \mu^{(1)} \sum_p F_{pp} f_p^- f_p^+  \nonumber\\
&& + \left( \mu^{(1)} \right)^2 \sum_p f_p^- f_p^+ - 2 \mu^{(1)} \langle E^{(1)}_I\rangle \bar{N} 
+ \left( \mu^{(1)} \right)^2 \bar{N}^2, \nonumber \\ \label{D1D1sq}
\end{eqnarray}
where Eqs.\ (\ref{NminusN}) and (\ref{E1N}) were used. 
The last two terms of Eq.\ (\ref{D1D1sq}) are unlinked, growing quadratically with size, which are, however, easily seen to be
cancelled by the corresponding contributions in the $\Omega^{(1)}\Omega^{(1)}$ term of Eq.\ (\ref{Omega2sq}). 

Let us evaluate $\langle E_I^{(2)} \rangle$ using the normal-ordered second-quantization rules for resolvent operators described in Appendix \ref{app:projector}. 
Using Einstein's convention of implied summations, we write
\begin{widetext}
\begin{eqnarray}
\langle E_I^{(2)} \rangle &=& \langle \langle I | \hat{V} \hat{R}_I \hat{V} | I \rangle \rangle \nonumber\\
&=& \langle \langle I | F_{pq} 
\contraction[1.0ex]{\{ }{\hat{p}}{^\dagger \hat{q} \} \{ \hat{a}^\dagger }{\hat{i}}
\contraction[0.5ex]{\{ \hat{p}^\dagger }{\hat{q}}{ \} \{ }{\hat{a}} 
\contraction[0.6ex]{\{ \hat{p}^\dagger \hat{q} \} \{ \hat{a}^\dagger \hat{i} \} |I \rangle (\epsilon_i - \epsilon_a)^{-1}  \langle I |\{ }{\hat{i}}{^\dagger \hat{a} \} \{ \hat{r}^\dagger }{\hat{s}} 
\contraction[0.5ex]{\{ \hat{p}^\dagger \hat{q} \} \{ \hat{a}^\dagger \hat{i} \} |I \rangle (\epsilon_i - \epsilon_a)^{-1}  \langle I |\{ \hat{i}^\dagger }{\hat{a}}{ \} \{ }{\hat{r}} 
\bcontraction[1.0ex]{\{ \hat{p}^\dagger \hat{q} \} \{}{ \hat{a}}{^\dagger \hat{i} \} |I \rangle (\epsilon_i - \epsilon_a)^{-1}  \langle I |\{ \hat{i}^\dagger}{ \hat{a} } 
\bcontraction[0.5ex]{\{ \hat{p}^\dagger \hat{q} \} \{ \hat{a}^\dagger }{\hat{i}}{ \} |I \rangle (\epsilon_i - \epsilon_a)^{-1}  \langle I |\{ }{\hat{i}} 
\{ \hat{p}^\dagger \hat{q} \} \{ \hat{a}^\dagger \hat{i} \} |I \rangle (\epsilon_i - \epsilon_a)^{-1}  \langle I |\{ \hat{i}^\dagger \hat{a} \} \{ \hat{r}^\dagger \hat{s} \} 
F_{rs} | I \rangle\rangle 
\nonumber\\&&
+ \left( \frac{1}{4}\right)^3 \langle \langle I | \langle pq || rs \rangle 
\contraction[0.5ex]{ \{ \hat{p}^\dagger \hat{q}^\dagger \hat{s}\,}{\hat{r}}{ \} \{ }{\hat{a}}
\contraction[1.0ex]{ \{ \hat{p}^\dagger \hat{q}^\dagger }{\hat{s}}{\,\hat{r} \} \{ \hat{a}^\dagger }{\hat{b}}
\contraction[1.5ex]{ \{ \hat{p}^\dagger }{\hat{q}}{^\dagger \hat{s}\,\hat{r} \} \{ \hat{a}^\dagger \hat{b}^\dagger }{\hat{j}} 
\contraction[2.0ex]{ \{}{ \hat{p}}{^\dagger \hat{q}^\dagger \hat{s}\,\hat{r} \} \{ \hat{a}^\dagger \hat{b}^\dagger \hat{j}\,}{\hat{i}}
\contraction[0.5ex]{ \{ \hat{p}^\dagger \hat{q}^\dagger \hat{s}\,\hat{r} \} \{ \hat{a}^\dagger \hat{b}^\dagger \hat{j}\,\hat{i} \} |I\rangle (\epsilon_i + \epsilon_j - \epsilon_a - \epsilon_b)^{-1} \langle I| \{ \hat{i}^\dagger \hat{j}^\dagger \hat{b}\,}{\hat{a}}{ \} \{ }{\hat{t}}
\contraction[0.6ex]{ \{ \hat{p}^\dagger \hat{q}^\dagger \hat{s}\,\hat{r} \} \{ \hat{a}^\dagger \hat{b}^\dagger \hat{j}\,\hat{i} \} |I\rangle (\epsilon_i + \epsilon_j - \epsilon_a - \epsilon_b)^{-1} \langle I| \{ \hat{i}^\dagger \hat{j}^\dagger }{\hat{b}}{\,\hat{a} \} \{ \hat{t}^\dagger}{ \hat{u}}
\contraction[1.1ex]{ \{ \hat{p}^\dagger \hat{q}^\dagger \hat{s}\,\hat{r} \} \{ \hat{a}^\dagger \hat{b}^\dagger \hat{j}\,\hat{i} \} |I\rangle (\epsilon_i + \epsilon_j - \epsilon_a - \epsilon_b)^{-1} \langle I| \{ \hat{i}^\dagger }{\hat{j}}{^\dagger \hat{b}\,\hat{a} \} \{ \hat{t}^\dagger\hat{u}^\dagger }{\hat{w}}
\contraction[1.6ex]{ \{ \hat{p}^\dagger \hat{q}^\dagger \hat{s}\,\hat{r} \} \{ \hat{a}^\dagger \hat{b}^\dagger \hat{j}\,\hat{i} \} |I\rangle (\epsilon_i + \epsilon_j - \epsilon_a - \epsilon_b)^{-1} \langle I| \{ }{\hat{i}}{^\dagger \hat{j}^\dagger \hat{b}\,\hat{a} \} \{ \hat{t}^\dagger \hat{u}^\dagger \hat{w}\,}{\hat{v}}
\bcontraction[0.5ex]{ \{ \hat{p}^\dagger \hat{q}^\dagger \hat{s}\,\hat{r} \} \{ \hat{a}^\dagger \hat{b}^\dagger \hat{j}\,}{\hat{i}}{ \} |I\rangle (\epsilon_i + \epsilon_j - \epsilon_a - \epsilon_b)^{-1} \langle I| \{ }{\hat{i}}
\bcontraction[1.0ex]{ \{ \hat{p}^\dagger \hat{q}^\dagger \hat{s}\,\hat{r} \} \{ \hat{a}^\dagger \hat{b}^\dagger }{\hat{j}}{\,\hat{i} \} |I\rangle (\epsilon_i + \epsilon_j - \epsilon_a - \epsilon_b)^{-1} \langle I| \{ \hat{i}^\dagger }{\hat{j}}
\bcontraction[1.5ex]{ \{ \hat{p}^\dagger \hat{q}^\dagger \hat{s}\,\hat{r} \} \{ \hat{a}^\dagger }{\hat{b}}{^\dagger \hat{j}\,\hat{i} \} |I\rangle (\epsilon_i + \epsilon_j - \epsilon_a - \epsilon_b)^{-1} \langle I| \{ \hat{i}^\dagger \hat{j}^\dagger }{\hat{b}}
\bcontraction[2.0ex]{ \{ \hat{p}^\dagger \hat{q}^\dagger \hat{s}\,\hat{r} \} \{ }{\hat{a}}{^\dagger \hat{b}^\dagger \hat{j}\,\hat{i} \} |I\rangle (\epsilon_i + \epsilon_j - \epsilon_a - \epsilon_b)^{-1} \langle I| \{ \hat{i}^\dagger \hat{j}^\dagger \hat{b}\,}{\hat{a}}
 \{ \hat{p}^\dagger \hat{q}^\dagger \hat{s}\,\hat{r} \} \{ \hat{a}^\dagger \hat{b}^\dagger \hat{j}\,\hat{i} \} |I\rangle (\epsilon_i + \epsilon_j - \epsilon_a - \epsilon_b)^{-1} \langle I| \{ \hat{i}^\dagger \hat{j}^\dagger \hat{b}\,\hat{a} \} \{ \hat{t}^\dagger \hat{u}^\dagger \hat{w}\,\hat{v} \}
 \langle tu || vw \rangle |I \rangle\rangle \times 16 \nonumber\\
 &=& \sum_{p,q}^{\text{denom.} \neq 0} \frac{|F_{pq}|^2f_p^- f_q^+ }{\epsilon_p - \epsilon_q} + \frac{1}{4} \sum_{p,q,r,s}  ^{\text{denom.} \neq 0} \frac{ | \langle pq || rs \rangle |^2 f_p^- f_q^- f_r^+ f_s^+ }{\epsilon_p +\epsilon_q - \epsilon_r- \epsilon_s} \equiv \langle E_I^{(2)} \rangle_{L}, \label{E2}
\end{eqnarray} 
\end{widetext}
where $i$, $j$, $a$, $b$ as well as $p$ through $w$ run over all spinorbitals and ``$\,\times\, 16$\,'' means that there are 16 distinct, equal-valued full contraction patterns.
The constant term $\langle E^{(1)}_I\rangle$ in $\hat{V}$ [see Eq.\ (\ref{V})] does not contribute because  the resolvent $\hat{R}_I$ erases all determinants in the degenerate
subspace of $I$ (see Appendix \ref{app:projector}). The latter is also responsible for the restriction (``$\text{denom.}\neq 0$'') of the summations to nonzero denominators. 
Although we did not examine the linkedness of the second-order HCPT energy corrections (be they eigenvalues or the trace
 of $\bm{E}^{(2)}$), we could nonetheless establish the linkedness of their thermal average, $\langle E_I^{(2)} \rangle$. 

Next, we evaluate $\langle E^{(1)}_I E^{(1)}_I \rangle$, recalling the rules for the inner projector $\hat{P}_I$ (see Appendix \ref{app:projector}).
\begin{widetext}
\begin{eqnarray}
\langle E^{(1)}_I E^{(1)}_I \rangle &=& \langle \langle I | \hat{V} \hat{P}_I \hat{V} | I \rangle \rangle 
\nonumber\\
&=& \langle E_I^{(1)} \rangle \langle E_I^{(1)} \rangle + \langle \langle I | F_{pq}  
\contraction[1.0ex]{\{ }{\hat{p}}{^\dagger \hat{q} \} \{ \hat{a}^\dagger }{\hat{i}}
\contraction[0.5ex]{\{ \hat{p}^\dagger }{\hat{q}}{ \} \{ }{\hat{a}} 
\contraction[0.6ex]{\{ \hat{p}^\dagger \hat{q} \} \{ \hat{a}^\dagger \hat{i} \} |I \rangle \langle I |\{ }{\hat{i}}{^\dagger \hat{a} \} \{ \hat{r}^\dagger }{\hat{s}} 
\contraction[0.5ex]{\{ \hat{p}^\dagger \hat{q} \} \{ \hat{a}^\dagger \hat{i} \} |I \rangle  \langle I |\{ \hat{i}^\dagger }{\hat{a}}{ \} \{ }{\hat{r}} 
\bcontraction[1.0ex]{\{ \hat{p}^\dagger \hat{q} \} \{}{ \hat{a}}{^\dagger \hat{i} \} |I \rangle   \langle I |\{ \hat{i}^\dagger}{ \hat{a} } 
\bcontraction[0.5ex]{\{ \hat{p}^\dagger \hat{q} \} \{ \hat{a}^\dagger }{\hat{i}}{ \} |I \rangle  \langle I |\{ }{\hat{i}} 
\{ \hat{p}^\dagger \hat{q} \} \{ \hat{a}^\dagger \hat{i} \} |I \rangle  \langle I |\{ \hat{i}^\dagger \hat{a} \} \{ \hat{r}^\dagger \hat{s} \} 
F_{rs} | I \rangle\rangle 
\nonumber\\&&
+ \left( \frac{1}{4}\right)^3 \langle \langle I | \langle pq || rs \rangle 
\contraction[0.5ex]{ \{ \hat{p}^\dagger \hat{q}^\dagger \hat{s}\,}{\hat{r}}{ \} \{ }{\hat{a}}
\contraction[1.0ex]{ \{ \hat{p}^\dagger \hat{q}^\dagger }{\hat{s}}{\,\hat{r} \} \{ \hat{a}^\dagger }{\hat{b}}
\contraction[1.5ex]{ \{ \hat{p}^\dagger }{\hat{q}}{^\dagger \hat{s}\,\hat{r} \} \{ \hat{a}^\dagger \hat{b}^\dagger }{\hat{j}} 
\contraction[2.0ex]{ \{}{ \hat{p}}{^\dagger \hat{q}^\dagger \hat{s}\,\hat{r} \} \{ \hat{a}^\dagger \hat{b}^\dagger \hat{j}\,}{\hat{i}}
\contraction[0.5ex]{ \{ \hat{p}^\dagger \hat{q}^\dagger \hat{s}\,\hat{r} \} \{ \hat{a}^\dagger \hat{b}^\dagger \hat{j}\,\hat{i} \} |I\rangle  \langle I| \{ \hat{i}^\dagger \hat{j}^\dagger \hat{b}\,}{\hat{a}}{ \} \{ }{\hat{t}}
\contraction[0.6ex]{ \{ \hat{p}^\dagger \hat{q}^\dagger \hat{s}\,\hat{r} \} \{ \hat{a}^\dagger \hat{b}^\dagger \hat{j}\,\hat{i} \} |I\rangle   \langle I| \{ \hat{i}^\dagger \hat{j}^\dagger }{\hat{b}}{\,\hat{a} \} \{ \hat{t}^\dagger}{ \hat{u}}
\contraction[1.1ex]{ \{ \hat{p}^\dagger \hat{q}^\dagger \hat{s}\,\hat{r} \} \{ \hat{a}^\dagger \hat{b}^\dagger \hat{j}\,\hat{i} \} |I\rangle  \langle I| \{ \hat{i}^\dagger }{\hat{j}}{^\dagger \hat{b}\,\hat{a} \} \{ \hat{t}^\dagger\hat{u}^\dagger }{\hat{w}}
\contraction[1.6ex]{ \{ \hat{p}^\dagger \hat{q}^\dagger \hat{s}\,\hat{r} \} \{ \hat{a}^\dagger \hat{b}^\dagger \hat{j}\,\hat{i} \} |I\rangle  \langle I| \{ }{\hat{i}}{^\dagger \hat{j}^\dagger \hat{b}\,\hat{a} \} \{ \hat{t}^\dagger \hat{u}^\dagger \hat{w}\,}{\hat{v}}
\bcontraction[0.5ex]{ \{ \hat{p}^\dagger \hat{q}^\dagger \hat{s}\,\hat{r} \} \{ \hat{a}^\dagger \hat{b}^\dagger \hat{j}\,}{\hat{i}}{ \} |I\rangle \langle I| \{ }{\hat{i}}
\bcontraction[1.0ex]{ \{ \hat{p}^\dagger \hat{q}^\dagger \hat{s}\,\hat{r} \} \{ \hat{a}^\dagger \hat{b}^\dagger }{\hat{j}}{\,\hat{i} \} |I\rangle \langle I| \{ \hat{i}^\dagger }{\hat{j}}
\bcontraction[1.5ex]{ \{ \hat{p}^\dagger \hat{q}^\dagger \hat{s}\,\hat{r} \} \{ \hat{a}^\dagger }{\hat{b}}{^\dagger \hat{j}\,\hat{i} \} |I\rangle \langle I| \{ \hat{i}^\dagger \hat{j}^\dagger }{\hat{b}}
\bcontraction[2.0ex]{ \{ \hat{p}^\dagger \hat{q}^\dagger \hat{s}\,\hat{r} \} \{ }{\hat{a}}{^\dagger \hat{b}^\dagger \hat{j}\,\hat{i} \} |I\rangle \langle I| \{ \hat{i}^\dagger \hat{j}^\dagger \hat{b}\,}{\hat{a}}
 \{ \hat{p}^\dagger \hat{q}^\dagger \hat{s}\,\hat{r} \} \{ \hat{a}^\dagger \hat{b}^\dagger \hat{j}\,\hat{i} \} |I\rangle \langle I| \{ \hat{i}^\dagger \hat{j}^\dagger \hat{b}\,\hat{a} \} \{ \hat{t}^\dagger \hat{u}^\dagger \hat{w}\,\hat{v} \}
 \langle tu || vw \rangle |I \rangle\rangle \times 16 \nonumber\\
 &=& \langle E_I^{(1)} \rangle \langle E_I^{(1)} \rangle +  \sum_{p,q}^{\text{denom.} = 0} {|F_{pq}|^2f_p^- f_q^+ } + \frac{1}{4} \sum_{p,q,r,s} ^{\text{denom.} = 0} { | \langle pq || rs \rangle |^2 f_p^- f_q^- f_r^+ f_s^+ } \equiv 
 \langle E_I^{(1)} \rangle_L \langle E_I^{(1)} \rangle_L + \langle E_I^{(1)}E_I^{(1)} \rangle_{L}. \label{E1E1}
\end{eqnarray} 
\end{widetext}
Unlike in $\langle E_I^{(2)} \rangle$, the inner projector $\hat{P}_I$ allows the constant part $\langle E_I^{(1)} \rangle$ of $\hat{V}$ to survive, resulting
in the unlinked contribution $\langle E_I^{(1)} \rangle_L \langle E_I^{(1)} \rangle_L$, which is cancelled by the corresponding contribution in the $\Omega^{(1)}\Omega^{(1)}$ term of Eq.\ (\ref{Omega2sq}). It is also responsible for the restrictions
(``$\text{denom.}=0$'')
to the summations to the cases whose fictitious denominator factor ($\epsilon_p - \epsilon_q$ in the first sum or $\epsilon_p + \epsilon_q - \epsilon_r - \epsilon_s$
in the second sum) is zero. See the discussion regarding Eq.\ (\ref{projector}) and Appendix \ref{app:projector} for the justification of this rule.

Substituting all these results into Eq.\ (\ref{Omega2sq}), we observe exact mutual cancellations
of all unlinked terms, reproducing the reduced analytical formula reported earlier:\cite{Hirata2ndorder}
\begin{eqnarray}
\Omega^{(2)} 
&=& \sum_{p,q}^{\text{denom.}\neq0} \frac{|F_{pq}|^2f_p^- f_q^+ }{\epsilon_p - \epsilon_q} 
 + \frac{1}{4} \sum_{p,q,r,s}^{\text{denom.}\neq0}\frac{|\langle pq||rs \rangle |^2f_p^- f_q^-f_r^+ f_s^+ }{\epsilon_p + \epsilon_q - \epsilon_r - \epsilon_s} 
 \nonumber \\&& 
 -\frac{\beta}{2} \sum_{p,q}^{\text{denom.}=0} {|F_{pq}|^2} f_p^- f_q^+ 
 - \frac{\beta}{8} \sum_{p,q,r,s}^{\text{denom.}=0}{|\langle pq||rs \rangle |^2} 
f_p^- f_q^-f_r^+ f_s^+ 
\nonumber\\&&
+ {\beta}  \mu^{(1)}\sum_{p} F_{pp} f_p^- f_p^+ 
 - \frac{\beta}{2} \left( \mu^{(1)}\right)^2 \sum_p f_p^- f_p^+ 
- \mu^{(2)} \bar{N} \label{Omega2reduced} \\
&\equiv& \langle D_I^{(2)} \rangle_{L} + \frac{(-\beta)}{2!}  \langle D_I^{(1)}D_I^{(1)} \rangle_{L}, \label{Omega2linked}
\end{eqnarray}
which is linked.

As per the $\mu^{(n)}$ recursion [Eq.\ (\ref{recursionmu})], we have
\begin{eqnarray}
\mu^{(2)} \langle N_I^2 - \bar{N}^2 \rangle &=& \langle E_I^{(2)} (N_I - \bar{N})\rangle \nonumber\\
&& + \frac{(-\beta)}{2!} \langle D_I^{(1)} D_I^{(1)} (N_I - \bar{N})\rangle, \label{mu2sq}
\end{eqnarray}
of which the left-hand side has already been evaluated in Eq.\ (\ref{NminusN}). The second term in the right-hand side is further expanded as
\begin{eqnarray}
\langle D_I^{(1)} D_I^{(1)} (N_I - \bar{N}) \rangle &=& \langle E_I^{(1)}E_I^{(1)} (N_I - \bar{N})\rangle 
\nonumber\\&&
- 2 \mu^{(1)} \langle E_I^{(1)} N_I (N_I - \bar{N})\rangle 
\nonumber\\&&
+ \left( \mu^{(1)} \right)^2 \langle N_I N_I (N_I - \bar{N})\rangle. \label{D1D1N} 
\end{eqnarray}
The first term of Eq.\ (\ref{mu2sq}) is evaluated as  
\begin{widetext}
\begin{eqnarray}
\langle E_I^{(2)} (N_I-\bar{N}) \rangle &=&  \langle \langle I | \hat{V}\hat{R}_I\hat{V} (\hat{N}-\bar{N})|I \rangle \rangle \nonumber \\
&=& \langle \langle I | F_{pq} 
\contraction[1.0ex]{ \{}{\hat{p}}{^\dagger \hat{q}\} \{\hat{a}^\dagger }{\hat{i}}
\contraction[0.5ex]{ \{\hat{p}^\dagger }{\hat{q}}{\} \{}{\hat{a}}
\bcontraction[1.0ex]{ \{\hat{p}^\dagger \hat{q}\} \{}{\hat{a}}{^\dagger \hat{i}\} | I \rangle(\epsilon_i - \epsilon_a)^{-1} \langle I | \{\hat{i}^\dagger }{\hat{a}} 
\bcontraction[0.5ex]{ \{\hat{p}^\dagger \hat{q}\} \{\hat{a}^\dagger }{\hat{i}}{\} | I \rangle(\epsilon_i - \epsilon_a)^{-1} \langle I | \{}{\hat{i}} 
\contraction[0.6ex]{ \{\hat{p}^\dagger \hat{q}\} \{\hat{a}^\dagger \hat{i}\} | I \rangle(\epsilon_i - \epsilon_a)^{-1} \langle I | \{}{\hat{i}}{^\dagger \hat{a}\} \{\hat{r}^\dagger \hat{s}\} F_{rs} \{\hat{t}^\dagger }{\hat{t}}   
\contraction[0.5ex]{ \{\hat{p}^\dagger \hat{q}\} \{\hat{a}^\dagger \hat{i}\} | I \rangle(\epsilon_i - \epsilon_a)^{-1} \langle I | \{\hat{i}^\dagger }{\hat{a}}{\} \{}{\hat{r}}
\contraction[0.5ex]{ \{\hat{p}^\dagger \hat{q}\} \{\hat{a}^\dagger \hat{i}\} | I \rangle(\epsilon_i - \epsilon_a)^{-1} \langle I | \{\hat{i}^\dagger \hat{a}\} \{\hat{r}^\dagger }{\hat{s}}{\} F_{rs}\{}{\hat{t}}
 \{\hat{p}^\dagger \hat{q}\} \{\hat{a}^\dagger \hat{i}\} | I \rangle(\epsilon_i - \epsilon_a)^{-1} \langle I | \{\hat{i}^\dagger \hat{a}\} \{\hat{r}^\dagger \hat{s}\} F_{rs} \{\hat{t}^\dagger \hat{t}\} | I \rangle \rangle 
 + \langle\langle I| F_{pq}  
\contraction[1.0ex]{ \{}{\hat{p}}{^\dagger \hat{q}\} \{\hat{a}^\dagger }{\hat{i}}
\contraction[0.5ex]{ \{\hat{p}^\dagger }{\hat{q}}{\} \{}{\hat{a}}
\bcontraction[1.0ex]{ \{\hat{p}^\dagger \hat{q}\} \{}{\hat{a}}{^\dagger \hat{i}\} |I \rangle (\epsilon_i - \epsilon_a)^{-1} \langle I|\{\hat{i}^\dagger }{\hat{a}} 
\bcontraction[0.5ex]{ \{\hat{p}^\dagger \hat{q}\} \{\hat{a}^\dagger }{\hat{i}}{\} |I \rangle (\epsilon_i - \epsilon_a)^{-1}\langle I| \{}{\hat{i}} 
\contraction[0.5ex]{ \{\hat{p}^\dagger \hat{q}\} \{\hat{a}^\dagger \hat{i}\} |I \rangle (\epsilon_i - \epsilon_a)^{-1}\langle I| \{\hat{i}^\dagger}{\hat{a}}{\} \{\hat{r}^\dagger \hat{s}\}F_{rs} \{}{\hat{t}}  
\contraction[0.6ex]{ \{\hat{p}^\dagger \hat{q}\} \{\hat{a}^\dagger \hat{i}\} |I \rangle (\epsilon_i - \epsilon_a)^{-1}\langle I| \{}{\hat{i}}{^\dagger \hat{a}\} \{\hat{r}^\dagger }{\hat{s}}
\contraction[1.5ex]{ \{\hat{p}^\dagger \hat{q}\} \{\hat{a}^\dagger \hat{i}\}|I  \rangle (\epsilon_i - \epsilon_a)^{-1}\langle I| \{\hat{i}^\dagger \hat{a}\} \{}{\hat{r}}{^\dagger \hat{s}\}F_{rs} \{\hat{t}^\dagger }{\hat{t}}
 \{\hat{p}^\dagger \hat{q}\} \{\hat{a}^\dagger \hat{i}\}|I \rangle (\epsilon_i - \epsilon_a)^{-1} \langle I| \{\hat{i}^\dagger \hat{a}\} \{\hat{r}^\dagger \hat{s}\} F_{rs}\{\hat{t}^\dagger \hat{t}\} |I \rangle \rangle  
\nonumber\\&& 
+ \frac{1}{4} \langle\langle I | F_{pq} 
\contraction[1.0ex]{ \{}{\hat{p}}{^\dagger \hat{q}\} \{\hat{a}^\dagger }{\hat{i}}
\contraction[0.5ex]{ \{\hat{p}^\dagger }{\hat{q}}{\} \{}{\hat{a}}
\bcontraction[1.0ex]{ \{\hat{p}^\dagger \hat{q}\} \{}{\hat{a}}{^\dagger \hat{i}\} |I \rangle (\epsilon_i - \epsilon_a)^{-1}\langle I| \{\hat{i}^\dagger }{\hat{a}} 
\bcontraction[0.5ex]{ \{\hat{p}^\dagger \hat{q}\} \{\hat{a}^\dagger }{\hat{i}}{\} |I \rangle (\epsilon_i - \epsilon_a)^{-1}\langle I| \{}{\hat{i}} 
\contraction[0.5ex]{ \{\hat{p}^\dagger \hat{q}\} \{\hat{a}^\dagger \hat{i}\} |I \rangle (\epsilon_i - \epsilon_a)^{-1}\langle I| \{}{\hat{i}}{^\dagger \hat{a}\} \{\hat{r}^\dagger \hat{s}^\dagger \hat{u}\, }{\hat{t}}
\contraction[0.5ex]{ \{\hat{p}^\dagger \hat{q}\} \{\hat{a}^\dagger \hat{i}\} |I \rangle (\epsilon_i - \epsilon_a)^{-1}\langle I| \{\hat{i}^\dagger }{\hat{a}}{\} \{}{\hat{r}}
\contraction[1.4ex]{ \{\hat{p}^\dagger \hat{q}\} \{\hat{a}^\dagger \hat{i}\} |I \rangle (\epsilon_i - \epsilon_a)^{-1}\langle I| \{\hat{i}^\dagger \hat{a}\} \{\hat{r}^\dagger }{\hat{s}}{^\dagger \hat{u}\, \hat{t}\} \langle rs || tu \rangle \{\hat{v}^\dagger }{\hat{v}}
\contraction[0.5ex]{ \{\hat{p}^\dagger \hat{q}\} \{\hat{a}^\dagger \hat{i}\} |I \rangle (\epsilon_i - \epsilon_a)^{-1}\langle I| \{\hat{i}^\dagger \hat{a}\} \{\hat{r}^\dagger \hat{s}^\dagger }{\hat{u} }{\,\hat{t}\}\langle rs || tu \rangle  \{}{\hat{v}}
 \{\hat{p}^\dagger \hat{q}\} \{\hat{a}^\dagger \hat{i}\} |I \rangle (\epsilon_i - \epsilon_a)^{-1}\langle I| \{\hat{i}^\dagger \hat{a}\} \{\hat{r}^\dagger \hat{s}^\dagger \hat{u} \,\hat{t}\} \langle rs || tu \rangle  \{\hat{v}^\dagger \hat{v}\} |I \rangle\rangle  \times  4 \nonumber\\
&& + \left( \frac{1}{4} \right)^2 \langle\langle I | \langle pq || rs \rangle 
\contraction[2.0ex]{ \{}{\hat{p}}{^\dagger \hat{q}^\dagger \hat{s}\, \hat{r}\} \{\hat{a}^\dagger\hat{b}^\dagger \hat{j} }{\hat{i}}
\contraction[1.5ex]{ \{\hat{p}^\dagger }{\hat{q}}{^\dagger \hat{s}\, \hat{r}\} \{\hat{a}^\dagger\hat{b}^\dagger }{\hat{j}}
\contraction[1.0ex]{ \{\hat{p}^\dagger \hat{q}^\dagger }{\hat{s}}{\, \hat{r}\} \{\hat{a}^\dagger}{\hat{b}}
\contraction[0.5ex]{ \{\hat{p}^\dagger \hat{q}^\dagger \hat{s}\, }{\hat{r}}{\} \{}{\hat{a}}
\bcontraction[2.0ex]{ \{\hat{p}^\dagger \hat{q}^\dagger \hat{s}\, \hat{r}\} \{}{\hat{a}}{^\dagger\hat{b}^\dagger \hat{j}\, \hat{i}\} |I \rangle (\epsilon_i + \epsilon_j- \epsilon_a- \epsilon_b)^{-1}\langle I| \{\hat{i}^\dagger \hat{j}^\dagger \hat{b}\,}{\hat{a}}
\bcontraction[1.5ex]{ \{\hat{p}^\dagger \hat{q}^\dagger \hat{s}\, \hat{r}\} \{\hat{a}^\dagger}{\hat{b}}{^\dagger \hat{j}\, \hat{i}\} |I \rangle (\epsilon_i + \epsilon_j- \epsilon_a- \epsilon_b)^{-1}\langle  I|\{\hat{i}^\dagger \hat{j}^\dagger }{\hat{b}}
\bcontraction[1.0ex]{ \{\hat{p}^\dagger \hat{q}^\dagger \hat{s}\, \hat{r}\} \{\hat{a}^\dagger\hat{b}^\dagger }{\hat{j}}{\, \hat{i}\} |I \rangle (\epsilon_i + \epsilon_j- \epsilon_a- \epsilon_b)^{-1}\langle I| \{\hat{i}^\dagger }{\hat{j}}
\bcontraction[0.5ex]{ \{\hat{p}^\dagger \hat{q}^\dagger \hat{s}\, \hat{r}\} \{\hat{a}^\dagger\hat{b}^\dagger \hat{j}\, }{\hat{i}}{\} |I \rangle (\epsilon_i + \epsilon_j- \epsilon_a- \epsilon_b)^{-1}\langle I| \{}{\hat{i}}
\contraction[1.5ex]{ \{\hat{p}^\dagger \hat{q}^\dagger \hat{s}\, \hat{r}\} \{\hat{a}^\dagger\hat{b}^\dagger \hat{j}\, \hat{i}\} |I \rangle (\epsilon_i + \epsilon_j- \epsilon_a- \epsilon_b)^{-1}\langle I| \{}{\hat{i}}{^\dagger \hat{j}^\dagger \hat{b}\,\hat{a}\} \{\hat{t}^\dagger }{\hat{u}}
\contraction[1.0ex]{ \{\hat{p}^\dagger \hat{q}^\dagger \hat{s}\, \hat{r}\} \{\hat{a}^\dagger\hat{b}^\dagger \hat{j}\, \hat{i}\} |I \rangle (\epsilon_i + \epsilon_j- \epsilon_a- \epsilon_b)^{-1}\langle I| \{\hat{i}^\dagger }{\hat{j}}{^\dagger \hat{b}\,\hat{a}\} \{\hat{t}^\dagger \hat{u}\} F_{tu} \{\hat{v}^\dagger }{\hat{v}}
\contraction[0.5ex]{ \{\hat{p}^\dagger \hat{q}^\dagger \hat{s}\, \hat{r}\} \{\hat{a}^\dagger\hat{b}^\dagger \hat{j}\, \hat{i}\} |I \rangle (\epsilon_i + \epsilon_j- \epsilon_a- \epsilon_b)^{-1}\langle I| \{\hat{i}^\dagger \hat{j}^\dagger }{\hat{b}}{\,\hat{a}\} \{\hat{t}^\dagger \hat{u}\} F_{tu} \{}{\hat{v}}
\contraction[0.5ex]{ \{\hat{p}^\dagger \hat{q}^\dagger \hat{s} \hat{r}\} \{\hat{a}^\dagger\hat{b}^\dagger \hat{j}\, \hat{i}\} |I \rangle (\epsilon_i + \epsilon_j- \epsilon_a- \epsilon_b)^{-1}\langle I| \{\hat{i}^\dagger \hat{j}^\dagger \hat{b}\,}{\hat{a}}{\} \{}{\hat{t}}
 \{\hat{p}^\dagger \hat{q}^\dagger \hat{s}\, \hat{r}\} \{\hat{a}^\dagger\hat{b}^\dagger \hat{j}\, \hat{i}\} |I \rangle (\epsilon_i + \epsilon_j- \epsilon_a- \epsilon_b)^{-1}\langle I| \{\hat{i}^\dagger \hat{j}^\dagger \hat{b}\,\hat{a}\} \{\hat{t}^\dagger \hat{u}\} F_{tu} \{\hat{v}^\dagger \hat{v}\} |I \rangle \rangle  \times 16 \nonumber \\
&& + \left( \frac{1}{4} \right)^3 \langle\langle I | \langle pq || rs \rangle  
\contraction[2.0ex]{ \{}{\hat{p}}{^\dagger \hat{q}^\dagger \hat{s}\, \hat{r}\} \{\hat{a}^\dagger\hat{b}^\dagger \hat{j}\, }{\hat{i}}
\contraction[1.5ex]{ \{\hat{p}^\dagger }{\hat{q}}{^\dagger \hat{s}\, \hat{r}\} \{\hat{a}^\dagger\hat{b}^\dagger }{\hat{j}}
\contraction[1.0ex]{ \{\hat{p}^\dagger \hat{q}^\dagger }{\hat{s}}{\, \hat{r}\} \{\hat{a}^\dagger}{\hat{b}}
\contraction[0.5ex]{ \{\hat{p}^\dagger \hat{q}^\dagger \hat{s}\, }{\hat{r}}{\} \{}{\hat{a}}
\bcontraction[2.0ex]{ \{\hat{p}^\dagger \hat{q}^\dagger \hat{s}\, \hat{r}\} \{}{\hat{a}}{^\dagger\hat{b}^\dagger \hat{j}\, \hat{i}\} |I \rangle (\epsilon_i + \epsilon_j- \epsilon_a- \epsilon_b)^{-1} \langle I|  \{\hat{i}^\dagger \hat{j}^\dagger \hat{b}\,}{\hat{a}}
\bcontraction[1.5ex]{ \{\hat{p}^\dagger \hat{q}^\dagger \hat{s} \,\hat{r}\} \{\hat{a}^\dagger}{\hat{b}}{^\dagger \hat{j} \,\hat{i}\} |I \rangle (\epsilon_i + \epsilon_j- \epsilon_a- \epsilon_b)^{-1} \langle I|  \{\hat{i}^\dagger \hat{j}^\dagger }{\hat{b}}
\bcontraction[1.0ex]{ \{\hat{p}^\dagger \hat{q}^\dagger \hat{s}\, \hat{r}\} \{\hat{a}^\dagger\hat{b}^\dagger }{\hat{j}}{\, \hat{i}\} |I \rangle (\epsilon_i + \epsilon_j- \epsilon_a- \epsilon_b)^{-1} \langle I|  \{\hat{i}^\dagger }{\hat{j}}
\bcontraction[0.5ex]{ \{\hat{p}^\dagger \hat{q}^\dagger \hat{s}\, \hat{r}\} \{\hat{a}^\dagger\hat{b}^\dagger \hat{j}\, }{\hat{i}}{\} |I \rangle (\epsilon_i + \epsilon_j- \epsilon_a- \epsilon_b)^{-1} \langle I|  \{}{\hat{i}}
\contraction[1.5ex]{ \{\hat{p}^\dagger \hat{q}^\dagger \hat{s}\, \hat{r}\} \{\hat{a}^\dagger\hat{b}^\dagger \hat{j}\, \hat{i}\} |I \rangle (\epsilon_i + \epsilon_j- \epsilon_a- \epsilon_b)^{-1} \langle I|  \{}{\hat{i}}{^\dagger \hat{j}^\dagger \hat{b}\,\hat{a}\} \{\hat{t}^\dagger\hat{u}^\dagger \hat{w}\, \hat{v}\} \langle tu || vw \rangle\{\hat{x}^\dagger }{\hat{x}}
\contraction[1.0ex]{ \{\hat{p}^\dagger \hat{q}^\dagger \hat{s}\, \hat{r}\} \{\hat{a}^\dagger\hat{b}^\dagger \hat{j}\, \hat{i}\} |I \rangle (\epsilon_i + \epsilon_j- \epsilon_a- \epsilon_b)^{-1} \langle I|  \{\hat{i}^\dagger }{\hat{j}}{^\dagger \hat{b}\,\hat{a}\} \{\hat{t}^\dagger\hat{u}^\dagger }{\hat{w}}
\contraction[0.5ex]{ \{\hat{p}^\dagger \hat{q}^\dagger \hat{s}\, \hat{r}\} \{\hat{a}^\dagger\hat{b}^\dagger \hat{j}\, \hat{i}\} |I \rangle (\epsilon_i + \epsilon_j- \epsilon_a- \epsilon_b)^{-1} \langle I|  \{\hat{i}^\dagger \hat{j}^\dagger }{\hat{b}}{\,\hat{a}\} \{\hat{t}^\dagger}{\hat{u}}
\contraction[0.5ex]{ \{\hat{p}^\dagger \hat{q}^\dagger \hat{s}\, \hat{r}\} \{\hat{a}^\dagger\hat{b}^\dagger \hat{j}\, \hat{i}\} |I \rangle (\epsilon_i + \epsilon_j- \epsilon_a- \epsilon_b)^{-1} \langle I|  \{\hat{i}^\dagger \hat{j}^\dagger \hat{b}\,}{\hat{a}}{\} \{}{\hat{t}}
\contraction[0.5ex]{ \{\hat{p}^\dagger \hat{q}^\dagger \hat{s}\, \hat{r}\} \{\hat{a}^\dagger\hat{b}^\dagger \hat{j}\, \hat{i}\} |I \rangle (\epsilon_i + \epsilon_j- \epsilon_a- \epsilon_b)^{-1} \langle I|  \{\hat{i}^\dagger \hat{j}^\dagger \hat{b}\,\hat{a}\} \{\hat{t}^\dagger\hat{u}^\dagger \hat{w}\, }{\hat{v}}{\} \langle tu || vw \rangle\{}{\hat{x}}
 \{\hat{p}^\dagger \hat{q}^\dagger \hat{s} \,\hat{r}\} \{\hat{a}^\dagger\hat{b}^\dagger \hat{j} \,\hat{i}\} |I \rangle (\epsilon_i + \epsilon_j- \epsilon_a- \epsilon_b)^{-1} \langle I| \{\hat{i}^\dagger \hat{j}^\dagger \hat{b}\,\hat{a}\} \{\hat{t}^\dagger\hat{u}^\dagger \hat{w} \,\hat{v}\} \langle tu || vw \rangle \{\hat{x}^\dagger \hat{x}\} |I \rangle \rangle  \times 32\nonumber \\
&& + \left( \frac{1}{4} \right)^3 \langle\langle I | \langle pq || rs \rangle 
\contraction[2.0ex]{ \{}{\hat{p}}{^\dagger \hat{q}^\dagger \hat{s}\, \hat{r}\} \{\hat{a}^\dagger\hat{b}^\dagger \hat{j}\, }{\hat{i}}
\contraction[1.5ex]{ \{\hat{p}^\dagger }{\hat{q}}{^\dagger \hat{s}\, \hat{r}\} \{\hat{a}^\dagger\hat{b}^\dagger }{\hat{j}}
\contraction[1.0ex]{ \{\hat{p}^\dagger \hat{q}^\dagger }{\hat{s}}{\, \hat{r}\} \{\hat{a}^\dagger}{\hat{b}}
\contraction[0.5ex]{ \{\hat{p}^\dagger \hat{q}^\dagger \hat{s}\, }{\hat{r}}{\} \{}{\hat{a}}
\bcontraction[2.0ex]{ \{\hat{p}^\dagger \hat{q}^\dagger \hat{s}\, \hat{r}\} \{}{\hat{a}}{^\dagger\hat{b}^\dagger \hat{j}\, \hat{i}\} |I \rangle (\epsilon_i + \epsilon_j- \epsilon_a- \epsilon_b)^{-1}\langle I| \{\hat{i}^\dagger \hat{j}^\dagger \hat{b}\,}{\hat{a}}
\bcontraction[1.5ex]{ \{\hat{p}^\dagger \hat{q}^\dagger \hat{s}\, \hat{r}\} \{\hat{a}^\dagger}{\hat{b}}{^\dagger \hat{j}\, \hat{i}\} |I \rangle (\epsilon_i + \epsilon_j- \epsilon_a- \epsilon_b)^{-1}\langle I| \{\hat{i}^\dagger \hat{j}^\dagger }{\hat{b}}
\bcontraction[1.0ex]{ \{\hat{p}^\dagger \hat{q}^\dagger \hat{s}\, \hat{r}\} \{\hat{a}^\dagger\hat{b}^\dagger }{\hat{j}}{\, \hat{i}\} |I \rangle (\epsilon_i + \epsilon_j- \epsilon_a- \epsilon_b)^{-1}\langle I| \{\hat{i}^\dagger }{\hat{j}}
\bcontraction[0.5ex]{ \{\hat{p}^\dagger \hat{q}^\dagger \hat{s}\, \hat{r}\} \{\hat{a}^\dagger\hat{b}^\dagger \hat{j}\, }{\hat{i}}{\} |I \rangle (\epsilon_i + \epsilon_j- \epsilon_a- \epsilon_b)^{-1}\langle I| \{}{\hat{i}}
\contraction[1.8ex]{ \{\hat{p}^\dagger \hat{q}^\dagger \hat{s}\, \hat{r}\} \{\hat{a}^\dagger\hat{b}^\dagger \hat{j}\, \hat{i}\} |I \rangle (\epsilon_i + \epsilon_j- \epsilon_a- \epsilon_b)^{-1}\langle I| \{}{\hat{i}}{^\dagger \hat{j}^\dagger \hat{b}\,\hat{a}\} \{\hat{t}^\dagger\hat{u}^\dagger \hat{w}\, }{\hat{v}}
\contraction[1.4ex]{ \{\hat{p}^\dagger \hat{q}^\dagger \hat{s}\, \hat{r}\} \{\hat{a}^\dagger\hat{b}^\dagger \hat{j}\, \hat{i}\} |I \rangle (\epsilon_i + \epsilon_j- \epsilon_a- \epsilon_b)^{-1}\langle I| \{\hat{i}^\dagger }{\hat{j}}{^\dagger \hat{b}\,\hat{a}\} \{\hat{t}^\dagger\hat{u}^\dagger }{\hat{w}}
\contraction[1.0ex]{ \{\hat{p}^\dagger \hat{q}^\dagger \hat{s}\, \hat{r}\} \{\hat{a}^\dagger\hat{b}^\dagger \hat{j}\, \hat{i}\} |I \rangle (\epsilon_i + \epsilon_j- \epsilon_a- \epsilon_b)^{-1}\langle I| \{\hat{i}^\dagger \hat{j}^\dagger }{\hat{b}}{\,\hat{a}\} \{\hat{t}^\dagger}{\hat{u}}
\contraction[1.1ex]{ \{\hat{p}^\dagger \hat{q}^\dagger \hat{s}\, \hat{r}\} \{\hat{a}^\dagger\hat{b}^\dagger \hat{j}\, \hat{i}\} |I \rangle (\epsilon_i + \epsilon_j- \epsilon_a- \epsilon_b)^{-1}\langle I| \{\hat{i}^\dagger \hat{j}^\dagger \hat{b}\,}{\hat{a}}{\} \{\hat{t}^\dagger\hat{u}^\dagger \hat{w} \,\hat{v}\} \langle tu || vw \rangle \{}{\hat{x}}
\contraction[0.4ex]{ \{\hat{p}^\dagger \hat{q}^\dagger \hat{s} \,\hat{r}\} \{\hat{a}^\dagger\hat{b}^\dagger \hat{j}\, \hat{i}\} |I \rangle (\epsilon_i + \epsilon_j- \epsilon_a- \epsilon_b)^{-1}\langle I| \{\hat{i}^\dagger \hat{j}^\dagger \hat{b}\,\hat{a}\} \{}{\hat{t}}{^\dagger\hat{u}^\dagger \hat{w}\, \hat{v}\} \langle tu || vw \rangle \{\hat{x}^\dagger }{\hat{x}}
 \{\hat{p}^\dagger \hat{q}^\dagger \hat{s}\, \hat{r}\} \{\hat{a}^\dagger\hat{b}^\dagger \hat{j} \,\hat{i}\} |I \rangle (\epsilon_i + \epsilon_j- \epsilon_a- \epsilon_b)^{-1}\langle I| \{\hat{i}^\dagger \hat{j}^\dagger \hat{b}\,\hat{a}\} \{\hat{t}^\dagger\hat{u}^\dagger \hat{w}\, \hat{v}\} \langle tu || vw \rangle \{\hat{x}^\dagger \hat{x}\} | I \rangle \rangle  \times  32 \\
&=& \sum_{p,q}^{\text{denom.}\neq 0} \frac{|F_{pq}|^2 f_p^- f_q^+ f_p^+}{\epsilon_p - \epsilon_q}  
-\sum_{p,q}^{\text{denom.}\neq 0} \frac{|F_{pq}|^2  f_p^- f_q^+ f_q^- }{\epsilon_p - \epsilon_q} 
\nonumber\\&& 
+ \sum_{p,q,s}^{\text{denom.}\neq 0} \frac{F_{pq} \langle qs || ps \rangle  f_p^- f_q^+ f_s^-  f_s^+}{\epsilon_p - \epsilon_q} 
+ \sum_{p,q,r}^{\text{denom.}\neq 0} \frac{\langle pq || rq \rangle F_{rp}  f_p^- f_q^- f_q^+  f_r^+ }{\epsilon_p - \epsilon_r}  
\nonumber\\&& 
+ \frac{1}{2}  \sum_{p,q,r,s}^{\text{denom.}\neq 0} \frac{|\langle pq || rs \rangle|^2 f_p^- f_q^- f_r^+ f_s^+ f_p^+ }{\epsilon_p + \epsilon_q- \epsilon_r- \epsilon_s}
- \frac{1}{2}  \sum_{p,q,r,s}^{\text{denom.}\neq 0} \frac{|\langle pq || rs \rangle|^2 f_p^- f_q^- f_r^+ f_s^+ f_r^- }{\epsilon_p + \epsilon_q- \epsilon_r- \epsilon_s} \label{E2NN} \\
&\equiv& \langle E_I^{(2)} (N_I-\bar{N}) \rangle_{L}.
\end{eqnarray}
The constant part [Eq.\ (\ref{V})] in the first $\hat{V}$ vanishes because the resolvent $\hat{R}_I$ annihilates it.  The constant part in
the second $\hat{V}$ does not contribute, either, because it cannot form a valid contraction with $\hat{N} - \bar{N} = \{ \hat{p}^\dagger \hat{p} \}$.
As a result, the above thermal average is linked. The first term of Eq.\ (\ref{D1D1N}) is evaluated similarly as
\begin{eqnarray}
\langle E_I^{(1)}E_I^{(1)} (N_I-\bar{N}) \rangle &=&  \langle \langle I | \hat{V}\hat{P}_I\hat{V} (\hat{N}-\bar{N})|I \rangle \rangle \nonumber \\
&=& \langle \langle I | \langle E^{(1)}_I\rangle | I \rangle \langle I | F_{pq} 
\contraction[1.0ex]{ \{ }{\hat{p}}{^\dagger \hat{q} \} \{ \hat{r}^\dagger }{\hat{r}}
\contraction[0.5ex]{ \{ \hat{p}^\dagger }{\hat{q}}{ \} \{ }{\hat{r}}
 \{ \hat{p}^\dagger \hat{q} \} \{ \hat{r}^\dagger \hat{r} \} | I \rangle \rangle 
+ \langle \langle I | F_{pq}  
\contraction[1.0ex]{\{ }{\hat{p}}{^\dagger \hat{q} \} \{ \hat{a}^\dagger }{\hat{i}} 
\contraction[0.5ex]{\{ \hat{p}^\dagger }{\hat{q}}{ \} \{ }{\hat{a}} 
\bcontraction[1.0ex]{\{ \hat{p}^\dagger \hat{q} \} \{ }{\hat{a}}{^\dagger \hat{i} \} | I \rangle \langle I | \{ \hat{i}^\dagger }{\hat{a}} 
\bcontraction[0.5ex]{\{ \hat{p}^\dagger \hat{q} \} \{ \hat{a}^\dagger }{\hat{i}}{ \} | I \rangle \langle I | \{ }{\hat{i}} 
\contraction[0.6ex]{\{ \hat{p}^\dagger \hat{q} \} \{ \hat{a}^\dagger \hat{i} \} | I \rangle \langle I | \{ }{\hat{i}}{^\dagger \hat{a} \} \langle E_I^{(1)} \rangle \{ \hat{r}^\dagger }{\hat{r}} 
\contraction[0.5ex]{\{ \hat{p}^\dagger \hat{q} \} \{ \hat{a}^\dagger \hat{i} \} | I \rangle \langle I | \{ \hat{i}^\dagger }{\hat{a}}{ \} \langle E_I^{(1)} \rangle \{ }{\hat{r}} 
\{ \hat{p}^\dagger \hat{q} \} \{ \hat{a}^\dagger \hat{i} \} | I \rangle \langle I | \{ \hat{i}^\dagger \hat{a} \} \langle E_I^{(1)} \rangle \{ \hat{r}^\dagger \hat{r} \} 
| I \rangle \rangle 
\nonumber\\&&
+ \langle \langle I | F_{pq} 
\contraction[1.0ex]{ \{}{\hat{p}}{^\dagger \hat{q}\} \{\hat{a}^\dagger }{\hat{i}}
\contraction[0.5ex]{ \{\hat{p}^\dagger }{\hat{q}}{\} \{}{\hat{a}}
\bcontraction[1.0ex]{ \{\hat{p}^\dagger \hat{q}\} \{}{\hat{a}}{^\dagger \hat{i}\} | I \rangle \langle I | \{\hat{i}^\dagger }{\hat{a}} 
\bcontraction[0.5ex]{ \{\hat{p}^\dagger \hat{q}\} \{\hat{a}^\dagger }{\hat{i}}{\} | I \rangle \langle I | \{}{\hat{i}} 
\contraction[0.6ex]{ \{\hat{p}^\dagger \hat{q}\} \{\hat{a}^\dagger \hat{i}\} | I \rangle \langle I | \{}{\hat{i}}{^\dagger \hat{a}\} \{\hat{r}^\dagger \hat{s}\} F_{rs} \{\hat{t}^\dagger }{\hat{t}}   
\contraction[0.5ex]{ \{\hat{p}^\dagger \hat{q}\} \{\hat{a}^\dagger \hat{i}\} | I \rangle \langle I | \{\hat{i}^\dagger }{\hat{a}}{\} \{}{\hat{r}}
\contraction[0.5ex]{ \{\hat{p}^\dagger \hat{q}\} \{\hat{a}^\dagger \hat{i}\} | I \rangle  \langle I | \{\hat{i}^\dagger \hat{a}\} \{\hat{r}^\dagger }{\hat{s}}{\} F_{rs}\{}{\hat{t}}
 \{\hat{p}^\dagger \hat{q}\} \{\hat{a}^\dagger \hat{i}\} | I \rangle \langle I | \{\hat{i}^\dagger \hat{a}\} \{\hat{r}^\dagger \hat{s}\} F_{rs} \{\hat{t}^\dagger \hat{t}\} | I \rangle \rangle 
 + \langle\langle I| F_{pq}  
\contraction[1.0ex]{ \{}{\hat{p}}{^\dagger \hat{q}\} \{\hat{a}^\dagger }{\hat{i}}
\contraction[0.5ex]{ \{\hat{p}^\dagger }{\hat{q}}{\} \{}{\hat{a}}
\bcontraction[1.0ex]{ \{\hat{p}^\dagger \hat{q}\} \{}{\hat{a}}{^\dagger \hat{i}\} |I \rangle  \langle I|\{\hat{i}^\dagger }{\hat{a}} 
\bcontraction[0.5ex]{ \{\hat{p}^\dagger \hat{q}\} \{\hat{a}^\dagger }{\hat{i}}{\} |I \rangle  \langle I| \{}{\hat{i}} 
\contraction[0.5ex]{ \{\hat{p}^\dagger \hat{q}\} \{\hat{a}^\dagger \hat{i}\} |I \rangle  \langle I| \{\hat{i}^\dagger}{\hat{a}}{\} \{\hat{r}^\dagger \hat{s}\}F_{rs} \{}{\hat{t}}  
\contraction[0.6ex]{ \{\hat{p}^\dagger \hat{q}\} \{\hat{a}^\dagger \hat{i}\} |I \rangle  \langle I| \{}{\hat{i}}{^\dagger \hat{a}\} \{\hat{r}^\dagger }{\hat{s}}
\contraction[1.5ex]{ \{\hat{p}^\dagger \hat{q}\} \{\hat{a}^\dagger \hat{i}\}|I  \rangle \langle I| \{\hat{i}^\dagger \hat{a}\} \{}{\hat{r}}{^\dagger \hat{s}\}F_{rs} \{\hat{t}^\dagger }{\hat{t}}
 \{\hat{p}^\dagger \hat{q}\} \{\hat{a}^\dagger \hat{i}\}|I \rangle  \langle I| \{\hat{i}^\dagger \hat{a}\} \{\hat{r}^\dagger \hat{s}\} F_{rs}\{\hat{t}^\dagger \hat{t}\} |I \rangle \rangle  
\nonumber\\&& 
+ \frac{1}{4} \langle\langle I | F_{pq} 
\contraction[1.0ex]{ \{}{\hat{p}}{^\dagger \hat{q}\} \{\hat{a}^\dagger }{\hat{i}}
\contraction[0.5ex]{ \{\hat{p}^\dagger }{\hat{q}}{\} \{}{\hat{a}}
\bcontraction[1.0ex]{ \{\hat{p}^\dagger \hat{q}\} \{}{\hat{a}}{^\dagger \hat{i}\} |I \rangle  \langle I| \{\hat{i}^\dagger }{\hat{a}} 
\bcontraction[0.5ex]{ \{\hat{p}^\dagger \hat{q}\} \{\hat{a}^\dagger }{\hat{i}}{\} |I \rangle  \langle I| \{}{\hat{i}} 
\contraction[0.5ex]{ \{\hat{p}^\dagger \hat{q}\} \{\hat{a}^\dagger \hat{i}\} |I \rangle \langle I| \{}{\hat{i}}{^\dagger \hat{a}\} \{\hat{r}^\dagger \hat{s}^\dagger \hat{u}\, }{\hat{t}}
\contraction[0.5ex]{ \{\hat{p}^\dagger \hat{q}\} \{\hat{a}^\dagger \hat{i}\} |I \rangle \langle I| \{\hat{i}^\dagger }{\hat{a}}{\} \{}{\hat{r}}
\contraction[1.4ex]{ \{\hat{p}^\dagger \hat{q}\} \{\hat{a}^\dagger \hat{i}\} |I \rangle \langle I| \{\hat{i}^\dagger \hat{a}\} \{\hat{r}^\dagger }{\hat{s}}{^\dagger \hat{u}\, \hat{t}\} \langle rs || tu \rangle \{\hat{v}^\dagger }{\hat{v}}
\contraction[0.5ex]{ \{\hat{p}^\dagger \hat{q}\} \{\hat{a}^\dagger \hat{i}\} |I \rangle \langle I| \{\hat{i}^\dagger \hat{a}\} \{\hat{r}^\dagger \hat{s}^\dagger }{\hat{u} }{\,\hat{t}\}\langle rs || tu \rangle  \{}{\hat{v}}
 \{\hat{p}^\dagger \hat{q}\} \{\hat{a}^\dagger \hat{i}\} |I \rangle \langle I| \{\hat{i}^\dagger \hat{a}\} \{\hat{r}^\dagger \hat{s}^\dagger \hat{u} \,\hat{t}\} \langle rs || tu \rangle  \{\hat{v}^\dagger \hat{v}\} |I \rangle\rangle  \times  4 \nonumber\\
&& + \left( \frac{1}{4} \right)^2 \langle\langle I | \langle pq || rs \rangle 
\contraction[2.0ex]{ \{}{\hat{p}}{^\dagger \hat{q}^\dagger \hat{s}\, \hat{r}\} \{\hat{a}^\dagger\hat{b}^\dagger \hat{j} }{\hat{i}}
\contraction[1.5ex]{ \{\hat{p}^\dagger }{\hat{q}}{^\dagger \hat{s}\, \hat{r}\} \{\hat{a}^\dagger\hat{b}^\dagger }{\hat{j}}
\contraction[1.0ex]{ \{\hat{p}^\dagger \hat{q}^\dagger }{\hat{s}}{\, \hat{r}\} \{\hat{a}^\dagger}{\hat{b}}
\contraction[0.5ex]{ \{\hat{p}^\dagger \hat{q}^\dagger \hat{s}\, }{\hat{r}}{\} \{}{\hat{a}}
\bcontraction[2.0ex]{ \{\hat{p}^\dagger \hat{q}^\dagger \hat{s}\, \hat{r}\} \{}{\hat{a}}{^\dagger\hat{b}^\dagger \hat{j}\, \hat{i}\} |I \rangle \langle I| \{\hat{i}^\dagger \hat{j}^\dagger \hat{b}\,}{\hat{a}}
\bcontraction[1.5ex]{ \{\hat{p}^\dagger \hat{q}^\dagger \hat{s}\, \hat{r}\} \{\hat{a}^\dagger}{\hat{b}}{^\dagger \hat{j}\, \hat{i}\} |I \rangle \langle  I|\{\hat{i}^\dagger \hat{j}^\dagger }{\hat{b}}
\bcontraction[1.0ex]{ \{\hat{p}^\dagger \hat{q}^\dagger \hat{s}\, \hat{r}\} \{\hat{a}^\dagger\hat{b}^\dagger }{\hat{j}}{\, \hat{i}\} |I \rangle \langle I| \{\hat{i}^\dagger }{\hat{j}}
\bcontraction[0.5ex]{ \{\hat{p}^\dagger \hat{q}^\dagger \hat{s}\, \hat{r}\} \{\hat{a}^\dagger\hat{b}^\dagger \hat{j}\, }{\hat{i}}{\} |I \rangle \langle I| \{}{\hat{i}}
\contraction[1.5ex]{ \{\hat{p}^\dagger \hat{q}^\dagger \hat{s}\, \hat{r}\} \{\hat{a}^\dagger\hat{b}^\dagger \hat{j}\, \hat{i}\} |I \rangle \langle I| \{}{\hat{i}}{^\dagger \hat{j}^\dagger \hat{b}\,\hat{a}\} \{\hat{t}^\dagger }{\hat{u}}
\contraction[1.0ex]{ \{\hat{p}^\dagger \hat{q}^\dagger \hat{s}\, \hat{r}\} \{\hat{a}^\dagger\hat{b}^\dagger \hat{j}\, \hat{i}\} |I \rangle \langle I| \{\hat{i}^\dagger }{\hat{j}}{^\dagger \hat{b}\,\hat{a}\} \{\hat{t}^\dagger \hat{u}\} F_{tu} \{\hat{v}^\dagger }{\hat{v}}
\contraction[0.5ex]{ \{\hat{p}^\dagger \hat{q}^\dagger \hat{s}\, \hat{r}\} \{\hat{a}^\dagger\hat{b}^\dagger \hat{j}\, \hat{i}\} |I \rangle \langle I| \{\hat{i}^\dagger \hat{j}^\dagger }{\hat{b}}{\,\hat{a}\} \{\hat{t}^\dagger \hat{u}\} F_{tu} \{}{\hat{v}}
\contraction[0.5ex]{ \{\hat{p}^\dagger \hat{q}^\dagger \hat{s} \hat{r}\} \{\hat{a}^\dagger\hat{b}^\dagger \hat{j}\, \hat{i}\} |I \rangle \langle I| \{\hat{i}^\dagger \hat{j}^\dagger \hat{b}\,}{\hat{a}}{\} \{}{\hat{t}}
 \{\hat{p}^\dagger \hat{q}^\dagger \hat{s}\, \hat{r}\} \{\hat{a}^\dagger\hat{b}^\dagger \hat{j}\, \hat{i}\} |I \rangle \langle I| \{\hat{i}^\dagger \hat{j}^\dagger \hat{b}\,\hat{a}\} \{\hat{t}^\dagger \hat{u}\} F_{tu} \{\hat{v}^\dagger \hat{v}\} |I \rangle \rangle  \times 16 \nonumber \\
&& + \left( \frac{1}{4} \right)^3 \langle\langle I | \langle pq || rs \rangle  
\contraction[2.0ex]{ \{}{\hat{p}}{^\dagger \hat{q}^\dagger \hat{s}\, \hat{r}\} \{\hat{a}^\dagger\hat{b}^\dagger \hat{j}\, }{\hat{i}}
\contraction[1.5ex]{ \{\hat{p}^\dagger }{\hat{q}}{^\dagger \hat{s}\, \hat{r}\} \{\hat{a}^\dagger\hat{b}^\dagger }{\hat{j}}
\contraction[1.0ex]{ \{\hat{p}^\dagger \hat{q}^\dagger }{\hat{s}}{\, \hat{r}\} \{\hat{a}^\dagger}{\hat{b}}
\contraction[0.5ex]{ \{\hat{p}^\dagger \hat{q}^\dagger \hat{s}\, }{\hat{r}}{\} \{}{\hat{a}}
\bcontraction[2.0ex]{ \{\hat{p}^\dagger \hat{q}^\dagger \hat{s}\, \hat{r}\} \{}{\hat{a}}{^\dagger\hat{b}^\dagger \hat{j}\, \hat{i}\} |I \rangle  \langle I|  \{\hat{i}^\dagger \hat{j}^\dagger \hat{b}\,}{\hat{a}}
\bcontraction[1.5ex]{ \{\hat{p}^\dagger \hat{q}^\dagger \hat{s} \,\hat{r}\} \{\hat{a}^\dagger}{\hat{b}}{^\dagger \hat{j} \,\hat{i}\} |I \rangle  \langle I|  \{\hat{i}^\dagger \hat{j}^\dagger }{\hat{b}}
\bcontraction[1.0ex]{ \{\hat{p}^\dagger \hat{q}^\dagger \hat{s}\, \hat{r}\} \{\hat{a}^\dagger\hat{b}^\dagger }{\hat{j}}{\, \hat{i}\} |I \rangle  \langle I|  \{\hat{i}^\dagger }{\hat{j}}
\bcontraction[0.5ex]{ \{\hat{p}^\dagger \hat{q}^\dagger \hat{s}\, \hat{r}\} \{\hat{a}^\dagger\hat{b}^\dagger \hat{j}\, }{\hat{i}}{\} |I \rangle  \langle I|  \{}{\hat{i}}
\contraction[1.5ex]{ \{\hat{p}^\dagger \hat{q}^\dagger \hat{s}\, \hat{r}\} \{\hat{a}^\dagger\hat{b}^\dagger \hat{j}\, \hat{i}\} |I \rangle  \langle I|  \{}{\hat{i}}{^\dagger \hat{j}^\dagger \hat{b}\,\hat{a}\} \{\hat{t}^\dagger\hat{u}^\dagger \hat{w}\, \hat{v}\} \langle tu || vw \rangle\{\hat{x}^\dagger }{\hat{x}}
\contraction[1.0ex]{ \{\hat{p}^\dagger \hat{q}^\dagger \hat{s}\, \hat{r}\} \{\hat{a}^\dagger\hat{b}^\dagger \hat{j}\, \hat{i}\} |I \rangle  \langle I|  \{\hat{i}^\dagger }{\hat{j}}{^\dagger \hat{b}\,\hat{a}\} \{\hat{t}^\dagger\hat{u}^\dagger }{\hat{w}}
\contraction[0.5ex]{ \{\hat{p}^\dagger \hat{q}^\dagger \hat{s}\, \hat{r}\} \{\hat{a}^\dagger\hat{b}^\dagger \hat{j}\, \hat{i}\} |I \rangle  \langle I|  \{\hat{i}^\dagger \hat{j}^\dagger }{\hat{b}}{\,\hat{a}\} \{\hat{t}^\dagger}{\hat{u}}
\contraction[0.5ex]{ \{\hat{p}^\dagger \hat{q}^\dagger \hat{s}\, \hat{r}\} \{\hat{a}^\dagger\hat{b}^\dagger \hat{j}\, \hat{i}\} |I \rangle  \langle I|  \{\hat{i}^\dagger \hat{j}^\dagger \hat{b}\,}{\hat{a}}{\} \{}{\hat{t}}
\contraction[0.5ex]{ \{\hat{p}^\dagger \hat{q}^\dagger \hat{s}\, \hat{r}\} \{\hat{a}^\dagger\hat{b}^\dagger \hat{j}\, \hat{i}\} |I \rangle  \langle I|  \{\hat{i}^\dagger \hat{j}^\dagger \hat{b}\,\hat{a}\} \{\hat{t}^\dagger\hat{u}^\dagger \hat{w}\, }{\hat{v}}{\} \langle tu || vw \rangle\{}{\hat{x}}
 \{\hat{p}^\dagger \hat{q}^\dagger \hat{s} \,\hat{r}\} \{\hat{a}^\dagger\hat{b}^\dagger \hat{j} \,\hat{i}\} |I \rangle  \langle I| \{\hat{i}^\dagger \hat{j}^\dagger \hat{b}\,\hat{a}\} \{\hat{t}^\dagger\hat{u}^\dagger \hat{w} \,\hat{v}\} \langle tu || vw \rangle \{\hat{x}^\dagger \hat{x}\} |I \rangle \rangle  \times 32\nonumber \\
&& + \left( \frac{1}{4} \right)^3 \langle\langle I | \langle pq || rs \rangle 
\contraction[2.0ex]{ \{}{\hat{p}}{^\dagger \hat{q}^\dagger \hat{s}\, \hat{r}\} \{\hat{a}^\dagger\hat{b}^\dagger \hat{j}\, }{\hat{i}}
\contraction[1.5ex]{ \{\hat{p}^\dagger }{\hat{q}}{^\dagger \hat{s}\, \hat{r}\} \{\hat{a}^\dagger\hat{b}^\dagger }{\hat{j}}
\contraction[1.0ex]{ \{\hat{p}^\dagger \hat{q}^\dagger }{\hat{s}}{\, \hat{r}\} \{\hat{a}^\dagger}{\hat{b}}
\contraction[0.5ex]{ \{\hat{p}^\dagger \hat{q}^\dagger \hat{s}\, }{\hat{r}}{\} \{}{\hat{a}}
\bcontraction[2.0ex]{ \{\hat{p}^\dagger \hat{q}^\dagger \hat{s}\, \hat{r}\} \{}{\hat{a}}{^\dagger\hat{b}^\dagger \hat{j}\, \hat{i}\} |I \rangle \langle I| \{\hat{i}^\dagger \hat{j}^\dagger \hat{b}\,}{\hat{a}}
\bcontraction[1.5ex]{ \{\hat{p}^\dagger \hat{q}^\dagger \hat{s}\, \hat{r}\} \{\hat{a}^\dagger}{\hat{b}}{^\dagger \hat{j}\, \hat{i}\} |I \rangle \langle I| \{\hat{i}^\dagger \hat{j}^\dagger }{\hat{b}}
\bcontraction[1.0ex]{ \{\hat{p}^\dagger \hat{q}^\dagger \hat{s}\, \hat{r}\} \{\hat{a}^\dagger\hat{b}^\dagger }{\hat{j}}{\, \hat{i}\} |I \rangle \langle I| \{\hat{i}^\dagger }{\hat{j}}
\bcontraction[0.5ex]{ \{\hat{p}^\dagger \hat{q}^\dagger \hat{s}\, \hat{r}\} \{\hat{a}^\dagger\hat{b}^\dagger \hat{j}\, }{\hat{i}}{\} |I \rangle \langle I| \{}{\hat{i}}
\contraction[1.8ex]{ \{\hat{p}^\dagger \hat{q}^\dagger \hat{s}\, \hat{r}\} \{\hat{a}^\dagger\hat{b}^\dagger \hat{j}\, \hat{i}\} |I \rangle \langle I| \{}{\hat{i}}{^\dagger \hat{j}^\dagger \hat{b}\,\hat{a}\} \{\hat{t}^\dagger\hat{u}^\dagger \hat{w}\, }{\hat{v}}
\contraction[1.4ex]{ \{\hat{p}^\dagger \hat{q}^\dagger \hat{s}\, \hat{r}\} \{\hat{a}^\dagger\hat{b}^\dagger \hat{j}\, \hat{i}\} |I \rangle \langle I| \{\hat{i}^\dagger }{\hat{j}}{^\dagger \hat{b}\,\hat{a}\} \{\hat{t}^\dagger\hat{u}^\dagger }{\hat{w}}
\contraction[1.0ex]{ \{\hat{p}^\dagger \hat{q}^\dagger \hat{s}\, \hat{r}\} \{\hat{a}^\dagger\hat{b}^\dagger \hat{j}\, \hat{i}\} |I \rangle \langle I| \{\hat{i}^\dagger \hat{j}^\dagger }{\hat{b}}{\,\hat{a}\} \{\hat{t}^\dagger}{\hat{u}}
\contraction[1.1ex]{ \{\hat{p}^\dagger \hat{q}^\dagger \hat{s}\, \hat{r}\} \{\hat{a}^\dagger\hat{b}^\dagger \hat{j}\, \hat{i}\} |I \rangle \langle I| \{\hat{i}^\dagger \hat{j}^\dagger \hat{b}\,}{\hat{a}}{\} \{\hat{t}^\dagger\hat{u}^\dagger \hat{w} \,\hat{v}\} \langle tu || vw \rangle \{}{\hat{x}}
\contraction[0.4ex]{ \{\hat{p}^\dagger \hat{q}^\dagger \hat{s} \,\hat{r}\} \{\hat{a}^\dagger\hat{b}^\dagger \hat{j}\, \hat{i}\} |I \rangle \langle I| \{\hat{i}^\dagger \hat{j}^\dagger \hat{b}\,\hat{a}\} \{}{\hat{t}}{^\dagger\hat{u}^\dagger \hat{w}\, \hat{v}\} \langle tu || vw \rangle \{\hat{x}^\dagger }{\hat{x}}
 \{\hat{p}^\dagger \hat{q}^\dagger \hat{s}\, \hat{r}\} \{\hat{a}^\dagger\hat{b}^\dagger \hat{j} \,\hat{i}\} |I \rangle \langle I| \{\hat{i}^\dagger \hat{j}^\dagger \hat{b}\,\hat{a}\} \{\hat{t}^\dagger\hat{u}^\dagger \hat{w}\, \hat{v}\} \langle tu || vw \rangle \{\hat{x}^\dagger \hat{x}\} | I \rangle \rangle  \times  32 \\
&=& 2 \langle E^{(1)}_I\rangle \sum_p F_{pp} f_p^- f_p^+ + \sum_{p,q}^{\text{denom.}= 0} {|F_{pq}|^2}  f_p^- f_q^+ f_p^+
-\sum_{p,q}^{\text{denom.}= 0} {|F_{pq}|^2}  f_p^- f_q^+ f_q^- 
\nonumber \\&& 
+ \sum_{p,q,s}^{\text{denom.}= 0} {F_{pq} \langle qs || ps \rangle } f_p^- f_q^+ f_s^-  f_s^+ 
+ \sum_{p,q,r}^{\text{denom.}= 0} {\langle pq || rq \rangle F_{rp}  } f_p^- f_q^- f_q^+  f_r^+  
\nonumber \\&& 
+ \frac{1}{2}  \sum_{p,q,r,s}^{\text{denom.}= 0} {|\langle pq || rs \rangle|^2 }
f_p^- f_q^- f_r^+ f_s^+ f_p^+ 
- \frac{1}{2}  \sum_{p,q,r,s}^{\text{denom.}= 0} {|\langle pq || rs \rangle|^2 }
f_p^- f_q^- f_r^+ f_s^+ f_r^-   \\
&\equiv& 2 \langle E_I^{(1)}\rangle_L\langle E_I^{(1)} (N_I-\bar{N}) \rangle_{L} + \langle E_I^{(1)}E_I^{(1)} (N_I-\bar{N}) \rangle_{L}.
\end{eqnarray}
\end{widetext}
The first term is unlinked, as each factor is extensive, making the product grow quadratically with size. 
This unlinked term will be cancelled  exactly (see below). 

Likewise, the thermal average in the second term of Eq.\ (\ref{D1D1N}) is evaluated as
\begin{eqnarray}
&& \langle E_I^{(1)}N_I (N_I-\bar{N}) \rangle \nonumber\\
&&=  \langle \langle I | \hat{V} \hat{N} (\hat{N}-\bar{N})|I \rangle \rangle \nonumber \\
&&= \langle \langle I | \langle E^{(1)}_I\rangle 
\contraction[1.0ex]{ \{ }{\hat{p}}{^\dagger \hat{p} \} \{ \hat{q}^\dagger }{\hat{q}}
\contraction[0.5ex]{ \{ \hat{p}^\dagger }{\hat{p}}{ \} \{ }{\hat{q}}
 \{ \hat{p}^\dagger \hat{p} \} \{ \hat{q}^\dagger \hat{q} \} | I \rangle \rangle 
%
+ \langle \langle I | F_{pq}  
\contraction[1.0ex]{\{ }{\hat{p}}{^\dagger \hat{q} \} \bar{N} \{ \hat{r}^\dagger }{\hat{r}} 
\contraction[0.5ex]{\{ \hat{p}^\dagger }{\hat{q}}{ \} \bar{N} \{ }{\hat{r}} 
\{ \hat{p}^\dagger \hat{q} \} \bar{N} \{ \hat{r}^\dagger \hat{r} \} 
| I \rangle \rangle 
\nonumber\\&&
+ \langle \langle I |F_{pq} 
\contraction[1.0ex]{\{ }{\hat{p}}{^\dagger \hat{q} \} \{ \hat{r}^\dagger \hat{r} \} \{ \hat{s}^\dagger }{\hat{s}}
\contraction[0.5ex]{\{ \hat{p}^\dagger }{\hat{q}}{ \} \{ }{\hat{r}}
\contraction[0.5ex]{\{ \hat{p}^\dagger \hat{q} \} \{ \hat{r}^\dagger }{\hat{r}}{ \} \{ }{\hat{s}}
\{ \hat{p}^\dagger \hat{q} \} \{ \hat{r}^\dagger \hat{r} \} \{ \hat{s}^\dagger \hat{s} \}
 | I \rangle \rangle 
%
+ \langle \langle I |F_{pq} 
\contraction[0.5ex]{\{ }{\hat{p}}{^\dagger \hat{q} \} \{ \hat{r}^\dagger }{\hat{r}}
\contraction[1.0ex]{\{ \hat{p}^\dagger }{\hat{q}}{ \} \{ \hat{r}^\dagger \hat{r} \} \{ }{\hat{s}}
\contraction[1.5ex]{\{ \hat{p}^\dagger \hat{q} \} \{ }{\hat{r}}{^\dagger \hat{r} \} \{ \hat{s}^\dagger }{\hat{s}}
\{ \hat{p}^\dagger \hat{q} \} \{ \hat{r}^\dagger \hat{r} \} \{ \hat{s}^\dagger \hat{s} \}
 | I \rangle \rangle 
\nonumber\\&&
+ \frac{1}{4} \langle \langle I |\langle pq || rs \rangle
\contraction[1.0ex]{\{ }{\hat{p}}{^\dagger \hat{q}^\dagger \hat{s}\,  \hat{r} \} \{ \hat{t}^\dagger }{\hat{t}}
\contraction[0.5ex]{\{ \hat{p}^\dagger \hat{q}^\dagger \hat{s}\, }{ \hat{r}}{ \} \{ }{\hat{t}}
\contraction[2.0ex]{\{ \hat{p}^\dagger }{\hat{q}}{^\dagger \hat{s}\,  \hat{r} \} \{ \hat{t}^\dagger \hat{t} \} \{ \hat{u}^\dagger }{\hat{u} }
\contraction[1.5ex]{\{ \hat{p}^\dagger \hat{q}^\dagger}{ \hat{s}}{\,  \hat{r} \} \{ \hat{t}^\dagger \hat{t} \} \{ }{\hat{u}}
\{ \hat{p}^\dagger \hat{q}^\dagger \hat{s}\,  \hat{r} \} \{ \hat{t}^\dagger \hat{t} \} \{ \hat{u}^\dagger \hat{u} \}
 | I \rangle \rangle \times 4
\nonumber \\
&&=  
\langle E^{(1)}_I\rangle \sum_{p} f_p^- f_p^+ 
+ \bar{N} \sum_{p} F_{pp} f_p^- f_p^+
+ \sum_{p} F_{pp} f_p^- f_p^+ f_p^+ 
\nonumber\\&&
- \sum_{p} F_{pp} f_p^- f_p^+  f_p^- 
+ \sum_{p,q} \langle pq || pq \rangle f_p^- f_p^+  f_q^- f_q^+  \label{verylong2} \\
&&\equiv \langle E_I^{(1)}\rangle_L\langle N_I^{2} -\bar{N}^2 \rangle_{L} 
+ \bar{N} \langle E_I^{(1)} (N_I-\bar{N}) \rangle_{L} 
\nonumber\\&&
+ \langle E_I^{(1)}N_I (N_I-\bar{N}) \rangle_{L}.
\end{eqnarray}
The first two terms are unlinked. 

The thermal average in the third term of Eq.\ (\ref{D1D1N}) becomes
\begin{eqnarray}
&& \langle N_I N_I (N_I-\bar{N}) \rangle \nonumber\\
&&=  \langle \langle I | \hat{N} \hat{N} (\hat{N}-\bar{N})|I \rangle \rangle \nonumber \\
&&= \langle \langle I | \bar{N} 
\contraction[1.0ex]{ \{ }{\hat{p}}{^\dagger \hat{p} \} \{ \hat{q}^\dagger }{\hat{q}}
\contraction[0.5ex]{ \{ \hat{p}^\dagger }{\hat{p}}{ \} \{ }{\hat{q}}
 \{ \hat{p}^\dagger \hat{p} \} \{ \hat{q}^\dagger \hat{q} \} | I \rangle \rangle 
%
+ \langle \langle I | 
\contraction[1.0ex]{\{ }{\hat{p}}{^\dagger \hat{p} \} \bar{N} \{ \hat{q}^\dagger }{\hat{q}} 
\contraction[0.5ex]{\{ \hat{p}^\dagger }{\hat{p}}{ \} \bar{N} \{ }{\hat{q}} 
\{ \hat{p}^\dagger \hat{p} \} \bar{N} \{ \hat{q}^\dagger \hat{q} \} 
| I \rangle \rangle 
\nonumber\\&&
+ \langle \langle I |
\contraction[1.0ex]{\{ }{\hat{p}}{^\dagger \hat{q} \} \{ \hat{r}^\dagger \hat{r} \} \{ \hat{s}^\dagger }{\hat{s}}
\contraction[0.5ex]{\{ \hat{p}^\dagger }{\hat{q}}{ \} \{ }{\hat{r}}
\contraction[0.5ex]{\{ \hat{p}^\dagger \hat{q} \} \{ \hat{r}^\dagger }{\hat{r}}{ \} \{ }{\hat{s}}
\{ \hat{p}^\dagger \hat{q} \} \{ \hat{r}^\dagger \hat{r} \} \{ \hat{s}^\dagger \hat{s} \}
 | I \rangle \rangle 
%
+ \langle \langle I |
\contraction[0.5ex]{\{ }{\hat{p}}{^\dagger \hat{q} \} \{ \hat{r}^\dagger }{\hat{r}}
\contraction[1.0ex]{\{ \hat{p}^\dagger }{\hat{q}}{ \} \{ \hat{r}^\dagger \hat{r} \} \{ }{\hat{s}}
\contraction[1.5ex]{\{ \hat{p}^\dagger \hat{q} \} \{ }{\hat{r}}{^\dagger \hat{r} \} \{ \hat{s}^\dagger }{\hat{s}}
\{ \hat{p}^\dagger \hat{q} \} \{ \hat{r}^\dagger \hat{r} \} \{ \hat{s}^\dagger \hat{s} \}
 | I \rangle \rangle 
\nonumber \\
&&=  
2 \bar{N} \sum_{p} f_p^- f_p^+ 
+ \sum_{p} f_p^- f_p^+ f_p^+ 
- \sum_{p}  f_p^- f_p^+  f_p^- \\
&&\equiv 2 \bar{N} \langle N_I^2 -\bar{N}^2 \rangle_{L} 
+ \langle N_I N_I (N_I-\bar{N}) \rangle_{L}.
\end{eqnarray}
The first  term is unlinked. 

Substituting these into Eq.\ (\ref{mu2sq}), we again
 observe systematic cancellations of unlinked terms with the aid of Eq.\ (\ref{mu1_linked}), leaving 
\begin{eqnarray}
&& \mu^{(2)} \langle N_I^2 - \bar{N}^2 \rangle_{L} \nonumber\\
&&=  \langle E_I^{(2)} (N_I - \bar{N})\rangle_{L} 
+ \frac{(-\beta)}{2!} \langle E_I^{(1)}E_I^{(1)} (N_I - \bar{N})\rangle_{L}
\nonumber\\&&
+ \beta \mu^{(1)} \langle E_I^{(1)} N_I (N_I - \bar{N})\rangle_{L}
+ \frac{(-\beta)}{2!} \left( \mu^{(1)} \right)^2 \langle N_I N_I (N_I - \bar{N})\rangle_{L} \nonumber \\
&&= \langle E_I^{(2)} (N_I - \bar{N})\rangle_{L} 
+ \frac{(-\beta)}{2!} \langle D_I^{(1)} D_I^{(1)} (N_I - \bar{N})\rangle_{L}, \label{mu2sq_linked}
\end{eqnarray}
which translates to 
\begin{eqnarray}
&& \mu^{(2)} \sum_p f_p^- f_p^+ \nonumber\\
&& = 
\sum_{p,q}^{\text{denom.}\neq 0} \frac{|F_{pq}|^2  f_p^- f_q^+ (f_p^+ - f_q^-)}{\epsilon_p - \epsilon_q} 
\nonumber\\&&
 + \sum_{p,q,r}^{\text{denom.}\neq 0} \frac{(F_{pq} \langle qr || pr \rangle + \langle pr || qr \rangle F_{qp})f_p^- f_q^+ f_r^-  f_r^+  }{\epsilon_p - \epsilon_q} 
 \nonumber \\
&& + \frac{1}{4}  \sum_{p,q,r,s}^{\text{denom.}\neq 0} \frac{|\langle pq || rs \rangle|^2 f_p^- f_q^- f_r^+ f_s^+ (f_p^+ + f_q^+ - f_r^- - f_s^-)}{\epsilon_p + \epsilon_q- \epsilon_r- \epsilon_s}
\nonumber\\
&& -\frac{\beta}{2} \sum_{p,q}^{\text{denom.}= 0} {|F_{pq}|^2}  f_p^- f_q^+ (f_p^+ - f_q^-)
\nonumber\\&&
 -\frac{\beta}{2} \sum_{p,q,r}^{\text{denom.}= 0} (F_{pq} \langle qr || pr \rangle + \langle pr || qr \rangle F_{qp}) f_p^- f_q^+ f_r^-  f_r^+ 
 \nonumber \\
&& -\frac{\beta}{8}  \sum_{p,q,r,s}^{\text{denom.}= 0} {|\langle pq || rs \rangle|^2 }f_p^- f_q^- f_r^+ f_s^+ (f_p^+ + f_q^+ - f_r^- - f_s^-) 
\nonumber\\&&
+\beta \mu^{(1)} \sum_{p} F_{pp} f_p^- f_p^+ (f_p^+ - f_p^-) 
+ \beta \mu^{(1)}  \sum_{p,q} \langle pq || pq \rangle f_p^- f_p^+  f_q^- f_q^+ 
\nonumber\\&& 
- \frac{\beta}{2} \left( \mu^{(1)} \right)^2 \sum_{p} f_p^- f_p^+ (f_p^+ - f_p^-).
\end{eqnarray} 
This is identified as Eq.\ (69) of Ref.\ \onlinecite{Hirata2ndorder}.

\subsection{Third order}

We shall derive the reduced analytical formulas for $\Omega^{(3)}$ using diagrams in Sec.\ \ref{sec:diagrams}. Here, we focus on its most important 
contribution, $\langle E^{(3)} \rangle$, since this is where a linked renormalization term appears for the first time, giving rise to a new, unique class of 
the renormalization diagrams. 

According to Eq.\ (\ref{HCPTrec3}), the third-order correction to the energy matrix of HCPT is written as
\begin{eqnarray}
{E}^{(3)}_{IJ} &=& \langle I | \hat{V}\hat{R}_I\hat{V}\hat{R}_I\hat{V} | J \rangle - 
\sum_{K}   \langle I | \hat{V} \hat{R}_I \hat{R}_I\hat{V} | K \rangle E_{KJ}^{(1)},  \label{E3}
\end{eqnarray}
where $K$ runs over all states that are degenerate with $I$ and $J$. The second term is the renormalization term
of the Rayleigh--Schr\"{o}dinger perturbation theory.\cite{shavitt} 

If $I$ is nondegenerate, HCPT reduces to MPPT, and $I = J = K$. 
The first term then consists of the linked and unlinked contributions, and is thus written as
\begin{eqnarray}
\langle I | \hat{V}\hat{R}_I\hat{V}\hat{R}_I\hat{V} | I \rangle &=& \langle I | \hat{V}\hat{R}_I\hat{V}\hat{R}_I\hat{V} | I \rangle_{L} \nonumber\\
&& + \langle I | \hat{V} \hat{R}_I \hat{R}_I\hat{V} | I \rangle_{L} E_I^{(1)} , \label{E3nondeg}
\end{eqnarray}
with the unlinked (second) term cancelling exactly the renormalization term [the second term of Eq.\ (\ref{E3})], leaving
\begin{eqnarray}
{E}^{(3)}_{I} &=& \langle I | \hat{V}\hat{R}_I\hat{V}\hat{R}_I\hat{V} | I \rangle_{L}. 
\end{eqnarray}

When $I$ is degenerate, the cancellation of unlinked terms may or may not be as straightforward in individual matrix elements, $E_{IJ}^{(3)}$.
However, the cancellation occurs completely analogously and straightforwardly in the thermal average, $\langle E_I^{(3)} \rangle$. Using the trace invariance of $\bm{E}^{(3)}$, we have
\begin{eqnarray}
\langle {E}^{(3)}_{I} \rangle &=& \langle {E}^{(3)}_{II} \rangle \nonumber\\
&=& \langle \langle I | \hat{V}\hat{R}_I\hat{V}\hat{R}_I\hat{V} | I \rangle \rangle - 
\langle  \langle I | \hat{V} \hat{R}_I \hat{R}_I\hat{V} | K \rangle E^{(1)}_{KI} \rangle \nonumber\\
&=& \langle \langle I | \hat{V}\hat{R}_I\hat{V}\hat{R}_I\hat{V} | I \rangle \rangle - 
\langle  \langle I | \hat{V} \hat{R}_I \hat{R}_I\hat{V} | K \rangle \langle K | \hat{V} | I \rangle \rangle \nonumber\\
&=& \langle \langle I | \hat{V}\hat{R}_I\hat{V}\hat{R}_I\hat{V} | I \rangle \rangle - 
\langle \langle I | \hat{V}  \hat{R}_I \hat{R}_I\hat{V} \hat{P}_I \hat{V} | I \rangle \rangle,  \label{E3sq}
\end{eqnarray}
where summation symbols are again suppressed.
Each term of the last line is then decomposed into linked and unlinked contributions. Recalling the 
normal-ordered form of $\hat{V}$ [Eq.\ (\ref{V})], we can decompose each term as
\begin{eqnarray}
\langle \langle I | \hat{V}\hat{R}_I\hat{V}\hat{R}_I\hat{V} | I \rangle \rangle 
&=& \langle \langle I | \hat{V}\hat{R}_I\hat{V}\hat{R}_I\hat{V} | I \rangle \rangle_{L} \nonumber\\
&& + 
\contraction[0.5ex]{\langle \langle I | }{\hat{V}}{}{\hat{R}}
\contraction[0.5ex]{\langle \langle I | \hat{V}\hat{R}_I \langle E^{(1)}_I\rangle }{\hat{R}}{_I}{\hat{V}}
\bcontraction[0.5ex]{\langle \langle I | \hat{V}}{\hat{R}}{_I \langle E^{(1)}_I\rangle }{\hat{R}}
\langle \langle I | \hat{V}\hat{R}_I \langle E^{(1)}_I\rangle \hat{R}_I\hat{V} | I \rangle \rangle, \label{unlinked_example1} \\
\langle \langle I | \hat{V}  \hat{R}_I \hat{R}_I\hat{V} \hat{P}_I \hat{V} | I \rangle \rangle &=&
\langle \langle I | \hat{V}  \hat{R}_I \hat{R}_I\hat{V} \hat{P}_I \hat{V} | I \rangle \rangle_{L} \nonumber\\
&& + \langle \langle I | 
\contraction[0.5ex]{}{\hat{V}}{}{ \hat{R}} 
\bcontraction[0.5ex]{\hat{V} }{\hat{R}}{_I }{\hat{R}}
\contraction[0.5ex]{\hat{V} \hat{R}_I }{\hat{R}}{_I}{\hat{V}}
\hat{V} \hat{R}_I \hat{R}_I\hat{V}
\langle E^{(1)}_I\rangle  | I \rangle \rangle, \label{unlinked_example2}
\end{eqnarray}
where the unlinked  contraction patterns are indicated by the staple symbols
in the second terms, which are found to have an equal value. 
They therefore cancel each other in Eq.\ (\ref{E3sq}), leaving linked $\langle E^{(3)}\rangle$:
\begin{eqnarray}
\langle {E}^{(3)}_{I} \rangle &=& \langle \langle I | \hat{V}\hat{R}_I\hat{V}\hat{R}_I\hat{V} | I \rangle \rangle_{L} - 
\langle \langle I | \hat{V} \hat{R}_I \hat{R}_I\hat{V} \hat{P}_I \hat{V} | I \rangle \rangle_{L}. \label{E3sqlinked}
\end{eqnarray}

Unlike in MPPT, the renormalization term in HCPT has both unlinked and linked contributions with the former cancelling exactly
the unlinked contribution in the parent term, while the latter is nonzero and therefore should not be overlooked or neglected. 
In the above example, this linked renormalized contribution 
has one resolvent ($\hat{R}_I$) operator
shifted from in between the last two $\hat{V}$ operators to in between the first two $\hat{V}$ operators. Diagrammatically,
it will correspond to a closed, connected diagram with no resolvent line between a pair of adjacent vertexes but with two resolvent lines crammed in between the other pair
of adjacent vertexes (see Sec.\ \ref{sec:Omega3}). This unusual diagram with a displaced resolvent occurs at $T > 0$, and should
be distinguished from the anomalous diagrams\cite{kohn,SANTRA} with missing resolvents.
In the next section, we will document the reduced analytical formula of $\langle E_I^{(3)} \rangle$ obtained with diagrams.

\section{Feynman diagrams\label{sec:diagrams}}

\begin{table}
\caption{Rules to generate all $\Omega^{(n)}$ diagrams in the Hugenholtz style ($n \geq 2$).
See the text for generating diagrammatic equations for $\mu^{(n)}$, and $S^{(n)}$  from 
the $\Omega^{(n)}$ equation.}
\label{tab:rules1}
\begin{ruledtabular}
\begin{tabular}{rl}
(1) & Place $n$ vertexes in an unambiguous vertical order, using \\
& three types of vertexes: filled-circle two-line ($F_{pq}$) vertex,  \\
& numbered two-line ($\mu$) vertex, and filled-circle four-line \\ 
& ($\langle pq|| rs \rangle$) vertex. Count an $\mu^{(m)}$ vertex as $m$ vertexes. \\
(2) & Connect all vertexes with lines to form a closed connected \\
& diagram in all topologically distinct manner. Each two-line \\
& (four-line) vertex should be connected with one outgoing \\
&   and one incoming (two outgoing and two incoming) lines. \\
(3) & Insert zero through $n-1$ resolvent (wiggly) lines into the  \\
& diagram in all $2^{n-1}$ ways with either zero or one resolvent \\
& in between each pair of adjacent vertexes. Call the diagram  \\
& with $n-1$ resolvents a normal diagram and the one with  \\
& $n-2$ or less resolvents an anomalous diagram. \\
(4) & Starting with the normal diagram, shift upwards one \\
& through $n-2$ resolvents in all possible ways. Call the  diagram \\
&  with at least one shifted resolvent a renormalization diagram.\\
\end{tabular}
\end{ruledtabular}
\end{table}

\begin{table}
\caption{Rules to transform an $\Omega^{(n)}$, $\mu^{(n)}$, or $S^{(n)}$ diagram in the Hugenholtz style to 
an algebraic formula ($n \geq 1$).}
\label{tab:rules2}
\begin{ruledtabular}
\begin{tabular}{rl}
(1) & Label lines with indexes $p$, $q$, $r$, $s$, etc. \\
(2) &  Associate each filled-circle two-line vertex with $F_{pq}$, where \\
&   $p$ ($q$) is the outgoing (incoming) line label. \\
(3) & Associate each vertex with number $n$ with $- \mu^{(n)}\delta_{pq}$, where \\
&  $p$ ($q$) is the outgoing (incoming) line label. \\
(4) & The $F_{pq}$ and $\mu^{(1)}$ vertexes can be consolidated into a shaded \\
& two-line vertex to reduce the number of diagrams. If this  \\
& simplification is used, associate the vertex with $F_{pq} - \mu^{(1)}\delta_{pq}$. \\
(5) & Associate each filled-circle four-line vertex with $\langle pq || rs \rangle$, \\
& where $p$, $q$, $r$, and $s$ are the left-outgoing, right-outgoing, \\
& left-incoming, and right-incoming line labels, respectively.\\
(6) & Associate a downgoing (upgoing) line with $f_p^-$ ($f_p^+$). \\
& The directionality of a bubble is judged at the farthest point \\
& from the vertex or from its point of contact. \\
(7) & Associate each dashed bubble with $\beta(\epsilon_p - \mu^{(0)})$ additionally.\\
(8) & Associate each resolvent line with $1/(\epsilon_p + \epsilon_q \dots - \epsilon_r - \epsilon_s \dots)$, \\
& where $p, q,\dots$ ($r, s,\dots$) are the labels of the downgoing \\
& (upgoing) lines intersecting  the resolvent line. \\
(9) & Sum over all line indexes. Restrict the summations to cases \\
& that existing denominator factors are nonzero and fictitious \\
& (due to missing resolvents) denominator factors are zero.\\
(10) & Multiply $1/n!$ for each set of $n$ equivalent lines. Two lines are \\
& equivalent when they start from a same vertex and end at a \\
& same vertex as well as have the identical line directions. \\
& Ignore attached bubbles when judging the equivalence.\\
(11) & Multiply $(-1)^{h+l}$ to the diagram with $h$ downgoing lines and \\
& $l$ loops. A bubble counts as a loop.\\
(12) & Multiply $(-\beta)^n / (n+1)!$ to the diagram with $n$ missing \\
& resolvents. Further multiply with $n$ to an $S^{(n)}$ diagram with  \\
& no dashed bubble. \\
(13) & Multiply $(-1)^n$ to the diagram with $n$ shifted resolvents.\\
\end{tabular}
\end{ruledtabular}
\end{table}

The diagrammatic rules are obtained by a systematic modification of the zero-temperature counterpart\cite{shavitt,mattuck1992guide,march} and justified by
the one-to-many correspondence with the foregoing normal-ordered second-quantization logic. The rules to generate diagrams and to interpret them
algebraically are given in Tables \ref{tab:rules1} and \ref{tab:rules2}, respectively. 

Each vertex graphically represents 
a matrix element of the normal-ordered Hamiltonian or number operator. 
Each line (edge) connecting two vertexes corresponds to a Wick contraction.
That only full contractions yield a nonzero thermal average translates to the diagrammatic rule that all lines must be terminated by vertexes at both ends 
to form a closed diagram (no dangling line). 
An internal contraction is depicted as a `bubble,' i.e., a contraction of a vertex with itself forming a short loop. 
In a $\Omega^{(n)}$ diagram, the bubbles are seen only as a part of the normal-ordered Hamiltonian or number 
operator such as $\langle E^{(1)}_I\rangle$, $F_{pq}$, and $\bar{N}$. This is because outside of these operator definitions, 
internal contractions are zero in a thermal average (see Appendix \ref{app:secondquantization}). 
An unlinked diagram (`unlinked' is synonymous with `disconnected' for a closed diagram) 
corresponds to a full contraction pattern that results in a product of two or more scalars with no shared summation index
[e.g., the second terms of Eqs.\ (\ref{unlinked_example1}) and (\ref{unlinked_example2})]. Section \ref{sec:linkeddiagramtheorem} proves
that all thermodynamic functions of the finite-temperature MBPT are represented by linked diagrams only. 

\subsection{First order} 

\begin{figure}
\includegraphics[scale=0.75]{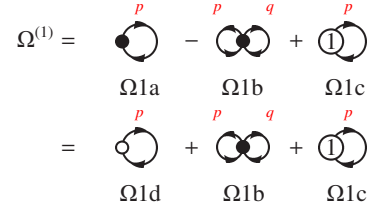}
\caption{\label{fig:Omega1}Diagrammatic equations of $\Omega^{(1)}$.}
\end{figure}

\begin{figure}
\includegraphics[scale=0.75]{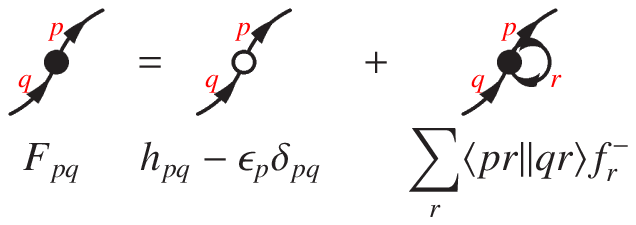}
\caption{\label{fig:fock}Definitions of vertexes. See also Appendix \ref{app:Hamiltonian}.}
\end{figure}

The diagrammatic rules are not so helpful at the zeroth and first orders, but
for the sake of completeness, we show in Fig.\ \ref{fig:Omega1} the diagrammatic formulation of $\Omega^{(1)}$. 
Figure \ref{fig:fock} defines the relevant vertexes. 

Using the interpretation rules in Table \ref{tab:rules2}, we can transform each diagram in Fig.\ \ref{fig:Omega1} as
\begin{eqnarray}
\Omega\text{1a} &=& (-1)^{1+1} \sum_{p} F_{pp} f_p^-, \\
\Omega\text{1b} &=& (-1)^{2+2} \left( \frac{1}{2} \right)^1 \sum_{p,q} \langle pq || pq \rangle f_p^-f_q^-, \\
\Omega\text{1c} &=& (-1)^{1+1} \sum_{p} (-\mu^{(1)}) f_p^- = -\mu^{(1)} \bar{N}.
\end{eqnarray}
Summing them with the signs indicated in the figure, we recover the reduced analytical formula of $\Omega^{(1)}$ [Eq.\ (\ref{Omega1reduced})].  

We can rewrite the diagrammatic equation in the first line of Fig.\ \ref{fig:Omega1} into the second line, which will turn out to be more convenient 
for the diagrammatic derivation of $\mu^{(1)}$ and $S^{(1)}$ (see below). An open-circle, two-line vertex stands for $h_{pq} - \epsilon_p \delta_{pq}$. 
The algebraic interpretation of $\Omega\text{1d}$ is then
\begin{eqnarray}
\Omega\text{1d} &=& (-1)^{1+1} \sum_{p} (h_{pp}-\epsilon_p) f_p^-,
\end{eqnarray}
leading to the same reduced analytical formula of $\Omega^{(1)}$.

\begin{figure}
\includegraphics[scale=0.75]{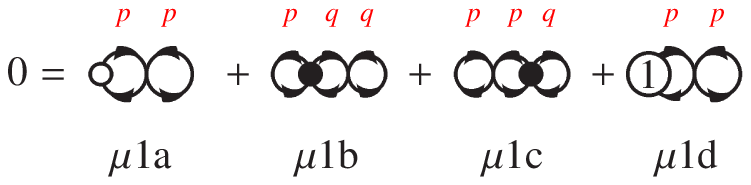}
\caption{\label{fig:mu1}Diagrammatic equation of $\mu^{(1)}$.}
\end{figure}

The diagrammatic rules for $\mu^{(n)}$ are based on Eq.\ (\ref{mu0deriv})  with
\begin{eqnarray}
\frac{\partial f_p^\pm}{\partial \mu^{(0)}} = \mp \beta f_p^- f_p^+. \label{dfdmu}
\end{eqnarray}
The diagrammatic equation for $\mu^{(n)}$ 
is therefore obtained by `differentiating' each line in the diagrammatic equation of 
$\Omega^{(n)}$ with respect to $\mu^{(0)}$. In view of Eq.\ (\ref{dfdmu}), this process
can be represented as attaching to each line a bubble with the opposite directionality (which 
is judged at the farthest point from its contact with the line). Note that the point of contact of the bubble 
is not a vertex, and therefore the bubble and line share the same spinorbital index. 

Figure \ref{fig:mu1} is the diagrammatic equation for $\mu^{(1)}$. It is based on the second line of Fig.\ \ref{fig:Omega1} rather than 
the first line. This is because the latter may hide the fact that $F_{pq}$ contains $\langle pr || qr \rangle$ that also needs to be differentiated with $\mu^{(0)}$.  
Each diagram is algebraically interpreted by the same rules of Table \ref{tab:rules2} as
\begin{eqnarray}
\mu\text{1a} &=& (-1)^{1+2} \sum_p (h_{pp}-\epsilon_p) f_p^- f_p^+, \\
\mu\text{1b} &=& (-1)^{2+3} \left(\frac{1}{2}\right)^1 \sum_{p,q} \langle pq || pq \rangle f_p^- f_q^-f_q^+, \\
\mu\text{1c} &=& (-1)^{2+3} \left(\frac{1}{2}\right)^1\sum_{p,q} \langle pq || pq \rangle f_p^- f_q^-f_p^+, \\
\mu\text{1d} &=& (-\mu^{(1)}) (-1)^{1+2} \sum_p f_p^- f_p^+.
\end{eqnarray}
They lead to the same reduced analytical equation for $\mu^{(1)}$ as Eq.\ (\ref{mu1reduced}).

\begin{figure}
\includegraphics[scale=0.75]{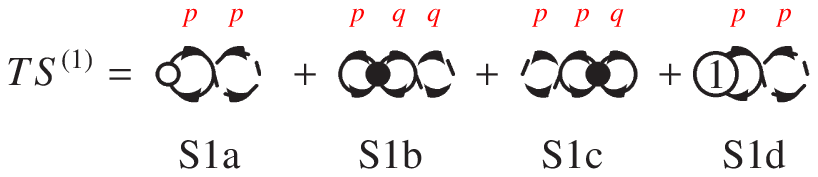}
\caption{\label{fig:S1}Diagrammatic equation of $TS^{(1)}$.}
\end{figure}

The diagrammatic equation for $S^{(n)}$ utilizes the following identity obtained by differentiating Eq.\ (\ref{S}) with $\lambda$ $n$ times:
\begin{eqnarray}
T S^{(n)} &=& \beta \left( \frac{\partial \Omega^{(n)}} {\partial \beta}\right)_{\mu^{(0)},\dots,\,\mu^{(n)}}
\end{eqnarray}
with
\begin{eqnarray}
\beta \frac{\partial f_p^\pm}{\partial \beta} = \pm \beta (\epsilon_p - \mu^{(0)}) f_p^- f_p^+. \label{dfdbeta}
\end{eqnarray}
Therefore, the diagrammatic rules for $S^{(n)}$ can be obtained by `differentiating' the $\Omega^{(n)}$ diagram equation 
with $\beta$, keeping in mind that anomalous diagrams also carry a power of $\beta$. 
A $\beta$-differentiation of a line is diagrammatically depicted as attaching to the line a dashed-bubble with the opposite directionality. 

At the first order, $\Omega^{(1)}$ does not have anomalous diagrams and so only lines (or more precisely their $f_p^-$ factors) 
need to be differentiated with 
$\beta$. The diagrammatic equation for $S^{(1)}$ is drawn in Fig.\ \ref{fig:S1}. Each diagram
is interpreted according to Table \ref{tab:rules2} as
\begin{eqnarray}
\text{S1a} &=& (-1)^{1+2} \sum_p (h_{pp}-\epsilon_p) \beta (\epsilon_p - \mu^{(0)}) f_p^- f_p^+, \\
\text{S1b} &=& (-1)^{2+3} \left(\frac{1}{2}\right)^1 \sum_{p,q} \langle pq || pq \rangle f_p^- f_q^-\beta (\epsilon_q - \mu^{(0)}) f_q^+, \\
\text{S1c} &=& (-1)^{2+3} \left(\frac{1}{2}\right)^1 \sum_{p,q} \langle pq || pq \rangle f_p^- f_q^-\beta (\epsilon_p - \mu^{(0)})f_p^+, \\
\text{S1d} &=& (-\mu^{(1)}) (-1)^{1+2} \sum_p f_p^- \beta (\epsilon_p - \mu^{(0)}) f_p^+,
\end{eqnarray}
leading to
\begin{eqnarray}
TS^{(1)}&=& -\beta \sum_p \left( F_{pp} - \mu^{(1)} \right) \left (\epsilon_p - \mu^{(0)}\right ) f_p^- f_p^+ \\
&=&  -\beta \sum_p \left( F_{pp} - \mu^{(1)} \right) \epsilon_p  f_p^- f_p^+. \label{S1reduced}
\end{eqnarray}
The second equality follows from Eq.\ (\ref{mu1reduced}). 
In general, for the purpose
of deriving the correct $S^{(n)}$ formulas, $\mu^{(0)}$ in Eq.\ (\ref{dfdbeta}) can be replaced by any constant (such as zero)
by virtue of Eq.\ (\ref{mu0deriv}).

The diagrammatic equation and rules for $U^{(n)}$ are a concatenation of 
those for $\Omega^{(n)}$, $\mu^{(n)}$, and $S^{(n)}$ given above because
\begin{eqnarray}
U^{(n)} = \Omega^{(n)} + \mu^{(n)} \bar{N} + TS^{(n)} .
\end{eqnarray}
The $U^{(1)}$ formula obtained as a combination of Eqs.\ (\ref{Omega1reduced}) and (\ref{S1reduced}) is the same
as Eq.\ (61) of Ref.\ \onlinecite{Hirata2ndorder}.

\subsection{Second order}

\begin{figure*}
\includegraphics[scale=0.75]{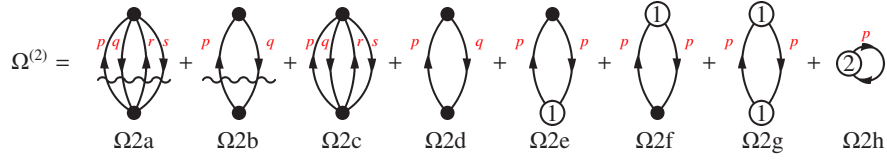}
\caption{\label{fig:Omega2}Diagrammatic equation of $\Omega^{(2)}$.}
\end{figure*}

Following the rules in Table \ref{tab:rules1}, we obtain eight diagrams that consist in $\Omega^{(2)}$, which are shown in Fig.\ \ref{fig:Omega2}.
They are interpreted algebraically according to Table \ref{tab:rules2}, leading to the same result as Eq.\ (\ref{Omega2reduced}).
 
The first two ($\Omega\text{2a}$ and $\Omega\text{2b}$) are isomorphic to the usual MBPT(2) energy diagrams ($\Omega\text{2b}$ being
the non-HF term\cite{shavitt}) with exactly one resolvent (wiggly) line in between the pair of vertexes. 
The third and fourth ($\Omega\text{2c}$ and $\Omega\text{2d}$) are the corresponding anomalous diagrams,\cite{kohn,SANTRA} whose
resolvent line is missing. In these diagrams, the summations over spinorbital indexes are limited to those cases whose fictitious denominators are zero.
Instead of being divided by the energy denominators, they are divided by $2 k_\text{B}T$ to restore the dimension of energy. 

Diagrams $\Omega\text{2e}$ through $\Omega\text{2g}$ are also unique to the finite-temperature MBPT.
The absence of a resolvent line implies that they are technically counted as anomalous diagrams.
The summations in these diagrams are therefore also limited to zero fictitious denominators, but this restriction is automatically fulfilled 
because the $\mu^{(n)}$ vertex is zero whenever $p \neq q$. 

\begin{figure}
\includegraphics[scale=0.75]{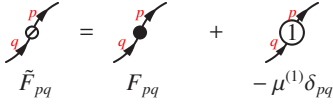}
\caption{\label{fig:fock2}Definitions of vertexes.}
\end{figure}

\begin{figure}
\includegraphics[scale=0.75]{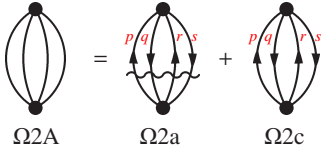}
\caption{\label{fig:Omega2skeleton1}A skeleton diagram of $\Omega^{(2)}$.}
\end{figure}

\begin{figure*}
\includegraphics[scale=0.75]{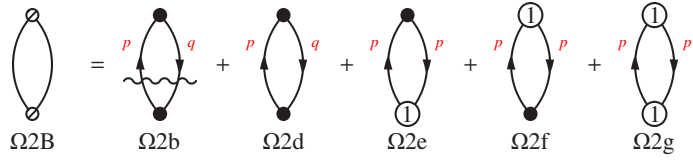}
\caption{\label{fig:Omega2skeleton2}Another skeleton diagram of $\Omega^{(2)}$.}
\end{figure*}

\begin{figure}
\includegraphics[scale=0.75]{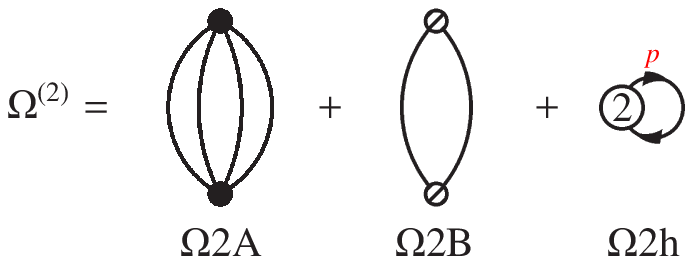}
\caption{\label{fig:Omega2skeleton3}Diagrammatic equation of $\Omega^{(2)}$ in terms of skeleton diagrams.}
\end{figure}

We can compress the diagrammatic equations of $\Omega^{(n)}$ by introducing skeleton diagrams\cite{shavitt} 
and a modified Fock operator defined in Fig.\ \ref{fig:fock2}.  A skeleton diagram is the one that is stripped of 
line directions, line labels, and resolvents. Single skeleton diagram $\Omega\text{2A}$ in Fig.\ \ref{fig:Omega2skeleton1}
stands for $\Omega\text{2a}$ plus $\Omega\text{2c}$, which differ from each other by the presence and absence of 
the resolvent. Skeleton diagram $\Omega\text{2B}$ in Fig.\ \ref{fig:Omega2skeleton2} is the sum of five diagrams.
A replacement of the $F_{pq}$ vertex by the $F_{pq} - \mu^{(1)}\delta_{pq}$ vertex has no effect on the normal diagram ($\Omega\text{2b}$)
because its summation is limited to $p\neq q$, but it consolidates the remaining four anomalous diagrams into one. This process is closely
related to the Luttinger--Ward prescription\cite{luttingerward} for the Kohn--Luttinger conundrum.\cite{kohn,Hirata_PRA} 
Together, we can simplify the diagrammatic equation of $\Omega^{(2)}$ into Fig.\ \ref{fig:Omega2skeleton3}.

\begin{figure*}
\includegraphics[scale=0.75]{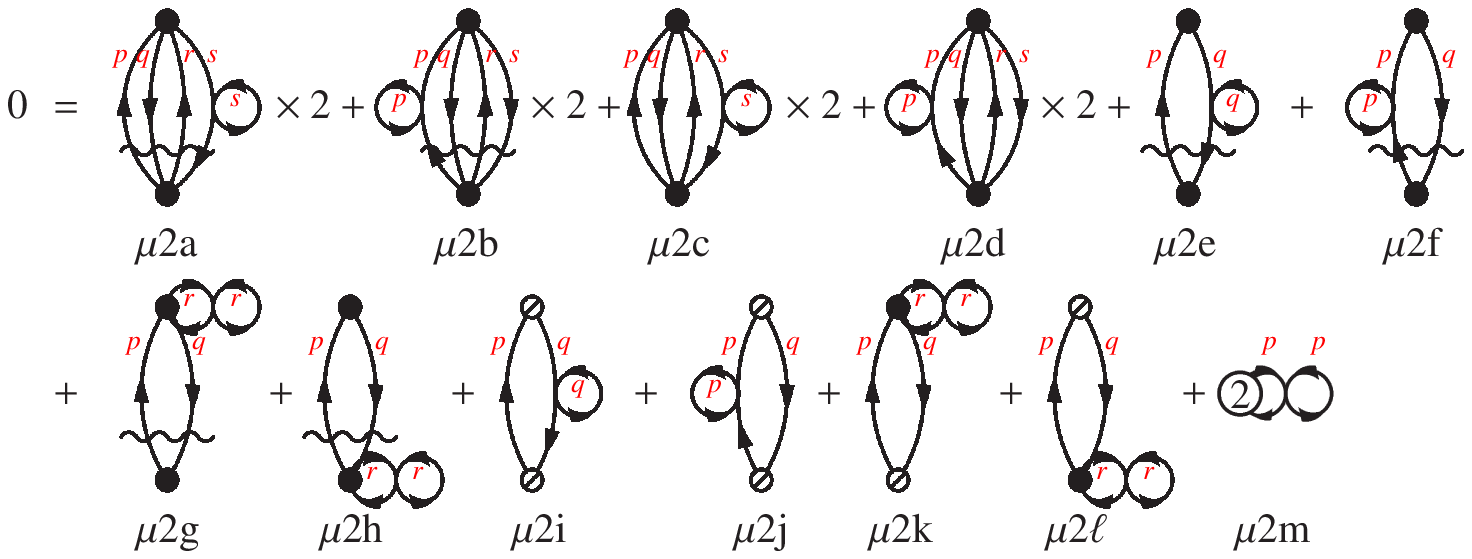}
\caption{\label{fig:mu2}Diagrammatic equation of $\mu^{(2)}$.}
\end{figure*}

The diagrammatic equation of $\mu^{(2)}$ is drawn in Fig.\ \ref{fig:mu2}. Translating them into algebraic formulas using 
the rules in Table \ref{tab:rules2}, we arrive at the same reduced analytical formula of $\mu^{(2)}$ as Eq.\ (69) of Ref.\ \onlinecite{Hirata2ndorder}.

\begin{figure*}
\includegraphics[scale=0.75]{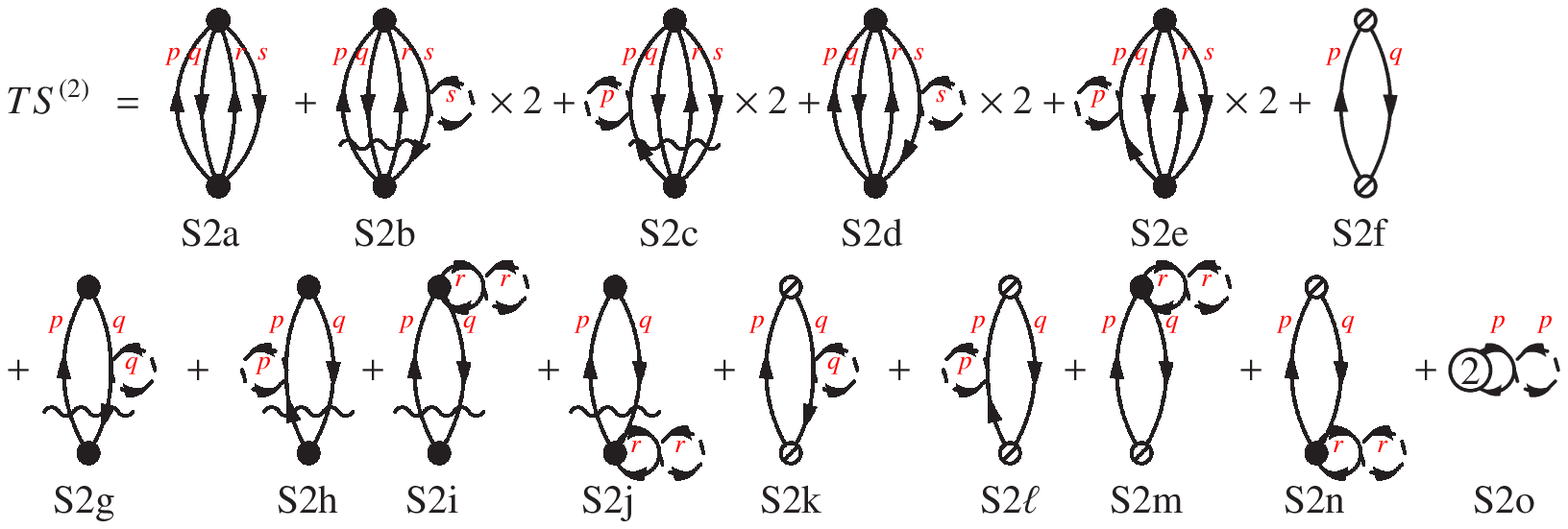}
\caption{\label{fig:S2}Diagrammatic equation of $S^{(2)}$.}
\end{figure*}

The diagrams defining $S^{(2)}$ are shown in Fig.\ \ref{fig:S2}. Evaluating them according to the rules in Table \ref{tab:rules2}, we
obtain the reduced analytical formula that is consistent with the $U^{(2)}$ formula, i.e., Eq.\ (74) of Ref.\ \onlinecite{Hirata2ndorder}. Note that the evaluation rules 
are slightly different between diagrams $\text{S2a}$ and $\text{S2f}$ and diagrams $\Omega\text{2c}$ and $\Omega\text{2d}$--$\Omega\text{2g}$ 
(cf.\ Rule 12 of Table \ref{tab:rules2}) despite the identical appearances of the diagrams. Care must be 
exercised when combining diagrammatic equations of $\Omega^{(n)}$ and $S^{(n)}$  
to form a concatenated  equation of $U^{(n)}$. 

\subsection{Third order\label{sec:Omega3}} 

\begin{figure*}
\includegraphics[scale=0.75]{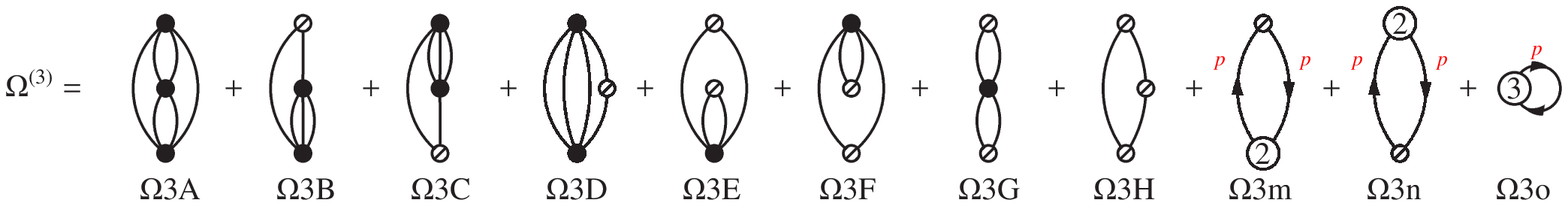}
\caption{\label{fig:Omega3}Diagrammatic equation of $\Omega^{(3)}$.}
\end{figure*}

\begin{figure}
\includegraphics[scale=0.75]{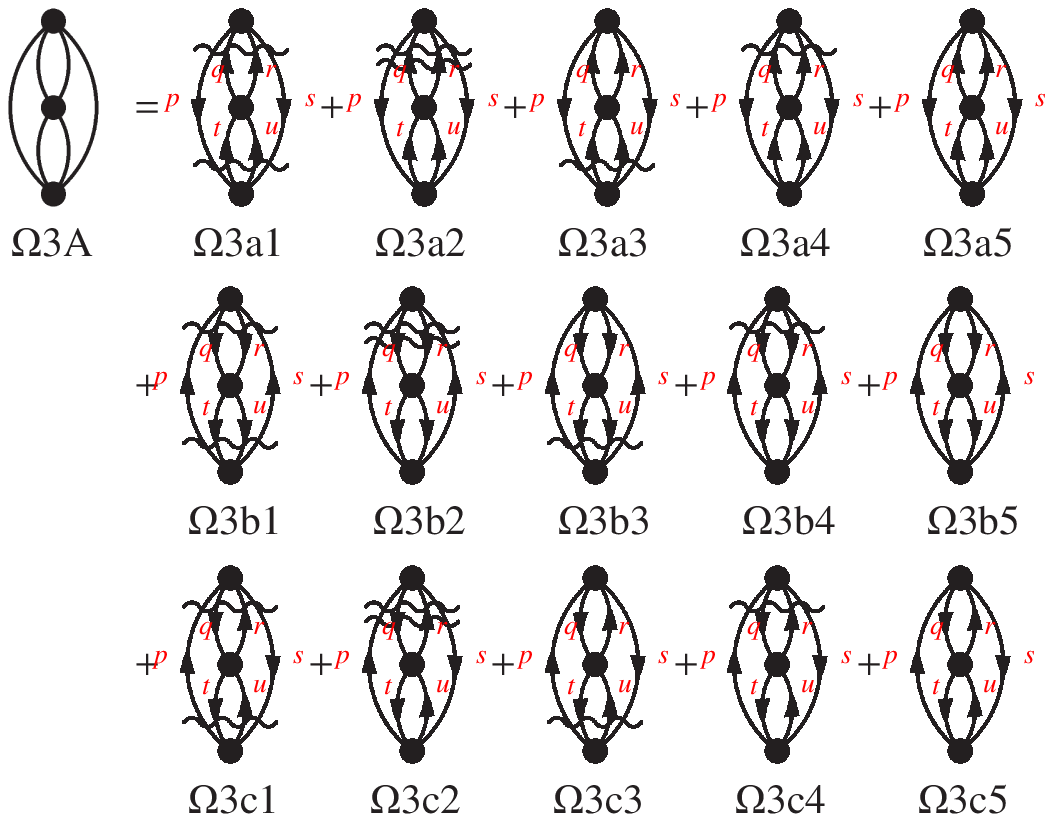}
\caption{\label{fig:Omega3A}Definition of the $\Omega\text{3A}$ skeleton diagram.}
\end{figure}

\begin{figure}
\includegraphics[scale=0.75]{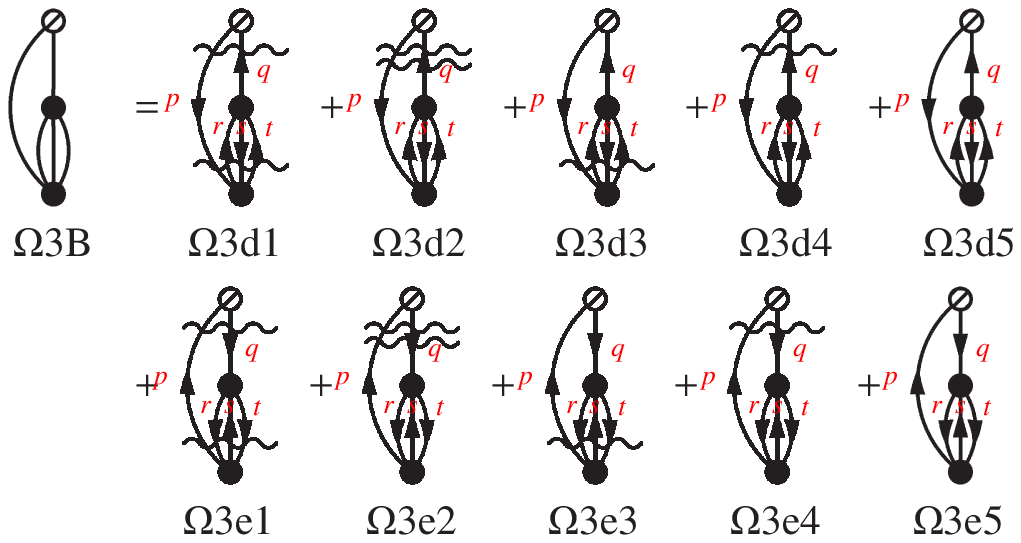}
\caption{\label{fig:Omega3B}Definition of the $\Omega\text{3B}$ skeleton diagram.}
\end{figure}

\begin{figure}
\includegraphics[scale=0.75]{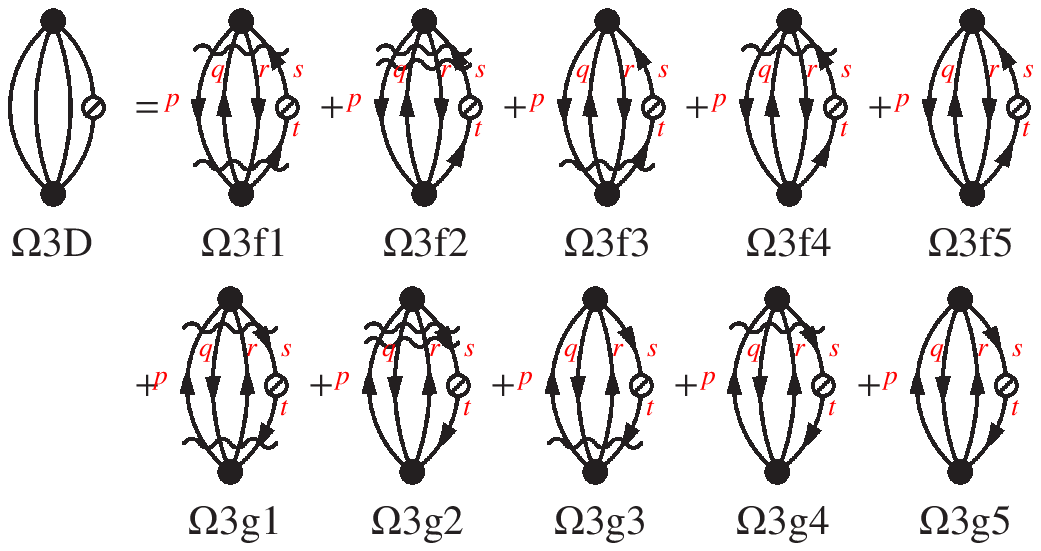}
\caption{\label{fig:Omega3D}Definition of the $\Omega\text{3D}$ skeleton diagram.}
\end{figure}

\begin{figure}
\includegraphics[scale=0.75]{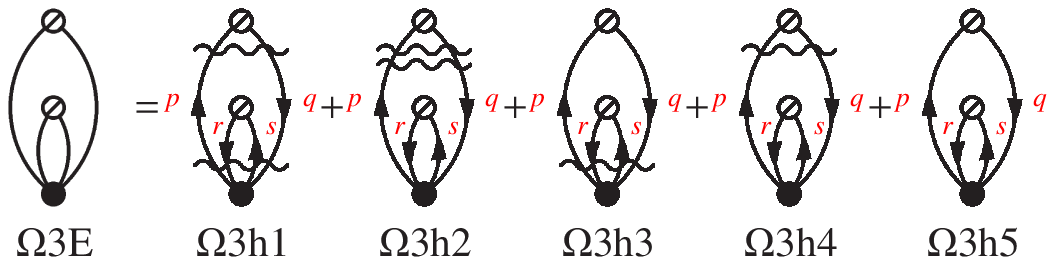}
\caption{\label{fig:Omega3E}Definition of the $\Omega\text{3E}$ skeleton diagram.}
\end{figure}

\begin{figure}
\includegraphics[scale=0.75]{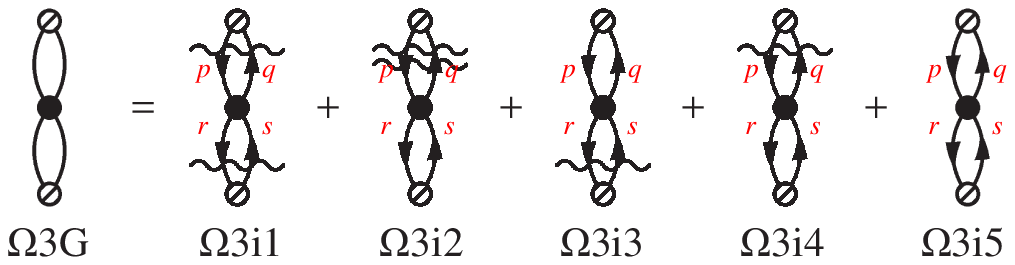}
\caption{\label{fig:Omega3G}Definition of the $\Omega\text{3G}$ skeleton diagram.}
\end{figure}

\begin{figure}
\includegraphics[scale=0.75]{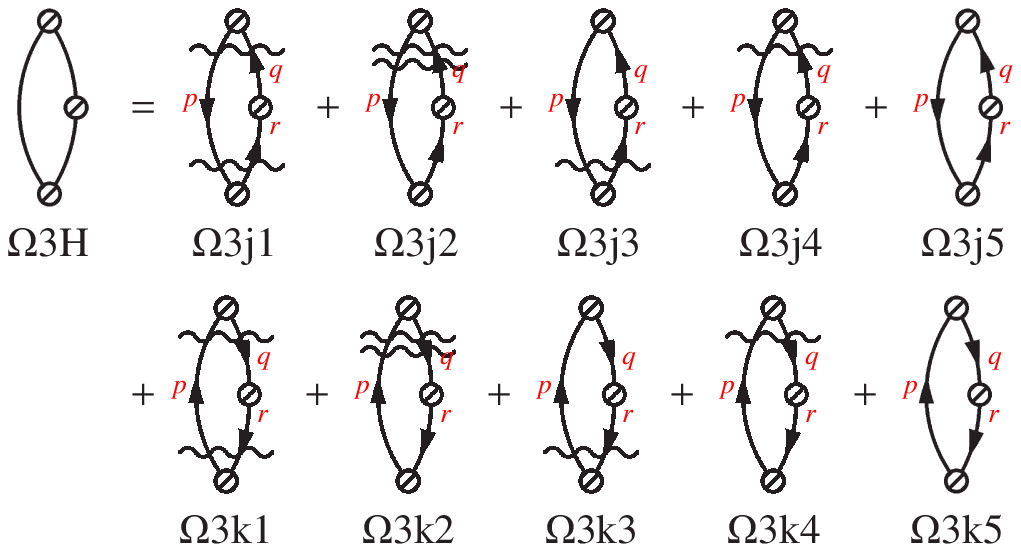}
\caption{\label{fig:Omega3H}Definition of the $\Omega\text{3H}$ skeleton diagram.}
\end{figure}

The diagrammatic equation of $\Omega^{(3)}$ is given in Fig.\ \ref{fig:Omega3}. 
It is the sum of eight skeleton diagrams [cf.\ zero-temperature MBPT(3) diagrams in page 133 of Shavitt and Bartlett\cite{shavitt}] plus two anomalous diagrams involving $\mu^{(2)}$ and 
a bubble diagram of $\mu^{(3)}$. Each skeleton diagram in turn consists of five to fifteen diagrams
differing from one another in line direction and/or placement of the resolvents, shown in Figs.\ \ref{fig:Omega3A} through \ref{fig:Omega3H}.
In each row of the right-hand sides of these diagram equations, the first diagram ($\Omega3\mathrm{x}1$) is the normal (parent) diagram.
The second diagram ($\Omega3\mathrm{x}2$) is the renormalization diagram with a resolvent shifted upwards. 
The third through fifth diagrams are an anomalous diagram with one or two missing resolvents.

The diagrams in the first row of the right-hand side of Fig.\ \ref{fig:Omega3A} are interpreted algebraically as
\begin{eqnarray}
&& \Omega\text{3a1} =(-1)^{2+2} \left(\frac{1}{2}\right)^3 
\nonumber\\&&\times\, 
\sum_{p,q,r,s,t,u}^{\substack{\text{denom.1}\neq 0 \\ \text{denom.2}\neq 0}} \frac{\langle ps || qr \rangle \langle qr || tu \rangle \langle tu || ps \rangle f_p^-  f_q^+   f_r^+ f_s^- f_t^+ f_u^+ }
{(\epsilon_p + \epsilon_s - \epsilon_q - \epsilon_r)(\epsilon_p + \epsilon_s - \epsilon_t - \epsilon_u)}, \label{3a1} 
\end{eqnarray}
\begin{eqnarray}
&& \Omega\text{3a2} =(-1)^1(-1)^{2+2} \left(\frac{1}{2}\right)^3 
\nonumber\\&&\times\, 
\sum_{p,q,r,s,t,u}^{\substack{\text{denom.1}\neq 0 \\ \text{denom.2}= 0}} \frac{\langle ps || qr \rangle \langle qr || tu \rangle \langle tu || ps \rangle f_p^-  f_q^+   f_r^+ f_s^- f_t^+ f_u^+ }
{(\epsilon_p + \epsilon_s - \epsilon_q - \epsilon_r)^2}, \label{3a2} 
\end{eqnarray}
\begin{eqnarray}
&& \Omega\text{3a3} =\frac{(-\beta)}{2!} (-1)^{2+2} \left(\frac{1}{2}\right)^3 
\nonumber\\&&\times\, 
\sum_{p,q,r,s,t,u}^{\substack{\text{denom.1}= 0 \\ \text{denom.2}\neq 0}} \frac{\langle ps || qr \rangle \langle qr || tu \rangle \langle tu || ps \rangle f_p^-  f_q^+   f_r^+ f_s^- f_t^+ f_u^+ }
{(\epsilon_p + \epsilon_s - \epsilon_t - \epsilon_u)}, 
\end{eqnarray}
\begin{eqnarray}
&& \Omega\text{3a4} =\frac{(-\beta)}{2!} (-1)^{2+2} \left(\frac{1}{2}\right)^3 
\nonumber\\&&\times\, 
\sum_{p,q,r,s,t,u}^{\substack{\text{denom.1}\neq 0 \\ \text{denom.2}= 0}} \frac{\langle ps || qr \rangle \langle qr || tu \rangle \langle tu || ps \rangle f_p^-  f_q^+   f_r^+ f_s^- f_t^+ f_u^+ }
{(\epsilon_p + \epsilon_s - \epsilon_q - \epsilon_r)}, 
\end{eqnarray}
and
\begin{eqnarray}
&& \Omega\text{3a5} =\frac{(-\beta)^2}{3!} (-1)^{2+2} \left(\frac{1}{2}\right)^3 
\nonumber\\&&\times\, 
\sum_{p,q,r,s,t,u}^{\substack{\text{denom.1}= 0 \\ \text{denom.2}= 0}} {\langle ps || qr \rangle \langle qr || tu \rangle \langle tu || ps \rangle f_p^-  f_q^+   f_r^+ f_s^- f_t^+ f_u^+ }, \label{3a5}
\end{eqnarray}
where `$\text{denom.1}$' and `$\text{denom.2}$' denote the first and second denominator factors, respectively, 
of the normal diagram [Eq.\ (\ref{3a1})]. Owing to this provision, none of the above summations encounters a division by zero.

Unlike in zero-temperature MBPT, where the renormalization term vanishes upon cancelling unlinked contribution 
in the parent diagram, in the finite-temperature MBPT, it has both linked and unlinked contributions, the latter cancelling exactly
the unlinked contribution in the parent diagram, while the former persists as the renormalization diagram $\Omega\text{3a2}$ [Eq.\ (\ref{3a2})]
with its resolvent shifted upwards. Unlike in anomalous diagrams, no resolvent is deleted in a renormalization diagram 
and therefore it does not have to be divided by $k_\text{B}T$ to restore the correct dimension of energy. 
Instead, it is multiplied by the parity of $(-1)$ raised to the power of the number of shifted
resolvents (which is equal to the number/nestedness of diagram insertions).\cite{shavitt} 
It is not clear if the  time-dependent, diagrammatic derivation\cite{matsubara,bloch,kohn,luttingerward,balian,blochbook} or the density-matrix formulation\cite{SANTRA} 
of the conventional finite-temperature MBPT properly accounts for such diagrams, the neglect of which
makes it erroneous at the third and higher orders and nonconvergent to the finite-temperature FCI.\cite{Kou} 
The renormalization diagrams may be related to the linked-disconnected diagrams in $\Delta$MP$n$.\cite{deltamp}

Diagrams $\Omega\text{3a3}$, $\Omega\text{3a4}$, and $\Omega\text{3a5}$ are anomalous diagrams of which one or two resolvents are missing.
The summation is correspondingly restricted to the cases where the missing denominator factors are zero. For each missing denominator factor,
the diagram is divided by $k_\text{B}T$ times some integer factor. The presence of these anomalous diagrams is considered well known.\cite{kohn,SANTRA} 

The remaining normal diagrams ($\Omega$3x1) are interpreted algebraically as
\begin{eqnarray}
&& \Omega\text{3b1} =(-1)^{4+2} \left(\frac{1}{2}\right)^3 
\nonumber\\ && \times\,
\sum_{p,q,r,s,t,u}^{\substack{\text{denom.1}\neq 0 \\ \text{denom.2}\neq 0}} \frac{\langle qr || ps \rangle \langle tu || qr \rangle \langle ps || tu \rangle f_p^+  f_q^-  f_r^-  f_s^+  f_t^- f_u^- }
{(\epsilon_q + \epsilon_r - \epsilon_p - \epsilon_s)(\epsilon_t + \epsilon_u - \epsilon_p - \epsilon_s)}, 
\end{eqnarray}
\begin{eqnarray}
&& \Omega\text{3c1} =(-1)^{3+2} 
\nonumber\\ && \times\,
\sum_{p,q,r,s,t,u}^{\substack{\text{denom.1}\neq 0 \\ \text{denom.2}\neq 0}} 
\frac{\langle qs || pr \rangle \langle tr || qu \rangle \langle pu || ts \rangle f_p^+ f_q^-  f_r^+ f_s^-  f_t^- f_u^+ }
{(\epsilon_q + \epsilon_s - \epsilon_p - \epsilon_r)(\epsilon_t + \epsilon_s - \epsilon_p - \epsilon_u)}, 
\end{eqnarray}
\begin{eqnarray}
&&\Omega\text{3d1} = (-1)^{2+2} \left(\frac{1}{2}\right)^1 \sum_{p,q,r,s,t}^{\substack{\text{denom.1}\neq 0 \\ \text{denom.2}\neq 0}}
\nonumber\\ && \times\,
\frac{( F_{pq} - \mu^{(1)}\delta_{pq} ) \langle qs || rt \rangle \langle rt || ps \rangle f_p^-  f_q^+   f_r^+ f_s^- f_t^+ }
{(\epsilon_p - \epsilon_q)(\epsilon_p + \epsilon_s - \epsilon_r - \epsilon_t)}, 
\end{eqnarray}
\begin{eqnarray}
&& \Omega\text{3e1} =(-1)^{3+2} \left(\frac{1}{2}\right)^1 \sum_{p,q,r,s,t}^{\substack{\text{denom.1}\neq 0 \\ \text{denom.2}\neq 0}}
\nonumber\\ && \times\,
\frac{( F_{qp} - \mu^{(1)}\delta_{qp} ) \langle rt || qs \rangle \langle ps || rt \rangle f_p^+  f_q^-   f_r^- f_s^+ f_t^- }
{(\epsilon_q - \epsilon_p)(\epsilon_r + \epsilon_t - \epsilon_p - \epsilon_s)}, 
\end{eqnarray}
\begin{eqnarray}
&&\Omega\text{3f1} =(-1)^{2+2} \left(\frac{1}{2}\right)^1 \sum_{p,q,r,s,t}^{\substack{\text{denom.1}\neq 0 \\ \text{denom.2}\neq 0}} 
\nonumber\\ && \times\,
\frac{\langle pr||qs \rangle (F_{st} - \mu^{(1)}\delta_{st} )\langle qt || pr \rangle f_p^-  f_q^+   f_r^- f_s^+ f_t^+ }
{(\epsilon_p+ \epsilon_r-\epsilon_q-\epsilon_s)(\epsilon_p+ \epsilon_r-\epsilon_q- \epsilon_t)}, 
\end{eqnarray}
\begin{eqnarray}
&&\Omega\text{3g1} =(-1)^{3+2} \left(\frac{1}{2}\right)^1 \sum_{p,q,r,s,t}^{\substack{\text{denom.1}\neq 0 \\ \text{denom.2}\neq 0}}
\nonumber\\ && \times\,
\frac{\langle qs||pr \rangle (F_{ts} - \mu^{(1)}\delta_{ts} )\langle pr || qt \rangle f_p^+  f_q^-   f_r^+ f_s^- f_t^- }
{(\epsilon_q+ \epsilon_s-\epsilon_p-\epsilon_r)(\epsilon_q+ \epsilon_t-\epsilon_p- \epsilon_r)}, 
\end{eqnarray}
\begin{eqnarray}
&&\Omega\text{3h1} = (-1)^{2+1} \sum_{p,q,r,s}^{\substack{\text{denom.1}\neq 0 \\ \text{denom.2}\neq 0}}
\nonumber\\ && \times\, 
\frac{(F_{qp} - \mu^{(1)}\delta_{qp} ) (F_{rs}  - \mu^{(1)}\delta_{rs} )\langle ps||rq \rangle f_p^+  f_q^-   f_r^- f_s^+ }
{(\epsilon_q-\epsilon_p)(\epsilon_q + \epsilon_r - \epsilon_p - \epsilon_s)}, 
\end{eqnarray}
\begin{eqnarray}
&&\Omega\text{3i1} =  (-1)^{2+1} \sum_{p,q,r,s}^{\substack{\text{denom.1}\neq 0 \\ \text{denom.2}\neq 0}}
 \nonumber\\ && \times\,
\frac{(F_{pq}  - \mu^{(1)}\delta_{pq} ) \langle rq||ps \rangle (F_{sr} - \mu^{(1)}\delta_{sr} ) f_p^-  f_q^+   f_r^- f_s^+ }
{(\epsilon_p-\epsilon_q)(\epsilon_r - \epsilon_s)}, 
\end{eqnarray}
\begin{eqnarray}
&&\Omega\text{3j1} =  (-1)^{1+1} \sum_{p,q,r}^{\substack{\text{denom.1}\neq 0 \\ \text{denom.2}\neq 0}}
\nonumber\\ && \times\, 
 \frac{(F_{pq} - \mu^{(1)}\delta_{pq} )( F_{qr} - \mu^{(1)}\delta_{qr} ) (F_{rp} - \mu^{(1)}\delta_{rp} )  f_p^-  f_q^+ f_r^+}
{(\epsilon_p - \epsilon_q)(\epsilon_p-\epsilon_r)}, \nonumber\\
\end{eqnarray}
and
\begin{eqnarray}
&&\Omega\text{3k1} =  (-1)^{2+1} \sum_{p,q,r}^{\substack{\text{denom.1}\neq 0 \\ \text{denom.2}\neq 0}}
\nonumber\\ && \times\, 
 \frac{(F_{qp} - \mu^{(1)}\delta_{qp} )( F_{rq} - \mu^{(1)}\delta_{rq} ) (F_{pr} - \mu^{(1)}\delta_{pr} )  f_p^+  f_q^- f_r^-}
{(\epsilon_q - \epsilon_p)(\epsilon_r-\epsilon_p)}. \nonumber\\
\end{eqnarray}
The algebraic formulas for the accompanying renormalization and anomalous 
diagrams can be inferred from these formulas
by a systematic modification of the denominators,
summation restrictions, and prefactors. 
Therefore, despite the numerousness of these third-order diagrams, 
their algebraic evaluations and computer implementation are not as horrendous as they may suggest.

In zero-temperature MBPT, $\Omega\text{3B}$ and $\Omega\text{3C}$  ($\Omega\text{3E}$ and $\Omega\text{3F}$ also) are a pair of conjugated diagrams,\cite{shavitt} whose 
values are the complex conjugate of each other. In the finite-temperature formalism, renormalization diagrams destroy such time-reversal
symmetry at a diagram level (not at the whole theory level), and we need to evaluate separately $\Omega\text{3B}$ and $\Omega\text{3C}$, which have values 
that are not simply related to each other.

The last three diagrams of Fig.\ \ref{fig:Omega3} are interpreted as
\begin{eqnarray}
\Omega\text{3m}  &=&(-1)^{1+1} \frac{(-\beta)}{2!} \sum_{p} {(F_{pp} - \mu^{(1)} ) (-\mu^{(2)}) f_p^-  f_p^+}, \label{3m} \\
\Omega\text{3o} &=&(-1)^{1+1} \sum_{p} (-\mu^{(3)}) f_p^- = -\mu^{(3)} \bar{N}.
\end{eqnarray}
Diagrams $\Omega\text{3m}$ and $\Omega\text{3n}$ are a pair of conjugate diagrams (since they are not renormalization diagrams), the algebraic formula of
the latter is the same as Eq.\ (\ref{3m}).

\subsection{Fourth order}

\begin{figure*}
\includegraphics[scale=0.75]{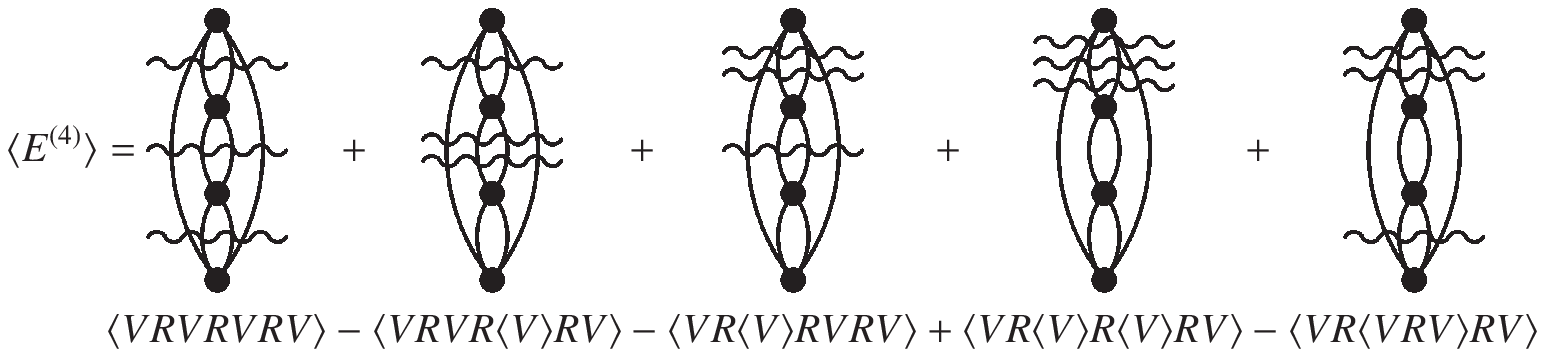}
\caption{\label{fig:E4}A normal diagram (the leftmost) and its renormalization diagrams in $\langle E_I^{(4)} \rangle$
and their corresponding Brueckner brackets.
Each wiggly line represents a resolvent. Line directions and line labels are suppressed. Signs 
in front of the Brueckner brackets are absorbed in the algebraic interpretations of the respective diagrams.}
\end{figure*}

At the fourth order, enumerating and interpreting diagrams manually is no longer practicable  and a computerized scheme\cite{tce} 
may be necessary (not pursued in this study). Here, we focus on how to generate renormalization diagrams, which may be the least trivial step
of the diagram enumeration. The $n$th-order anomalous diagrams are straightforwardly spawned by simply eliminating resolvents in all $2^{n-1}-1$ ways. 

Figure \ref{fig:E4} illustrates how all renormalization diagrams are generated systematically from a normal diagram. The latter has exactly one
resolvent line in between every pair of adjacent vertexes.  The corresponding renormalization diagrams are obtained by shifting one or more (up to $n-2$) 
resolvent lines upwards in all possible ways. However, it may be safer to use the method of the Brueckner brackets,\cite{brueckner,Manne} which enumerates 
all renormalization terms by indicating how various operators are Wick contracted in such a way that the result is unlinked (in the case of MPPT). 
In our case, many reference states are degenerate and these brackets 
are not necessarily unlinked, as emphasized earlier. 

For instance, the first renormalization diagram (the second diagram) in the figure corresponds to the Brueckner bracket,
\begin{eqnarray}
\langle VRVR\langle V \rangle RV \rangle &=&
\contraction[0.5ex]{\langle VR}{V}{}{R}
\contraction[0.5ex]{\langle VRVR\langle V \rangle }{R}{}{V}
\contraction[0.5ex]{\langle }{V}{}{R}
\bcontraction[0.5ex]{\langle VRV}{R}{\langle V \rangle }{R}
\bcontraction[0.5ex]{\langle V}{R}{}{V}
\langle VRVR\langle V \rangle RV \rangle \nonumber\\
&=&
\contraction[0.5ex]{\langle VR}{V}{}{R}
\contraction[0.5ex]{\langle VRVR }{R}{}{V}
\contraction[0.5ex]{\langle }{V}{}{R}
\bcontraction[0.5ex]{\langle VRV}{R}{}{R}
\bcontraction[0.5ex]{\langle V}{R}{}{V}
\langle VRVRRVPV \rangle, \label{VRVRVRV}
\end{eqnarray}
where $V$ stands for $\hat{V}$, $P$ for $\hat{P}_I$, $R$ for $\hat{R}_I$, and $\langle \dots \rangle$ is a thermal average, i.e.,
$\langle \langle I | \dots | I \rangle\rangle$.
Associating $V$'s from left to right with the vertexes from top to bottom, Eq.\ (\ref{VRVRVRV}) is the diagram in which 
there is one resolvent in between the first (top)
and second vertexes, two in between the second and third, and none in between the third and fourth (bottom). 
In other words, the bottom resolvent has been shifted up to the middle. 
The parity associated with this bracket is $(-1)$ because there is one insertion (one inner bracket),\cite{Manne,shavitt} and this factor is properly accounted for by
Rule 13 of Table \ref{tab:rules2}. The fourth diagram in Fig.\ \ref{fig:E4} is an example of double insertions,\cite{Manne,shavitt} whose
parity is $(-1)^2$. Generally, an $n$-tuple insertion or $n$-fold nested insertion carries a factor of $(-1)^n$. 

There should also be anomalous renormalization diagrams.

\section{Linked-diagram theorem\label{sec:linkeddiagramtheorem}}

We have established in Eqs.\ (\ref{Omega1linked}) and (\ref{Omega2linked}) that $\Omega^{(1)}$ and $\Omega^{(2)}$
are diagrammatically linked and thus size-consistent. We may summarize this as
\begin{eqnarray}
\Omega^{(1)} &=& \langle D_I^{(1)} \rangle = \langle D_I^{(1)} \rangle_{L},  \label{Omega1_linked}  \\
\Omega^{(2)} &=& \langle D_I^{(2)} \rangle + \frac{(-\beta)}{2!}  \left(  \langle D_I^{(1)}D_I^{(1)} \rangle - \Omega^{(1)} \Omega^{(1)} \right) \nonumber \\
&=& \langle D_I^{(2)} \rangle_{L} + \frac{(-\beta)}{2!}  \left(  \langle D_I^{(1)}D_I^{(1)} \rangle - \langle D_I^{(1)}\rangle_{L}\langle D_I^{(1)}\rangle_{L} \right) \nonumber \\
&=& \langle D_I^{(2)} \rangle_{L} + \frac{(-\beta)}{2!}  \langle D_I^{(1)}D_I^{(1)} \rangle_{L},
\label{Omega2_linked}
\end{eqnarray}
where subscript $L$ means linked.
These can be used to show that $\Omega^{(3)}$ is also linked.
\begin{eqnarray}
\Omega^{(3)} 
&=& \langle D_I^{(3)} \rangle + \frac{(-\beta)}{2!} \left(  \langle D_I^{(1)}D_I^{(2)} \rangle - \Omega^{(1)} \Omega^{(2)} \right)
\nonumber\\&& 
+ \frac{(-\beta)}{2!} \left(  \langle D_I^{(2)}D_I^{(1)} \rangle - \Omega^{(2)} \Omega^{(1)} \right) \nonumber\\
&& + \frac{(-\beta)^2}{3!} \left( \langle D_I^{(1)}D_I^{(1)} D_I^{(1)} \rangle - \Omega^{(1)}\Omega^{(1)}\Omega^{(1)} \right) \\
&=& \langle D_I^{(3)} \rangle_{L} + \frac{(-\beta)}{2!} \left( \langle D_I^{(1)}D_I^{(2)} \rangle - \langle D_I^{(1)} \rangle _{L} \langle D_I^{(2)} \rangle_{L}  \right) 
\nonumber\\&&
+ \frac{(-\beta)}{2!} \left( \langle D_I^{(2)}D_I^{(1)} \rangle -  \langle D_I^{(2)} \rangle_{L} \langle D_I^{(1)}\rangle_{L} \right) \nonumber\\
&& + \frac{(-\beta)^2}{3!} \left( \langle D_I^{(1)}D_I^{(1)} D_I^{(1)} \rangle -3 \langle D_I^{(1)}\rangle_{L} \langle D_I^{(1)} D_I^{(1)} \rangle_{L} \right. 
\nonumber\\&&
\left.  - \langle D_I^{(1)} \rangle_{L} \langle D_I^{(1)} \rangle_{L} \langle D_I^{(1)} \rangle_{L} \right)  \\
&=& \langle D_I^{(3)} \rangle_{L} + \frac{(-\beta)}{2!} \langle D_I^{(1)}D_I^{(2)} \rangle_{L}
+ \frac{(-\beta)}{2!} \langle D_I^{(2)}D_I^{(1)} \rangle_{L} \nonumber\\
&& + \frac{(-\beta)^2}{3!} \langle D_I^{(1)}D_I^{(1)} D_I^{(1)} \rangle_{L}. \label{Omega3_linked}
\end{eqnarray}
They lead us to speculate that the recursion of $\Omega^{(n)}$ [Eq.\ (\ref{recursionOmega2})] is always reducible to a linked form:
\begin{eqnarray}
\Omega^{(n)} 
&=& \langle D_I^{(n)} \rangle + \frac{(-\beta)}{2!} \sum_{i=1}^{n-1} \left(  \langle D_I^{(i)}D_I^{(n-i)} \rangle - \Omega^{(i)} \Omega^{(n-i)} \right) \nonumber\\
&& + \frac{(-\beta)^2}{3!} \sum_{i=1}^{n-2}\sum_{j=1}^{n-i-1} \left( \langle D_I^{(i)}D_I^{(j)} D_I^{(n-i-j)} \rangle - \Omega^{(i)}\Omega^{(j)}\Omega^{(n-i-j)} \right) 
\nonumber\\&& + \dots + \frac{(-\beta)^{n-1}}{n!} \left(  \langle (D_I^{(1)})^n\rangle -  (\Omega^{(1)})^n \right)  \label{recursionOmegalinked0} \\
&=& \langle D_I^{(n)} \rangle_{L} + \frac{(-\beta)}{2!} \sum_{i=1}^{n-1} \langle D_I^{(i)}D_I^{(n-i)} \rangle_{L}
\nonumber\\&&
+ \frac{(-\beta)^2}{3!} \sum_{i=1}^{n-2}\sum_{j=1}^{n-i-1} \langle D_I^{(i)}D_I^{(j)} D_I^{(n-i-j)} \rangle_{L} 
\nonumber\\&& + \dots + \frac{(-\beta)^{n-1}}{n!}  \langle (D_I^{(1)})^n\rangle_{L}. \label{recursionOmegalinked}
\end{eqnarray}
This is indeed the case, and has already been implicit in the diagrammatic rules presented in Sec.\ \ref{sec:diagrams}.
A proof of this assertion  is given here as the linked-diagram theorem at finite temperature. 

The proof consists of two parts: First, we show that $\langle D^{(n)} \rangle$ is always linked by relying heavily on the linked-diagram theorem
at zero temperature.\cite{GellmannLow,brueckner,goldstone,hugenholtz,Frantz,Manne,Harris,shavitt}
Second, we prove that the unlinked terms in the subsequent sums in Eq.\ (\ref{recursionOmegalinked0}) are cancelled out exactly. 
Note that the cancellation does not complete within each parenthesis, and we have to consider the recursion
holistically. 

Using the $\langle E^{(n)}\rangle$ recursion in Sec.\ \ref{sec:HCPT}, we can rewrite $\langle D^{(n)} \rangle$ as
\begin{eqnarray}
\langle D^{(n)} \rangle &=& \langle E^{(n)} \rangle - \mu^{(n)} \bar{N} \\
&=& \langle \langle \Phi_I^{(0)} | \hat{V} | \Phi_I^{(n-1)} \rangle \rangle - (\mu^{(n)} \bar{N} )_{L}\\ 
&=& \langle \langle \Phi_I^{(0)} | \hat{V} \hat{R}_I \hat{V} | \Phi_I^{(n-2)} \rangle \rangle 
\nonumber\\&&
- \sum_{i=1}^{n-2} \langle \sum_J \langle \Phi_I^{(0)} | \hat{V} \hat{R}_I  | \Phi_J^{(n-1-i)} \rangle E_{JI}^{(i)}  \rangle - (\mu^{(n)} \bar{N} )_{L}, \nonumber\\ \label{127}
\end{eqnarray}
where $\mu^{(n)} \bar{N}$ is linked despite the fact that it is a product of two scalars. This is because $\mu^{(n)}$ is intensive.\cite{HirataTCA}

As emphasized, at finite temperature
the second term (the renormalization term) consists of both unlinked and linked contributions. Factors 
${E}_{JI}^{(i)}$ and $\langle \Phi_I^{(0)} | \hat{V} \hat{R}_I  | \Phi_J^{(n-1-i)} \rangle$ are proven by induction (see below) to be individually linked,
but not necessarily to each other. 
The unlinked contribution comes from the diagonal elements ${E}_{II}^{(i)}$, making the corresponding summand a product of two extensive scalars. 
Its off-diagonal elements ${E}_{JI}^{(i)}$ ($J \neq I$) are linked to 
the $\langle \Phi_I^{(0)} | \hat{V} \hat{R}_I  | \Phi_J^{(n-1-i)} \rangle$ part via $J$, which is degenerate with but distinct from $I$. 

Therefore, we can further rewrite the salient portion of the right-hand side of Eq. (\ref{127}) as
\begin{eqnarray}
&& \langle \langle \Phi_I^{(0)} | \hat{V} \hat{R}_I \hat{V} | \Phi_I^{(n-2)} \rangle \rangle 
- \sum_{i=1}^{n-2} \langle \sum_J  \langle \Phi_I^{(0)} | \hat{V} \hat{R}_I  | \Phi_J^{(n-1-i)} \rangle E_{JI}^{(i)} \rangle 
\nonumber\\&&
= \langle \langle \Phi_I^{(0)} | \hat{V} \hat{R}_I \hat{V} | \Phi_I^{(n-2)} \rangle \rangle_{L}
- \sum_{i=1}^{n-2} \langle \sum_{J}  \langle \Phi_I^{(0)} | \hat{V} \hat{R}_I  | \Phi_J^{(n-1-i)} \rangle E_{JI}^{(i)} \rangle_{L} 
\nonumber\\ &&
+ \langle \langle \Phi_I^{(0)} | \hat{V} \hat{R}_I \hat{V} | \Phi_I^{(n-2)} \rangle \rangle_{U} 
- \sum_{i=1}^{n-2} \langle \sum_{J} \langle \Phi_I^{(0)} | \hat{V} \hat{R}_I  | \Phi_J^{(n-1-i)} \rangle  E_{JI}^{(i)} \rangle_{U} 
\nonumber\\&&
= \langle \langle \Phi_I^{(0)} | \hat{V} \hat{R}_I \hat{V} | \Phi_I^{(n-2)} \rangle \rangle_{L}
- \sum_{i=1}^{n-2} \langle \sum_{J \neq I}  \langle \Phi_I^{(0)} | \hat{V} \hat{R}_I  | \Phi_J^{(n-1-i)} \rangle E_{JI}^{(i)} \rangle_{L}
\nonumber\\ &&
+ \langle \langle \Phi_I^{(0)} | \hat{V} \hat{R}_I \hat{V} | \Phi_I^{(n-2)} \rangle \rangle_{U} 
- \sum_{i=1}^{n-2} \langle \langle \Phi_I^{(0)} | \hat{V} \hat{R}_I  | \Phi_I^{(n-1-i)} \rangle  E_{II}^{(i)} \rangle_{U}
, \nonumber\\
\end{eqnarray}
where subscript $U$ means unlinked. The identical line of logic proving the linked-diagram theorem at zero temperature  
can be used to show that the last two terms cancel each other out, leaving only the linked contributions, i.e.,
\begin{eqnarray}
&& \langle \langle \Phi_I^{(0)} | \hat{V} \hat{R}_I \hat{V} | \Phi_I^{(n-2)} \rangle \rangle 
- \sum_{i=1}^{n-2} \langle \sum_J  \langle \Phi_I^{(0)} | \hat{V} \hat{R}_I  | \Phi_J^{(n-1-i)} \rangle E_{JI}^{(i)} \rangle 
\nonumber\\&&
= \langle \langle \Phi_I^{(0)} | \hat{V} \hat{R}_I \hat{V} | \Phi_I^{(n-2)} \rangle \rangle_{L}
- \sum_{i=1}^{n-2} \langle \sum_{J \neq I} \langle \Phi_I^{(0)} | \hat{V} \hat{R}_I  | \Phi_J^{(n-1-i)} \rangle E_{JI}^{(i)}  \rangle_{L}. \nonumber\\
\end{eqnarray}
The reader is referred to Chapter 6 of Shavitt and Bartlett\cite{shavitt} or Manne\cite{Manne} for the time-independent proof 
of the linked-diagram theorem of zero-temperature MBPT, which can be easily seen to be applicable to thermal averages with no modification.

Next, we prove that the second and subsequent terms in Eq.\ (\ref{recursionOmegalinked0}) are linked by 
virtue of a systematic cancellation of all unlinked terms between  
$\langle D_I^{(i_1)} \dots D_I^{(i_k)} \rangle$ ($2 \leq k \leq m$) and the products of $\Omega^{(i)}$'s. We use equations in Appendix \ref{app:StirlingS2} only for 
notational clarity.  

Let us define $A^{(n)}$ and $B^{(n)}$ by
\begin{eqnarray}
A^{(n)} &=& (-\beta) \Omega^{(n)}, \\
B^{(n)} &=& (-\beta) \langle D_I^{(n)} \rangle + \frac{(-\beta)^2}{2!} \sum_{i=1}^{n-1} \langle D_I^{(i)}D_I^{(n-i)} \rangle  \nonumber\\
&& + \frac{(-\beta)^3}{3!} \sum_{i=1}^{n-2}\sum_{j=1}^{n-i-1} \langle D_I^{(i)}D_I^{(j)} D_I^{(n-i-j)} \rangle 
\nonumber\\&& + \dots + \frac{(-\beta)^{n}}{n!} \langle (D_I^{(1)})^n\rangle. 
\end{eqnarray}
Equation (\ref{recursionOmegalinked0}) can then be identified as Eq.\ (\ref{A2}) with these definitions of $A^{(n)}$ and $B^{(n)}$. 
Furthermore, if we divide $B^{(n)}$ into the linked and unlinked parts,
\begin{eqnarray}
B^{(n)} = \left( B^{(n)} \right)_L + \left( B^{(n)} \right)_U,
\end{eqnarray}
 the unlinked part is a sum of products of the linked part, having the following exponential structure:\cite{VCC}
\begin{eqnarray}
\left( B^{(n)} \right)_U &=& \frac{1}{2!}\sum_{i = 1}^{n-1} \left( B^{(i)} \right)_L \left( B^{(n-i)} \right)_L \nonumber\\
&& + \frac{1}{3!}\sum_{i = 1}^{n-2}\sum_{j=1}^{n-i-1}  \left( B^{(i)} \right)_L \left( B^{(j)} \right)_L \left( B^{(n-i-j)} \right)_L \nonumber\\
&& + \dots + \frac{1}{n!} \left( B^{(1)} \right)_L^n.
\end{eqnarray}
This means
\begin{eqnarray}
B^{(n)}  &=&  \left( B^{(n)} \right)_L+ \frac{1}{2!}\sum_{i = 1}^{n-1} \left( B^{(i)} \right)_L \left( B^{(n-i)} \right)_L \nonumber\\
&& + \frac{1}{3!}\sum_{i = 1}^{n-2}\sum_{j=1}^{n-i-1}  \left( B^{(i)} \right)_L \left( B^{(j)} \right)_L \left( B^{(n-i-j)} \right)_L \nonumber\\
&& + \dots + \frac{1}{n!} \left( B^{(1)} \right)_L^n.
\end{eqnarray}
Comparing this with Eq.\ (\ref{B}), we immediately surmise $A^{(n)} = \left( B^{(n)} \right)_L$, proving the linkedness of $\Omega^{(n)}$. 

\section{Numerical benchmarks}

\subsection{General-order algorithms} 

The recursion relationships for the grand canonical ensemble in Sec.\ \ref{sec:recursions} 
and those for the canonical ensemble in Appendix \ref{app:canonical} have been implemented in a determinant-based FCI program.
The algorithms are literal translations of those equations. 
They enable calculations of $\mu^{(n)}$, $\Omega^{(n)}$, and $U^{(n)}$ in the grand canonical ensemble
and $F^{(n)}$ and $U^{(n)}$ in the canonical ensemble for an ideal gas of the smallest molecules at any arbitrary perturbation order, while
$S^{(n)}$ can be inferred from them. 
These calculations are more expensive than FCI itself and are not intended for realistic applications. Instead, they are meant to furnish 
benchmark data\cite{knowles_mp,Hirata2017,JhaHirata,hirata_cc,kallay_cc,olsen_cc,hirata_eomcc,hirata_ipeomcc} to assess the validity of the theories and the correctness of 
more efficient algorithms\cite{Shimazaki,hirata_qp,logarithm,hybrid,willow_qp} based on reduced analytical formulas to be developed in the future. The latter 
algorithms should have the same cost scaling as their respective zero-temperature counterparts, and be applicable to solids
in the grand canonical ensemble.\cite{He2014,Hermes2015}
Practical utility of the canonical ensemble, which does not lend itself to reduced analytical formulas, is doubtful at this point. 

We have also implemented the reduced analytical formulas of $\Omega^{(n)}$, $\mu^{(n)}$, $U^{(n)}$, and $S^{(n)}$ in the grand canonical ensemble
in the range of $0 \leq n \leq 3$. We already implemented the analytical formulas for $n \leq 2$ in our previous studies.\cite{Hirata1storder,Hirata2ndorder} 
 That of $\Omega^{(3)}$ derived
in Sec.\ \ref{sec:Omega3} has been manually implemented. 
The analytical formulas for $\mu^{(3)}$, $U^{(3)}$, and $S^{(3)}$ have then been directly computer-programmed
by systematic modifications of the $\Omega^{(3)}$ code. 
The derivations at higher $n$ would be extremely tedious and need to be computerized.\cite{tce}

We have repeated the $\lambda$-variation calculations
in both ensembles up to a higher order than previously reported.\cite{JhaHirata,JhaHirata_canonical} 
A $\lambda$-variation calculation\cite{Hirata2017,JhaHirata} determines the several lowest-order perturbation corrections
by numerical differentiation with respect to $\lambda$ of the corresponding quantity obtained by the finite-temperature FCI (Ref.\ \onlinecite{Kou})
with a perturbation-scaled Hamiltonian,
$\hat{H} = \hat{H}_0 + \lambda\hat{V}$.
The lower-order data already reported in our previous studies
are reproduced by the present calculations. The new data at higher orders are not necessarily precise 
because of the round-off and finite-difference errors, which have been detected by the disagreements among 
different finite-difference formulas (not shown). 

\subsection{Grand canonical ensemble\label{sec:grandcanonicaldata}}

Tables \ref{tab:grandcanonical5}, \ref{tab:grandcanonical6}, and \ref{tab:grandcanonical7} compile the thermodynamic functions 
in the grand canonical ensemble  at $T = 10^5$, $10^6$, and $10^7$\,K, respectively, 
 of an ideal gas of the identical hydrogen fluoride
molecules ($0.9168$\,\AA) in the STO-3G basis set.\cite{Kou,JhaHirata,JhaHirata_canonical} 
The gas consists of infinitely many nonrotating rigid hydrogen fluoride molecules that are not interacting with one another
except to exchange electrons. The system and temperatures considered are unrealistic, only serving
as a convenient test case for formalisms and computer codes, although a similar atomic calculation might be relevant to the Saha ionization equation.\cite{Saha1,Saha2}
The zero-temperature canonical HF wave function for the singlet ground state has been used as the reference. 

\begin{table*}
\caption{Perturbation corrections to the thermodynamic functions of an ideal gas of the hydrogen fluoride molecules ($0.9168$\,\AA\ in STO-3G) in the grand canonical ensemble at $T=10^5\,\text{K}$.}
\label{tab:grandcanonical5}
\begin{ruledtabular}
\begin{tabular}{rddddddddd}
& \multicolumn{3}{c}{$\Omega^{(n)}\,/\,E_{\text{h}}$} 
& \multicolumn{3}{c}{$\mu^{(n)}\,/\,E_{\text{h}}$} 
& \multicolumn{3}{c}{$U^{(n)}\,/\,E_{\text{h}}$} 
\\ \cline{2-4}\cline{5-7}\cline{8-10}
$n$ 
& \multicolumn{1}{r}{Recursion\footnotemark[1]} & \multicolumn{1}{r}{Analytical\footnotemark[2]} & \multicolumn{1}{r}{$\lambda$-variation\footnotemark[3]} 
& \multicolumn{1}{r}{Recursion\footnotemark[1]} & \multicolumn{1}{r}{Analytical\footnotemark[2]} & \multicolumn{1}{r}{$\lambda$-variation\footnotemark[3]} 
& \multicolumn{1}{r}{Recursion\footnotemark[1]} & \multicolumn{1}{r}{Analytical\footnotemark[2]} & \multicolumn{1}{r}{$\lambda$-variation\footnotemark[3]} 
\\ \hline
 0&  -55.63656&-55.63656&-55.63656&  0.27224& 0.27224& 0.27224& -52.01659&-52.01659&-52.01659\\
 1&  -45.26843&-45.26843&-45.26843& -0.07519&-0.07519&-0.07519& -45.94786&-45.94786&-45.94786\\
 2&   -2.58148& -2.58148& -2.58147&  0.23198& 0.23198& 0.23198&   0.09841&  0.09841&  0.09841\\
 3&    4.41331&  4.41331&  4.41356& -0.42177&-0.42177&-0.42180&  -0.14604& -0.14604& -0.14602\\
 4&   -9.72934&         & -9.73864&  0.92740&        & 0.92833&  -0.17127&         & -0.17196\\
 5&   22.27604&         & 22.18099& -2.13089&        &-2.12063&   0.72367&         &  0.71965\\
 6&  -53.38526&         &         &  5.11603&        &        &  -2.94946&         &         \\
 7&  129.97820&         &         &-12.46129&        &        &  10.60352&         &         \\
 8& -311.27100&         &         & 29.81298&        &        & -38.     &         &         \\
 9&  713.88281&         &         &-67.96933&        &        & 121.     &         &         \\
10&-1661.6&         &         &147.87702&        &        &3888.     &         &         \\
$\sum_0^{10}$       &-1268.9 & & &101.17918& & & 3881.    \\
FCI\footnotemark[4] & -102.10659 & & &  0.29568& & & -98.04938\\
\end{tabular}
\footnotetext[1]{MBPT($n$) recursions. The entries with fewer significant figures suffer from large round-off errors.}
\footnotetext[2]{MBPT($n$) reduced analytical formulas.}
\footnotetext[3]{Central seven-point formula with $\Delta\lambda = 0.01$. See also Ref.\ \onlinecite{JhaHirata}}
\footnotetext[4]{Finite-temperature full configuration interaction. See also Ref.\ \onlinecite{Kou}}
\end{ruledtabular}
\end{table*}

\begin{table*}
\caption{Same as Table \ref{tab:grandcanonical5} but at $T=10^6\,\text{K}$.}
\label{tab:grandcanonical6}
\begin{ruledtabular}
\begin{tabular}{rddddddddd}
& \multicolumn{3}{c}{$\Omega^{(n)}\,/\,E_{\text{h}}$} 
& \multicolumn{3}{c}{$\mu^{(n)}\,/\,E_{\text{h}}$} 
& \multicolumn{3}{c}{$U^{(n)}\,/\,E_{\text{h}}$} 
\\ \cline{2-4}\cline{5-7}\cline{8-10}
$n$ 
& \multicolumn{1}{r}{Recursion\footnotemark[1]} & \multicolumn{1}{r}{Analytical\footnotemark[2]} & \multicolumn{1}{r}{$\lambda$-variation\footnotemark[3]} 
& \multicolumn{1}{r}{Recursion\footnotemark[1]} & \multicolumn{1}{r}{Analytical\footnotemark[2]} & \multicolumn{1}{r}{$\lambda$-variation\footnotemark[3]} 
& \multicolumn{1}{r}{Recursion\footnotemark[1]} & \multicolumn{1}{r}{Analytical\footnotemark[2]} & \multicolumn{1}{r}{$\lambda$-variation\footnotemark[3]} 
\\ \hline
  0 &  -105.94753 &-105.94753 & -105.94753& 3.96130   &  3.96130 &  3.96130& -50.59635 &-50.59635  & -50.59635\\
  1 &   -44.52564 & -44.52564 &  -44.52564&-0.16896   & -0.16896 & -0.16896& -46.17665 &-46.17665  & -46.17665\\
  2 &    -0.96431 &  -0.96431 &   -0.96425& 0.08509   &  0.08509 &  0.08509&  -0.21984 & -0.21984  &  -0.21983\\
  3 &     0.24939 &   0.24939 &    0.25774&-0.02270   & -0.02270 & -0.02353&   0.06464 &  0.06464  &   0.06648\\
  4 &    -0.07381 &           &   -0.45286& 0.00676   &          &  0.04467&  -0.02389 &           &  -0.10720\\
  5 &     0.02296 &           &  -15.64135&-0.00210   &          &  1.56433&   0.00945 &           &  -3.44242\\
  6 &    -0.00699 &           &           & 0.00063   &          &         &  -0.00373 &           &          \\
  7 &     0.00187 &           &           &-0.00017   &          &         &   0.00140 &           &          \\
  8 &    -0.00032 &           &           & 0.00003   &          &         &  -0.00048 &           &          \\
  9 &    -0.00005 &           &           & 0.00001   &          &         &   0.00014 &           &          \\
 10 &     0.00009 &           &           &-0.00001   &          &         &  -0.00002 &           &          \\
$\sum_0^{10}$   & -151.24436 & & & 3.85989 & & & -96.94533 \\
FCI\footnotemark[4] & -151.24440 & & & 3.85990 & & & -96.94534\\
\end{tabular}
\footnotetext[1]{MBPT($n$) recursions.}
\footnotetext[2]{MBPT($n$) reduced analytical formulas.}
\footnotetext[3]{Central seven-point formula with $\Delta\lambda = 0.01$. See also Ref.\ \onlinecite{JhaHirata}}
\footnotetext[4]{Finite-temperature full configuration interaction. See also Ref.\ \onlinecite{Kou}}
\end{ruledtabular}
\end{table*}

\begin{table*}
\caption{Same as Table \ref{tab:grandcanonical5} but at $T=10^7\,\text{K}$.}
\label{tab:grandcanonical7}
\begin{ruledtabular}
\begin{tabular}{rddddddddd}
& \multicolumn{3}{c}{$\Omega^{(n)}\,/\,E_{\text{h}}$} 
& \multicolumn{3}{c}{$\mu^{(n)}\,/\,E_{\text{h}}$} 
& \multicolumn{3}{c}{$U^{(n)}\,/\,E_{\text{h}}$} 
\\ \cline{2-4}\cline{5-7}\cline{8-10}
$n$ 
& \multicolumn{1}{r}{Recursion\footnotemark[1]} & \multicolumn{1}{r}{Analytical\footnotemark[2]} & \multicolumn{1}{r}{$\lambda$-variation\footnotemark[3]} 
& \multicolumn{1}{r}{Recursion\footnotemark[1]} & \multicolumn{1}{r}{Analytical\footnotemark[2]} & \multicolumn{1}{r}{$\lambda$-variation\footnotemark[3]} 
& \multicolumn{1}{r}{Recursion\footnotemark[1]} & \multicolumn{1}{r}{Analytical\footnotemark[2]} & \multicolumn{1}{r}{$\lambda$-variation\footnotemark[3]} 
\\ \hline
  0 & -686.70814 &-686.70814 &-686.70814 &  47.15012  & 47.15012 & 47.15012& -45.78911 &-45.78911  & -45.78911\\
  1 &  -43.19911 & -43.19911 & -43.19911 &  -0.29811  & -0.29811 & -0.29811& -46.23554 &-46.23554  & -46.23554\\
  2 &   -0.19696 &  -0.19696 &  -0.19691 &   0.01774  &  0.01774 &  0.01774&  -0.03260 & -0.03260  &  -0.03258\\
  3 &    0.00951 &   0.00951 &   0.05365 &  -0.00088  & -0.00088 & -0.00529&   0.00179 &  0.00179  &   0.01333\\
  4 &   -0.00053 &           &   1.45717 &   0.00005  &          & -0.14572&  -0.00013 &           &   0.38065\\
  5 &    0.00003 &           & -44.48327 &  -0.00000  &          &  4.44833&   0.00001 &           & -11.63101\\
  6 &   -0.00000 &           &           &   0.00000  &          &         &  -0.00000 &           &          \\
  7 &    0.00000 &           &           &  -0.00000  &          &         &   0.00000 &           &          \\
  8 &   -0.00000 &           &           &   0.00000  &          &         &  -0.00000 &           &          \\
  9 &    0.00000 &           &           &  -0.00000  &          &         &   0.00000 &           &          \\
 10 &   -0.00000 &           &           &   0.00000  &          &         &  -0.00000 &           &          \\
$\sum_0^{10}$   &-730.09519 & & & 46.86892 & & & -92.05557 \\
FCI\footnotemark[4] &-730.09519 & & & 46.86892 & & & -92.05557\\
\end{tabular}
\footnotetext[1]{MBPT($n$) recursions.}
\footnotetext[2]{MBPT($n$) reduced analytical formulas.}
\footnotetext[3]{Central seven-point formula with $\Delta\lambda = 0.01$. See also Ref.\ \onlinecite{JhaHirata}}
\footnotetext[4]{Finite-temperature full configuration interaction. See also Ref.\ \onlinecite{Kou}}
\end{ruledtabular}
\end{table*}

In all cases, $\Omega^{(n)}$, $\mu^{(n)}$, and $U^{(n)}$ computed by the sum-over-states (recursion) and sum-over-orbitals (reduced analytical) 
formulas agree with each other for at least ten decimal places,  verifying the formulas and computer programs mutually.

At $T=10^5$\,K, they display 
a clear sign of divergence. This may not be surprising in light of the fact that HCPT for many
states are already divergent.\cite{Hirata2017} This is a different type of divergence than discovered by Kohn and Luttinger\cite{kohn} and analyzed by us.\cite{Hirata_PRA}
The latter is concerned with the zero-temperature limit, at which
the radius of convergence is zero under some circumstances,\cite{Hirata_PRA} 
whereas the former should have a finite radius of convergence, the rate of which may be accelerated 
by, e.g., the Pad\'{e} approximant.\cite{Pade,Benderbook,laidig,hirata_cc}
At lower temperatures (not shown), $U^{(n)}$ reproduces $E^{(n)}$ 
for the reference state computed by zero-temperature MBPT, and is much less prone to divergence (see, however, Ref.\ \onlinecite{Olsen}).
At $T=10^6$ and $10^7$\,K, the thermodynamic functions are rapidly convergent at the respective finite-temperature FCI values.
At $10^7$\,K, the tenth-order approximations are within $10^{-7}\,E_\text{h}$ of FCI, testifying that the finite-temperature MBPT presented here
is a converging series of approximation towards exactness. This would not be the case if the renormalization diagrams were overlooked or neglected. 

At $T=10^5$\,K, the $\lambda$-variation method turns out to be surprisingly accurate. It remains somewhat useful up to the
fifth order. 
At $T=10^6$ and $10^7$\,K,  it breaks down at lower orders, hardly serving as a benchmark even for the second-order corrections. 
The stability of the $\lambda$-variation method may be anticorrelated with the 
convergence of the perturbation series. 
At lower temperatures (not shown), the finite-temperature FCI (and hence  the $\lambda$-variation also) becomes unstable
because of the inherent difficulty of precisely determining $\mu$. At $T=0$, 
any value of $\mu$ falling in between the highest-occupied (HOMO) and lowest-unoccupied molecular orbital (LUMO) energies 
satisfies the electroneutrality condition, making it increasingly difficult for $\mu$ to reach the correct zero-temperature limit,\cite{Kou,Hirata_PRA} which is the midpoint
of HOMO and LUMO. Nevertheless, for the temperatures used in these tables, 
analytical results and the $\lambda$-variation benchmarks are in good agreement.


\begin{figure}
\includegraphics[scale=0.65]{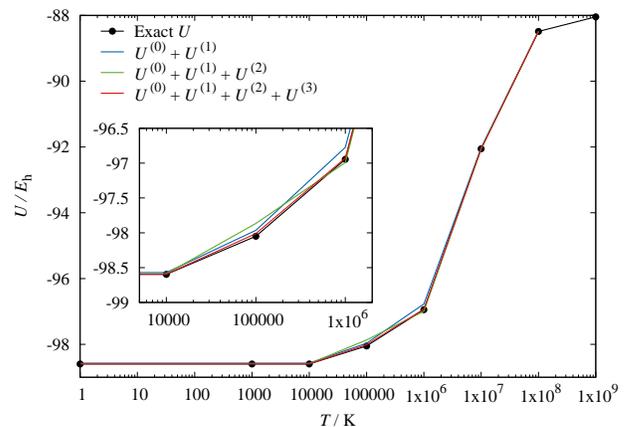}
\caption{\label{fig:U_plot}$U^{(0)} + \dots + U^{(n)}$ ($1 \leq n \leq 3$) as a function of temperature for the same system as Table \ref{tab:grandcanonical5}.}
\end{figure}

\begin{figure}
\includegraphics[scale=0.65]{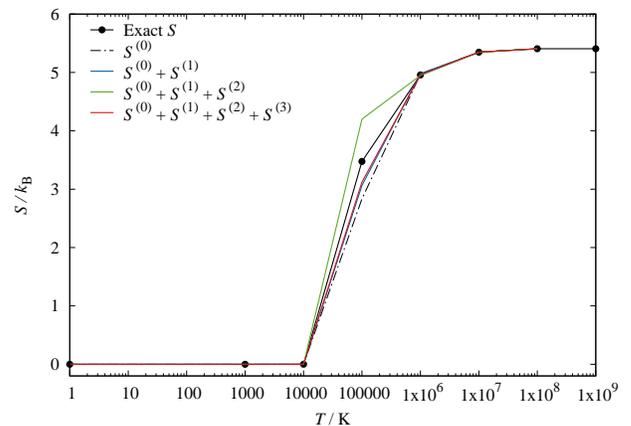}
\caption{\label{fig:S_plot}$S^{(0)} + \dots + S^{(n)}$ ($0 \leq n \leq 3$) as a function of temperature for the same system as Table \ref{tab:grandcanonical5}.}
\end{figure}

Figures \ref{fig:U_plot} and \ref{fig:S_plot} plot the perturbation approximations to $U$ and $S$ and their FCI values as a function of temperature. 
It can be seen that the perturbation approximations are so accurate that they are indiscernible from the FCI curves at both low and high temperatures.
Only at mid-temperatures ($10^4 \leq T/\text{K} \leq 10^6$) are the differences noticeable. This is the same temperature domain where these thermodynamic
functions exhibit strong temperature dependence and the perturbation series tend to diverge. Still, low-order perturbation theories seem to work well,
correcting the vast majority of the discrepancies between the zeroth-order (Fermi--Dirac) theory and FCI in $U$ ($U^{(0)}$ is not shown 
in Fig.\ \ref{fig:U_plot} because  it is far outside the graph). For $S$, the first-order approximation overcorrects the errors in $S^{(0)}$ in this temperature domain,
while the second- and third-order approximations roughly halve the errors. Given the divergence of the underlying perturbation series at these temperatures,  
the lower-order perturbation theories are promising.

\subsection{Canonical ensemble\label{sec:canonicaldata}}

\begin{table}
\caption{Perturbation corrections to the thermodynamic functions of an ideal gas of the hydrogen fluoride molecules ($0.9168$\,\AA\ in STO-3G) in the canonical ensemble at $T=10^5\,\text{K}$.}
\label{tab:canonical5}
\begin{ruledtabular}
\begin{tabular}{rdddd}
& \multicolumn{2}{c}{$F^{(n)}\,/\,E_{\text{h}}$} 
& \multicolumn{2}{c}{$U^{(n)}\,/\,E_{\text{h}}$} 
\\ \cline{2-3}\cline{4-5}
$n$ 
& \multicolumn{1}{r}{Recursion\footnotemark[1]} & \multicolumn{1}{r}{$\lambda$-variation\footnotemark[2]} 
& \multicolumn{1}{r}{Recursion\footnotemark[1]} & \multicolumn{1}{r}{$\lambda$-variation\footnotemark[2]} 
\\ \hline
  0 & -52.67166 &-52.67166&  -52.26452&-52.26452\\
  1 & -46.16315 &-46.16315&  -45.69442&-45.69442\\
  2 &  -0.14657 & -0.14657&   -0.02151& -0.02151\\
  3 &  -0.05239 & -0.05239&   -0.16648& -0.16648\\
  4 &   0.00162 &  0.00162&   -0.08826& -0.08826\\
  5 &   0.01117 &  0.01121&    0.01989&  0.02029\\
  6 &   0.00359 &         &    0.05664&         \\
  7 &  -0.00346 &         &   -0.5    &         \\
  8 &   0.04590 &         &   16.     &         \\
  9 &  -0.39063 &         & -562.     &         \\
 10 & -36.6   &         &22464.     &         \\
$\sum_0^{10}$   & -136.0 & & 21820.& \\
FCI\footnotemark[3] & -99.02043& &  -98.17836\\
\end{tabular}
\footnotetext[1]{MBPT($n$) recursions. The entries with fewer significant figures suffer from large round-off errors.}
\footnotetext[2]{Central seven-point formula with $\Delta\lambda = 0.01$. See also Ref.\ \onlinecite{JhaHirata_canonical}.}
\footnotetext[3]{Finite-temperature full configuration interaction. See also Ref.\ \onlinecite{Kou}.}
\end{ruledtabular}
\end{table}

\begin{table}
\caption{Same as Table \ref{tab:canonical5} but at $T=10^6\,\text{K}$.}
\label{tab:canonical6}
\begin{ruledtabular}
\begin{tabular}{rdddd}
& \multicolumn{2}{c}{$F^{(n)}\,/\,E_{\text{h}}$} 
& \multicolumn{2}{c}{$U^{(n)}\,/\,E_{\text{h}}$} 
\\ \cline{2-3}\cline{4-5}
$n$ 
& \multicolumn{1}{r}{Recursion\footnotemark[1]} & \multicolumn{1}{r}{$\lambda$-variation\footnotemark[2]} 
& \multicolumn{1}{r}{Recursion\footnotemark[1]} & \multicolumn{1}{r}{$\lambda$-variation\footnotemark[2]} 
\\ \hline
  0 &-62.55550 &-62.55550&-50.62284&-50.62284\\
  1 &-46.77859 &-46.77859&-46.71663&-46.71663\\
  2 & -0.01647 & -0.01647& -0.03420& -0.03420\\
  3 &  0.00030 &  0.00030&  0.00090&  0.00090\\
  4 &  0.00000 &  0.00000&  0.00000&  0.00000\\
  5 & -0.00000 &  0.00001& -0.00000&  0.00002\\
  6 &  0.00000 &&  0.00000&\\
  7 &  0.00000 && -0.00000&\\
  8 & -0.00000 &&  0.00000&\\
  9 & -0.00000 && -0.00000&\\
 10 &  0.00000 && -0.00000&\\
$\sum_0^{10}$   &  -109.35026   & &  -97.37278   & \\
FCI\footnotemark[3] & -109.35026    & & -97.37278     \\
\end{tabular}
\footnotetext[1]{MBPT($n$) recursions.}
\footnotetext[2]{Central seven-point formula with $\Delta\lambda = 0.01$. See also Ref.\ \onlinecite{JhaHirata_canonical}.}
\footnotetext[3]{Finite-temperature full configuration interaction. See also Ref.\ \onlinecite{Kou}.}
\end{ruledtabular}
\end{table}

\begin{table}
\caption{Same as Table \ref{tab:canonical5} but  at $T=10^7\,\text{K}$.}
\label{tab:canonical7}
\begin{ruledtabular}
\begin{tabular}{rdddd}
& \multicolumn{2}{c}{$F^{(n)}\,/\,E_{\text{h}}$} 
& \multicolumn{2}{c}{$U^{(n)}\,/\,E_{\text{h}}$} 
\\ \cline{2-3}\cline{4-5}
$n$ 
& \multicolumn{1}{r}{Recursion\footnotemark[1]} & \multicolumn{1}{r}{$\lambda$-variation\footnotemark[2]} 
& \multicolumn{1}{r}{Recursion\footnotemark[1]} & \multicolumn{1}{r}{$\lambda$-variation\footnotemark[2]} 
\\ \hline
  0 &-176.80350 &-176.80350&-46.00278&-46.00278\\
  1 & -46.85743 & -46.85743&-46.84516&-46.84516\\
  2 &  -0.00243 &  -0.00243& -0.00371& -0.00371\\
  3 &   0.00003 &   0.00003&  0.00005&  0.00005\\
  4 &  -0.00000 &  -0.00000& -0.00000& -0.00000\\
  5 &   0.00000 &   0.00001&  0.00000&  0.00001\\
  6 &  -0.00000 && -0.00000&\\
  7 &   0.00000 &&  0.00000&\\
  8 &   0.00000 &&  0.00000&\\
  9 &   0.00000 &&  0.00000&\\
 10 &   0.00000 &&  0.00000&\\
$\sum_0^{10}$   &  -223.66334   & &  -92.85159   & \\
FCI\footnotemark[3] & -223.66334    & & -92.85159     \\
\end{tabular}
\footnotetext[1]{MBPT($n$) recursions.}
\footnotetext[2]{Central seven-point formula with $\Delta\lambda = 0.01$. See also Ref.\ \onlinecite{JhaHirata_canonical}.}
\footnotetext[3]{Finite-temperature full configuration interaction. See also Ref.\ \onlinecite{Kou}.}
\end{ruledtabular}
\end{table}

Thermodynamic functions  in the canonical ensemble have been calculated  for the same system as Sec.\ \ref{sec:grandcanonicaldata}.
The results  at $T = 10^5$, $10^6$, and $10^7$\,K are summarized in Tables \ref{tab:canonical5}, \ref{tab:canonical6}, and \ref{tab:canonical7}, respectively.
Only the sum-over-states (recursion) formulas and $\lambda$-variation method are available for this ensemble.

Owing to the much fewer number of states involved in the canonical ensemble, the thermodynamic functions are less prone to divergence
and the $\lambda$-variation method tends to be more accurate. At $T=10^6$ and $10^7$\,K, the tenth-order perturbation theory agrees with FCI for five and ten decimal places, respectively, which attests to the convergence of the perturbation series to exactness at any temperature unless they diverge.

\section{Conclusions} 

We have fully developed the finite-temperature MBPT that expands all thermodynamic functions in uniform perturbation series.
Both grand canonical and canonical ensembles are considered, although only the former lends itself to more drastic mathematical simplifications, making 
it  useful for condensed-matter applications.
The Rayleigh--Schr\"{o}dinger-like algebraic recursions have been obtained for both, yielding their sum-over-states analytical formulas of all thermodynamic functions
at any arbitrary order. 

For the grand canonical ensemble, the sum-over-states formulas have been transformed to sum-over-orbitals analytical formulas 
by the method of normal-ordered second quantization at finite temperature. 
The rules of Wick contractions and various operators in the normal-ordered form 
have been derived (as opposed to postulated) and 
how they are practically applied to integral and thermal-average evaluations has been illustrated. Feynman diagrammatic rules are then
introduced as a straightforward graphical representation of Wick contractions as lines and of molecular integrals as vertexes.
Nowhere in this formulation do we resort to time-dependent logic\cite{matsubara,bloch,kohn,luttingerward,balian,blochbook} or human 
intuitions to exhaustively enumerate all valid diagrams, which include the renormalization, anomalous, and anomalous renormalization diagrams.

Normal ordering  regroups the Hamiltonian operator into the finite-temperature HF energy, finite-temperature
Fock matrix elements times one-electron operators, and two-electron integrals times two-electron operators. This natural partitioning
underscores the significance of the finite-temperature HF theory as the foundation of all converging finite-temperature electron-correlation theories. 
Its energy expression is identified with $\langle E^{(0)}_I\rangle + \langle E^{(1)}_I\rangle$, which differs from either $U^{(0)}+ U^{(1)}$ or $\Omega^{(0)} + \Omega^{(1)}$
at $T > 0$. 

In the nondegenerate, zero-temperature MBPT, the renormalization terms cancel 
exactly the unlinked contribution of the equal magnitude in the parent term, 
ensuring the size-consistency of MBPT at any order, as proven by the linked-diagram theorem.\cite{GellmannLow,brueckner,goldstone,hugenholtz,Frantz,Manne}
In the finite-temperature MBPT, which sums over 
all possible states many of which are degenerate, the renormalization terms consist of both linked and unlinked 
contributions. The unlinked contribution cancels the same in the parent term, restoring the linkedness 
of the perturbation corrections at any order, while nonzero linked renormalization terms persist.
They correspond to diagrams whose resolvent lines are displaced upwards in all possible ways so that
there are two or more resolvents in between some pairs of adjacent vertexes and none in between others. 
These renormalization diagrams are distinguished from 
the well-known anomalous diagrams whose resolvent lines are deleted. They both are essential numerically, and 
had it not for them, the perturbation series would not converge at the finite-temperature FCI limit.  
It is unclear whether the conventional time-dependent, diagrammatic formulation\cite{matsubara,bloch,kohn,luttingerward,balian,blochbook} or 
density-matrix derivation\cite{SANTRA} are aware of such diagrams.

We have proven the linked-diagram theorem of the finite-temperature MBPT. The theorem is based on the 
linked-diagram theorem at zero temperature\cite{GellmannLow,brueckner,goldstone,hugenholtz,Frantz,Manne} and the systematic cancellation of unlinked anomalous diagrams.

We have derived reduced analytical formulas for the perturbation corrections to thermodynamic functions
in the grand canonical ensemble up to the third order. We have also implemented general-order algorithms
in both ensemble, and calculated the benchmark perturbation corrections up to the tenth order.
At intermediate temperatures, where thermodynamic functions vary with temperature considerably,
the perturbation series tend to diverge, but elsewhere they converge rapidly towards the finite-temperature FCI limits,
establishing the correctness of the theory.

In a separate study,\cite{Hirata_PRA} we showed that the finite-temperature MBPT at the first and second orders 
does not converge at the correct zero-temperature limit when the reference wave function is qualitatively different from the exact one, confirming
Kohn and Luttinger.\cite{kohn} 
This originates from the nonanalyticity of the Boltzmann factor and is therefore expected to plague most any finite-temperature MBPT, 
e.g., the one in the canonical ensemble also.\cite{JhaHirata_canonical} Nonconvergence should persist at third and all higher orders
under the same condition.

This study and another reported in Ref.\ \onlinecite{Hirata2017} have been 
conducted to lay a firm mathematical foundation of perturbation theories,\cite{bartlett_arpc,hirata_mp2} which
continue to be a workhorse for efficient, size-consistent,\cite{HirataTCA,conjecture} converging {\it ab initio} electronic structure calculations\cite{Willow2012, Willow2013, Willow2013b, Willow2014, Willow2014a, Willow2014b, Doran2016, Johnson2016, Johnson2018, Doran2019, Doran2020a, Doran2020b,Doran2021} for large complex molecules and solids.\cite{tenno_review}
Unlike time-dependent, diagrammatic expositions,\cite{march,thouless1972quantum,mattuck1992guide,Fetter} 
which are often incomplete and intractable, our derivations follow the transparent, linear, time-independent, algebraic 
logic, involving only elementary calculus and combinatorial identities. They justify the second-quantized and diagrammatic rules just as tools for expediency
rather than rely on them as a graphical method of derivations that are often based on human intuitions detached from mathematics. 
Our derivation strategy can therefore be universally applied to any perturbation theory insofar as the exact limit is known,
reliably leading to recursions, order-by-order analytical formulas, and general-order algorithms.

\section{Data Availability Statement}
The data that supports the findings of this study are available within the article.

\acknowledgments
The author thanks Professor Rodney J. Bartlett for teaching him the quantum-field-theoretical techniques used in this study.
This work was supported by the Center for Scalable, Predictive methods for Excitation and Correlated phenomena (SPEC), which is funded by 
the U.S. Department of Energy, Office of Science, Office of Basic Energy Sciences, 
Chemical Sciences, Geosciences, and Biosciences Division, as a part of the Computational Chemical Sciences Program
and also by the U.S. Department of Energy, Office of Science, Office of Basic Energy Sciences under Grant No.\ DE-SC0006028.

\appendix

\section{Recursions in the canonical ensemble\label{app:canonical}} 

Thermodynamic functions in the canonical ensemble considered are the canonical partition function ($Z$),
Helmholtz energy ($F$), internal energy ($U$), and entropy ($S$), which are related to one another by
\begin{eqnarray}
Z &=& \sum_I e^{-\beta E_I}, \\
F &=& -\frac{1}{\beta} \ln Z, \\
U&=& -\frac{\partial}{\partial \beta} \ln Z,
\end{eqnarray}
and
\begin{eqnarray}
F &=& U - TS,
\end{eqnarray}
where the $I$-summation goes over all spin states with a fixed total number ($\bar{N}$) of electrons.
The recursion for $Z^{(n)}$ is given by 
\begin{eqnarray}
\frac{Z^{(n)}}{Z^{(0)}} &=& ( -\beta) \langle  E^{(n)} \rangle + \frac{(-\beta)^2}{2!} \sum_{i=1}^{n-1} \langle E^{(i)}E^{(n-i)} \rangle
\nonumber\\&& 
+ \frac{(-\beta)^3}{3!} \sum_{i=1}^{n-2}\sum_{j=1}^{n-i-1} \langle E^{(i)}E^{(j)} E^{(n-i-j)} \rangle \nonumber\\
&&+ \dots + \frac{(-\beta)^n}{n!}  \langle (E^{(1)})^n\rangle,
\end{eqnarray}
where $\langle \dots \rangle$ is defined differently from Eq.\ (\ref{X}) only in this appendix as
\begin{eqnarray}
\langle X_I \rangle \equiv \frac{\sum_I X_I e^{-\beta E_I^{(0)}}}{\sum_I e^{-\beta E_I^{(0)}}}.
\end{eqnarray}
The recursion for $F^{(n)}$ is inferred from Eq.\ (\ref{recursionOmega2}) as
\begin{eqnarray}
F^{(n)} &=& \langle E^{(n)} \rangle + \frac{(-\beta)}{2!} \sum_{i=1}^{n-1} \left( \langle E^{(i)}E^{(n-i)} \rangle - F^{(i)}F^{(n-i)} \right) 
\nonumber\\&& 
+ \frac{(-\beta)^2}{3!} \sum_{i=1}^{n-2}\sum_{j=1}^{n-i-1} \left( \langle E^{(i)}E^{(j)} E^{(n-i-j)} \rangle - F^{(i)}F^{(j)} F^{(n-i-j)}\right) \nonumber\\&&
+ \dots + \frac{(-\beta)^{n-1}}{n!}  \left\{ \langle (E^{(1)})^n\rangle - (F^{(1)})^n  \right\}.  \label{recursionF}
\end{eqnarray}
Likewise, $U^{(n)}$ is defined recursively by
\begin{eqnarray}
U^{(n)}  &=&
 \langle  E_I^{(n)} \rangle + {(-\beta)}\sum_{i=1}^{n} \langle E_I^{(i)}E_I^{(n-i)} \rangle 
- (-\beta) \sum_{i=1}^{n} F^{(i)} U^{(n-i)}  
 \nonumber\\&&
+ \frac{(-\beta)^2}{2!} \sum_{i=1}^{n-1}\sum_{j=1}^{n-i} \langle E_I^{(i)}E_I^{(j)} E_I^{(n-i-j)} \rangle 
\nonumber\\&&
- \frac{(-\beta)^2}{2!} \sum_{i=1}^{n-1}\sum_{j=1}^{n-i} F^{(i)} F^{(j)} U^{(n-i-j)} 
\nonumber\\&&
+ \dots 
+ \frac{(-\beta)^n}{n!}  \langle (E_I^{(1)})^n E_I^{(0)} \rangle 
- \frac{(-\beta)^n}{n!}   (F^{(1)})^n U^{(0)} , \label{recursionU}
\end{eqnarray}
which implies the recursion for $S^{(n)} = k_{\text{B}}\beta ( U^{(n)} - F^{(n)} ) $.

These have been obtained by restricting the summations to $\bar{N}$-electron states 
and setting $\mu^{(n)} = 0$ for all $n$ in the corresponding recursions for the grand canonical ensemble.

\section{A proof of equivalence of Eqs.\ (\ref{recursionOmega0}) and (\ref{recursionOmega})\label{app:StirlingS2}}

Generally, we prove the equivalence of two recursions,
\begin{eqnarray}
A^{(n)} &=& B^{(n)} - \frac{1}{2} \sum_{i=1}^{n-1} B^{(i)}B^{(n-i)} 
\nonumber\\&&
+ \frac{1}{3} \sum_{i=1}^{n-2}\sum_{j=1}^{n-i-1} B^{(i)}B^{(j)}B^{(n-i-j)} 
\nonumber\\&&
- \dots - \frac{(-1)^n}{n} (B^{(1)})^n, \label{A1}
\end{eqnarray}
and
\begin{eqnarray}
A^{(n)} &=& B^{(n)} - \frac{1}{2!} \sum_{i=1}^{n-1} A^{(i)}A^{(n-i)} 
\nonumber\\&&
- \frac{1}{3!} \sum_{i=1}^{n-2}\sum_{j=1}^{n-i-1} A^{(i)}A^{(j)}A^{(n-i-j)} 
\nonumber\\&&
- \dots - \frac{1}{n!} (A^{(1)})^n. \label{A2} 
\end{eqnarray}
Equations (\ref{recursionOmega0}) and (\ref{recursionOmega}) are a special case of these in which
$A^{(n)} = -\beta \Omega^{(n)}$ and $B^{(n)} = \Xi^{(n)}/\Xi^{(0)}$. We can rewrite the second equation as
\begin{eqnarray}
B^{(n)} &=& A^{(n)} + \frac{1}{2!} \sum_{i=1}^{n-1} A^{(i)}A^{(n-i)} 
\nonumber\\&&
+ \frac{1}{3!} \sum_{i=1}^{n-2}\sum_{j=1}^{n-i-1} A^{(i)}A^{(j)}A^{(n-i-j)} 
\nonumber\\&&
+ \dots 
+ \frac{1}{n!} (A^{(1)})^n \label{B}.
\end{eqnarray}
Substituting this into the right-hand side of Eq.\ (\ref{A1}), we can convert it into a form expressed entirely in terms of $A^{(i)}$ as
\begin{eqnarray}
A^{(n)} &=& C_{A^{(n)}} A^{(n)} + \sum_{i=1}^{n-1}  C_{A^{(i)}A^{(n-i)}} A^{(i)}A^{(n-i)} 
\nonumber\\&&
+ \sum_{i=1}^{n-2}\sum_{j=1}^{n-i-1} C_{A^{(i)}A^{(j)}A^{(n-i-j)}} A^{(i)}A^{(j)}A^{(n-i-j)} 
\nonumber\\&&
+ \dots + C_{(A^{(1)})^n} (A^{(1)})^n. \label{A1_2}
\end{eqnarray} 
We shall show that $C_{A^{(n)}} = 1$ and $C_{A^{(i_1)}A^{(i_2)} \cdots A^{(i_m)}} = 0$ ($m \geq 2$). 

By inspection, we find $C_{A^{(n)}} = 1$, $C_{A^{(i)}A^{(n-i)}} = 1/2! - 1/2 = 0$, and $C_{A^{(i)}A^{(j)}A^{(n-i-j)}} = 1/3! - (1/2)(1/2!) - (1/2)(1/2!) + 1/3= 0$.
In general, the coefficient multiplying $A^{(i_1)}A^{(i_2)} \cdots A^{(i_m)}$ ($m \geq 2$) is given as
\begin{eqnarray}
&&C_{A^{(i_1)}A^{(i_2)} \cdots A^{(i_m)}}  \nonumber\\
&&= \frac{1}{m!} -\frac{1}{2} \sum_{i=1}^{m-1} \frac{1}{i! (m-i)!} 
+ \frac{1}{3} \sum_{i=1}^{m-2}\sum_{j=1}^{m-i-1} \frac{1}{i! j! (m-i-j)!} 
\nonumber\\&&
+ \dots 
- \frac{(-1)^m}{m} \left(\frac{1}{1!}\right)^m \nonumber \\
&&= \frac{1}{m!}\left( \frac{1!}{1}  \stirling{m}{1} - \frac{2!}{2}  \stirling{m}{2}  + \frac{3!}{3} \stirling{m}{3}  
+ \dots - \frac{(-1)^mm!}{m}   \stirling{m}{m} \right) , \nonumber\\ \label{zero}
\end{eqnarray}
where $\stirling{m}{k}$ is the Stirling number of the second kind (the number of ways $m$ objects are partitioned 
into $k$ nonempty sets).\cite{KaoZetterberg} It is defined as
\begin{eqnarray}
\stirling{m}{k} = \frac{1}{k!} \sum_{m_1,\dots,m_k} \frac{m!}{\prod_{i=1}^k m_i !},
\end{eqnarray}
where the summation goes over all natural numbers $m_1, \dots, m_k$ that satisfy $m_1 + \dots + m_k = m$.
There is an identity involving the Stirling number, which reads
\begin{eqnarray}
\sum_{k=1}^{m} (-1)^k (k-1)! \stirling{m}{k} = 0, \label{stirlingidentity}
\end{eqnarray}
which implies $C_{A^{(i_1)}A^{(i_2)} \cdots A^{(i_m)}} = 0$ ($m \geq 2$), proving the equivalence.

The identity (\ref{stirlingidentity}) can in turn be proven by induction using the recursion:
\begin{eqnarray}
\stirling{n+1}{k} = k\stirling{n}{k}+ \stirling{n}{k-1} . \label{Stirlingrecursion}
\end{eqnarray} 
Equation (\ref{stirlingidentity}) is true for $m=2$. If it is true for $m = 2, \dots, n$, 
it also holds true for $m = n+1$ and therefore for any $m \geq 2$ because
\begin{eqnarray}
&&\sum_{k=1}^{n+1} (-1)^k (k-1)! \stirling{n+1}{k} 
\nonumber\\
&&= \sum_{k=1}^{n+1} (-1)^k k! \stirling{n}{k}
+ \sum_{k=1}^{n+1} (-1)^k (k-1)! \stirling{n}{k-1} \nonumber\\
&&= \sum_{k=1}^{n+1} (-1)^k k! \stirling{n}{k}
+ \sum_{k=0}^{n} (-1)^{k+1} k! \stirling{n}{k}\nonumber\\
&&= \sum_{k=1}^{n} (-1)^k k! \stirling{n}{k}
+ \sum_{k=1}^{n} (-1)^{k+1} k! \stirling{n}{k} = 0,
\end{eqnarray}
where we used $\stirling{n}{0} = \stirling{n}{n+1} = 0$. 

Finally, the recursion for the Stirling number [Eq.\ (\ref{Stirlingrecursion})] can  be  justified as follows: 
The left-hand side counts the number of ways $n+1$ objects are partitioned into $k$ nonempty sets. 
The first term in the right-hand side is the number of ways in which 
the $(n+1)$th object is absorbed into one of the existing $k$ nonempty sets of the first $n$ objects.
The second term is the number of ways in which the $(n+1)$th object forms a new ($k$th) set containing only it
and the first $n$ objects are partitioned into $k-1$ nonempty sets.

One may view Maclaurin series of the right-hand sides of the following equations
as a mnemonic for Eqs.\ (\ref{A1}) and (\ref{B}), respectively:
\begin{eqnarray}
A^{(n)} &=& \ln \left(B^{(n)} + 1\right), \\
B^{(n)} &=& \exp\left(A^{(n)}\right) - 1,
\end{eqnarray}
which, however, may not be taken literally. 

\section{Normal-ordered second quantization at finite temperature\label{app:secondquantization}}

In this appendix, we fully develop the normal-ordered second quantization at finite temperature\cite{march,sanyal} 
within the time-independent framework in a transparent and pedagogical manner. 
It serves as a basis of the diagrammatic derivation and the linked-diagram theorem.
The reader is referred to Shavitt and Bartlett \cite{shavitt} for normal-ordered second quantization at zero temperature
and to March, Young, and Sampanthar\cite{march} for an abridged exposition of its finite-temperature counterpart.
The following expounds on the latter with enough details to be actually used in deriving the second-order formulas in Sec.\ \ref{sec:second}.

\subsection{Normal ordering at finite temperature}

A product of creation ($\hat{p}^\dagger$) and annihilation ($\hat{p}$) operators in the normal order (denoted by $\{ \dots \}$) is defined [cf.\ Eq.\ (9.3.2) of Ref.\ \onlinecite{march}] as
\begin{eqnarray}
\{ \hat{p}^\dagger \hat{p} \} &\equiv& f_p^+ \hat{p}^\dagger \hat{p} - f_p^- \hat{p}\,\hat{p}^\dagger, \\
\{ \hat{p} \,\hat{p}^\dagger \} &\equiv& f_p^- \hat{p}\, \hat{p}^\dagger - f_p^+ \hat{p}^\dagger\hat{p},
\end{eqnarray} 
where $f_p^-$ and $f_p^+$ are the Fermi--Dirac distribution functions given by Eqs.\ (\ref{occupancy}) and (\ref{vacancy}).
When $p \neq q$, 
\begin{eqnarray}
\{ \hat{p}^\dagger \hat{q} \} &\equiv& \hat{p}^\dagger \hat{q} , \\
\{ \hat{p}\, \hat{q}^\dagger \} &\equiv& \hat{p}\, \hat{q}^\dagger .
\end{eqnarray} 
A normal-ordered product of two creation operators or of two annihilation operators is also the same 
as the original order. 

In analogy to the zero-temperature case, the normal ordering at finite temperature 
is designed so that the thermal average of a normal-ordered product is always zero.
Using Einstein's convention of implied summations of repeated spinorbital indexes, we confirm
\begin{eqnarray}
\langle \langle I | x_{pq} \{ \hat{p}^\dagger \hat{q} \} | I \rangle \rangle &\equiv& 
\frac{\sum_I \langle I | x_{pq} \{ \hat{p}^\dagger \hat{q} \} | I \rangle e^{-\beta D_I^{(0)}} }{\sum_I e^{-\beta D_I^{(0)}}}  \nonumber\\
&=& \frac{\sum_I \langle I | x_{pp} \{ \hat{p}^\dagger \hat{p} \} | I \rangle e^{-\beta D_I^{(0)}} }{\sum_I e^{-\beta D_I^{(0)}}} \nonumber\\
&=& \frac{\sum_I \langle I | x_{pp} f_p^+ \hat{p}^\dagger \hat{p} | I \rangle e^{-\beta D_I^{(0)}} }{\sum_I e^{-\beta D_I^{(0)}}} 
\nonumber \\ &&
 - \frac{\sum_I \langle I | x_{pp} f_p^-  \hat{p}\, \hat{p}^\dagger | I \rangle e^{-\beta D_I^{(0)}} }{\sum_I e^{-\beta D_I^{(0)}}} \nonumber\\
&=& \sum_p x_{pp} f_p^+ f_p^- - \sum_p x_{pp} f_p^- f_p^+ \label{Boltzmann_sum1}\nonumber \\
&=& 0,
\end{eqnarray}
where we used the Boltzmann sum identities I and V of Ref.\ \onlinecite{Hirata2ndorder} in the penultimate equality.

Next, we define a Wick contraction (denoted by a staple symbol) as the difference between the operator product and its normal-ordered counterpart.
\begin{eqnarray}
\contraction[0.5ex]{}{\hat{p}}{^\dagger}{\hat{p}}
\hat{p}^\dagger \hat{p} &\equiv& \hat{p}^\dagger \hat{p} - \{ \hat{p}^\dagger \hat{p} \} = f_p^- (\hat{p}^\dagger \hat{p} + \hat{p}\, \hat{p}^\dagger) = f_p^- , \\
\contraction[0.5ex]{}{\hat{p}}{\,}{\hat{p}}
\hat{p} \,\hat{p}^\dagger &\equiv& \hat{p}\, \hat{p}^\dagger - \{ \hat{p} \,\hat{p}^\dagger \} = f_p^+ (\hat{p} \,\hat{p}^\dagger + \hat{p}^\dagger \hat{p}) = f_p^+,
\end{eqnarray}
where we used the anticommutation rules of fermion creation and annihilation operators. 
When $p \neq q$,
\begin{eqnarray}
\contraction[0.5ex]{}{\hat{p}}{^\dagger}{\hat{q}}
\hat{p}^\dagger \hat{q} &\equiv& \hat{p}^\dagger \hat{q} - \{ \hat{p}^\dagger \hat{q} \} = 0, \\
\contraction[0.5ex]{}{\hat{p}}{\,}{\hat{q}}
\hat{p} \,\hat{q}^\dagger &\equiv& \hat{p} \,\hat{q}^\dagger - \{ \hat{p}\, \hat{q}^\dagger \} = 0.
\end{eqnarray}
A Wick contraction of two creation or of two annihilation operators is also zero.

The concept of the normal ordering can be generalized to any even number of operators. For instance, a product of four operators in the normal order is
defined as ($p \neq q$),
\begin{eqnarray}
\{ \hat{p}^\dagger \hat{q}^\dagger \hat{q}\, \hat{p} \} &=& 
 - \{ \hat{p}^\dagger  \hat{q}^\dagger \hat{p}\, \hat{q}  \}
= \{ \hat{p}^\dagger \hat{p}\, \hat{q}^\dagger \hat{q}  \}  
\\
&\equiv& 
\left( f_p^+ \hat{p}^\dagger \hat{p} - f_p^- \hat{p}\,\hat{p}^\dagger \right) 
\left( f_q^+ \hat{q}^\dagger \hat{q} - f_q^- \hat{q}\,\hat{q}^\dagger  \right) 
\\
&=& f_p^+ f_q^+ \hat{p}^\dagger \hat{q}^\dagger \hat{q}\, \hat{p} 
- f_p^+ f_q^- \hat{p}^\dagger  \hat{q}\, \hat{q}^\dagger \hat{p}
 - f_p^- f_q^+ \hat{p} \, \hat{q}^\dagger \hat{q}\, \hat{p}^\dagger
 \nonumber\\&& 
+ f_p^- f_q^- \hat{p} \,\hat{q}\, \hat{q}^\dagger \hat{p}^\dagger.
\end{eqnarray}
When $pq \neq rs$ or $pq \neq sr$,
\begin{eqnarray}
\{ \hat{p}^\dagger \hat{q}^\dagger \hat{r}\, \hat{s} \} &\equiv&  \hat{p}^\dagger \hat{q}^\dagger \hat{r}\, \hat{s}. \label{pqrs}
\end{eqnarray}
The thermal average of a normal-ordered product of any number of operators is also always zero by construction.
This can be verified for the above product as follows:
\begin{eqnarray}
&& \langle \langle I |x_{pqpq} \{ \hat{p}^\dagger\hat{q}^\dagger\hat{q}\,\hat{p} \} |I \rangle \rangle 
\nonumber\\&&
= \langle \langle I |x_{pqpq} f_p^+ f_q^+ \hat{p}^\dagger \hat{q}^\dagger \hat{q}\, \hat{p} |I \rangle \rangle  
- \langle \langle I |x_{pqpq} f_p^+ f_q^- \hat{p}^\dagger  \hat{q}\, \hat{q}^\dagger \hat{p} |I \rangle \rangle  
\nonumber\\&&
- \langle \langle I |x_{pqpq} f_p^- f_q^+ \hat{p} \, \hat{q}^\dagger \hat{q}\, \hat{p}^\dagger |I \rangle \rangle  
+ \langle \langle I |x_{pqpq} f_p^- f_q^- \hat{p} \,\hat{q}\, \hat{q}^\dagger \hat{p}^\dagger |I \rangle \rangle \nonumber\\ 
&& = \sum_{p,q}  x_{pqpq}f_p^+ f_q^+f_p^- f_q^- - \sum_{p,q}x_{pqpq}f_p^+ f_q^-f_p^- f_q^+ \nonumber\\&&
- \sum_{p,q}x_{pqpq} f_p^- f_q^+ f_p^+ f_q^- + \sum_{p,q} x_{pqpq} f_p^- f_q^- f_p^+ f_q^+ \nonumber \\
&&= 0,
\end{eqnarray}
where Boltzmann sum identities III, V, and IX of Ref.\ \onlinecite{Hirata2ndorder} were used. 

For a product of more than two operators, there are two types of contractions:\ partial and full contractions.
Together, they are defined as 
\begin{eqnarray}
\contraction[0.5ex]{\{ \dots }{\hat{p}}{^\dagger \dots \hat{q} \dots }{\hat{p}}
\contraction[1.0ex]{\{ \dots \hat{p}^\dagger \dots }{\hat{q}}{ \dots \hat{p} \dots }{\hat{q}}
\{ \dots \hat{p}^\dagger \dots \hat{q} \dots \hat{p} \dots \hat{q}^\dagger \dots \} 
= (-1)^P 
\contraction[0.5ex]{}{\hat{p}}{^\dagger }{\hat{p}} 
\hat{p}^\dagger \hat{p} \,\,
\contraction[0.5ex]{}{\hat{q}}{\, }{\hat{q}}
\hat{q}\, \hat{q}^\dagger \{ \dots \}, \nonumber\\
\end{eqnarray}
where $\{ \dots \}$ in the right-hand side gathers the uncontracted operators (kept in the original order), and $P$ is the number of permutations
necessary to reorder the whole operator sequence from the left- to right-hand side. If 
there are no operators left in $\{ \dots \}$, it is a full contraction; otherwise it is a partial contraction. 
A full contraction is a real number, while a partial contraction is a normal-ordered operator multiplied by a real number. 
The parity of a full contraction is $+1$ ($-1$) for an even (odd) number of intersections of the staple symbols of Wick contractions.\cite{shavitt}

\subsection{Wick's theorem at finite temperature}

Wick's theorem states that a product of operators is the sum of its normal-ordered product and all of its 
partial and full contractions.\cite{shavitt} This holds true at zero and nonzero temperatures. 
For a product of two operators, the theorem is just a restatement of the definitions of Wick contractions:
\begin{eqnarray}
\contraction[0.5ex] {\hat{p}^\dagger \hat{q} = \{ \hat{p}^\dagger \hat{q} \} + }{\hat{p}}{^\dagger}{\hat{q}}
\hat{p}^\dagger \hat{q} &=& \{ \hat{p}^\dagger \hat{q} \} + \hat{p}^\dagger\hat{q}, \\
\contraction[0.5ex]{\hat{p}\, \hat{q}^\dagger = \{ \hat{p}\, \hat{q}^\dagger \} + }{\hat{p}}{\,}{\hat{q}}
\hat{p}\, \hat{q}^\dagger &=& \{ \hat{p}\, \hat{q}^\dagger \} + \hat{p}\,\hat{q}^\dagger.
\end{eqnarray}
For a product of four operators, the theorem asserts
\begin{eqnarray}
\hat{p}^\dagger  \hat{q}\, \hat{r}^\dagger \hat{s} &=& \{ \hat{p}^\dagger \hat{q}\, \hat{r}^\dagger  \hat{s} \} 
\contraction[0.5ex]{+ \{ }{\hat{p}}{^\dagger \hat{q}\,\hat{r}^\dagger  }{\hat{s}}
+ \{ \hat{p}^\dagger \hat{q}\,  \hat{r}^\dagger \hat{s} \} 
\contraction[0.5ex]{+ \{ \hat{p}^\dagger }{\hat{q}}{\,}{\hat{r}}
+ \{ \hat{p}^\dagger \hat{q}\, \hat{r}^\dagger  \hat{s} \} 
\contraction[0.5ex]{+ \{ }{\hat{p}}{^\dagger }{\hat{q}} 
+ \{ \hat{p}^\dagger \hat{q}\,  \hat{r}^\dagger \hat{s} \} 
\nonumber\\&&
\contraction[0.5ex]{+ \{ \hat{p}^\dagger \hat{q}\, }{\hat{r}}{^\dagger }{\hat{s}} 
+ \{ \hat{p}^\dagger \hat{q}\,  \hat{r}^\dagger \hat{s} \} 
\contraction[0.5ex]{+ \{ \hat{p}^\dagger }{\hat{q}}{\,}{\hat{r}}
\contraction[1.0ex]{+ \{ }{\hat{p}}{^\dagger \hat{q}\,\hat{r}^\dagger  }{\hat{s}}
+ \{ \hat{p}^\dagger \hat{q}\, \hat{r}^\dagger  \hat{s} \} 
\contraction[0.5ex]{+ \{ }{\hat{p}}{^\dagger }{\hat{q}} 
\contraction[0.5ex]{+ \{ \hat{p}^\dagger \hat{q}\, }{\hat{r}}{^\dagger  }{\hat{s}} 
+ \{ \hat{p}^\dagger \hat{q}\, \hat{r}^\dagger  \hat{s} \} . \label{pq,rs}
\end{eqnarray}
We can verify this for the following simpler, but nontrivial case by inspection:
\begin{eqnarray}
\hat{p}^\dagger \hat{q}^\dagger \hat{q}\, \hat{p} &=& \{ \hat{p}^\dagger \hat{q}^\dagger \hat{q}\, \hat{p} \} 
\contraction[0.5ex]{+ \{ }{\hat{p}}{^\dagger \hat{q}^\dagger \hat{q}\, }{\hat{p}}
+ \{ \hat{p}^\dagger \hat{q}^\dagger \hat{q}\, \hat{p} \} 
\contraction[0.5ex]{+ \{ \hat{p}^\dagger }{\hat{q}}{^\dagger }{\hat{q}}
+ \{ \hat{p}^\dagger \hat{q}^\dagger \hat{q} \,\hat{p} \}  
\nonumber\\&&
\contraction[0.5ex]{+ \{ \hat{p}^\dagger }{\hat{q}}{^\dagger }{\hat{q}}
\contraction[1.0ex]{+ \{ }{\hat{p}}{^\dagger \hat{q}^\dagger \hat{q}\,}{\hat{p}} 
+ \{ \hat{p}^\dagger \hat{q}^\dagger \hat{q}\, \hat{p} \} \\
&=& \{ \hat{p}^\dagger \hat{q}^\dagger \hat{q}\, \hat{p} \} 
\contraction[0.5ex]{+}{\hat{p}}{^\dagger}{\hat{p}} 
+ \hat{p}^\dagger \hat{p}\, \{ \hat{q}^\dagger \hat{q} \} 
\contraction[0.5ex]{+}{\hat{q}}{^\dagger}{\hat{q}} 
+ \hat{q}^\dagger \hat{q}\, \{ \hat{p}^\dagger \hat{p} \} 
\nonumber\\&&
\contraction[0.5ex]{+ }{\hat{p}}{^\dagger}{\hat{p}}
\contraction[0.5ex]{+ \hat{p}^\dagger \hat{p}\,}{\hat{q}}{^\dagger}{\hat{q}} 
+ \hat{p}^\dagger \hat{p}\,\hat{q}^\dagger \hat{q} \\
&=& f_p^+ f_q^+ \hat{p}^\dagger \hat{q}^\dagger \hat{q}\, \hat{p} 
- f_p^+ f_q^- \hat{p}^\dagger  \hat{q}\, \hat{q}^\dagger \hat{p}
 - f_p^- f_q^+ \hat{p}\,\hat{q}^\dagger \hat{q}\, \hat{p}^\dagger 
 \nonumber\\&& 
+ f_p^- f_q^- \hat{p}\, \hat{q}\, \hat{q}^\dagger \hat{p}^\dagger 
+ f_p^- ( f_q^+ \hat{q}^\dagger \hat{q} - f_q^- \hat{q}\, \hat{q}^\dagger ) 
\nonumber\\&&
+ f_q^- ( f_p^+ \hat{p}^\dagger \hat{p} - f_p^- \hat{p}\, \hat{p}^\dagger )  + f_p^- f_q^- \\
&=& f_p^+ f_q^+ \hat{p}^\dagger \hat{q}^\dagger \hat{q}\, \hat{p} 
- f_p^+ f_q^- \hat{p}^\dagger  (1 - \hat{q}^\dagger \hat{q})\, \hat{p}
\nonumber\\&&
 - f_p^- f_q^+ \hat{q}^\dagger (1 - \hat{p}^\dagger \hat{p} )\, \hat{q} 
+ f_p^- f_q^- (1 - \hat{p}^\dagger \hat{p})(1- \hat{q}^\dagger\hat{q} ) 
\nonumber\\&&
+ f_p^- ( f_q^+ \hat{q}^\dagger \hat{q} - f_q^- \hat{q} \,\hat{q}^\dagger ) 
+ f_q^- ( f_p^+ \hat{p}^\dagger \hat{p} - f_p^- \hat{p}\, \hat{p}^\dagger )  
\nonumber \\&& 
+ f_p^- f_q^- \\
&=& (f_p^+ f_q^+ + f_p^+ f_q^- + f_p^- f_q^+ + f_p^- f_q^-) \hat{p}^\dagger \hat{q}^\dagger \hat{q}\, \hat{p} \nonumber\\
&& +  f_p^- f_q^- - f_p^- f_q^- (\hat{p}^\dagger \hat{p} + \hat{p}\, \hat{p}^\dagger) 
\nonumber \\&& 
- f_p^- f_q^- (\hat{q}^\dagger \hat{q} + \hat{q} \,\hat{q}^\dagger) + f_p^- f_q^-  \\
&=& \hat{p}^\dagger \hat{q}^\dagger \hat{q} \,\hat{p}.
\end{eqnarray}
When $pq \neq rs$ or $pq \neq sr$, Eq.\ (\ref{pqrs}) embodies the theorem. 
In general situations, the theorem can be proven by induction. See Ref.\ \onlinecite{shavitt} for an outline 
of the proof for the zero-temperature case.

Since the thermal average of any normal-ordered product of operators vanishes, a nonzero thermal average 
arises solely from full contractions, e.g.,
\begin{eqnarray}
\langle \langle I | x_{pqrs} \hat{p}^\dagger \hat{q}\, \hat{r}^\dagger \hat{s} | I \rangle \rangle  &=& 
 x_{pqrs} 
\contraction[0.5ex]{\{ \hat{p}^\dagger }{\hat{q}}{\,}{\hat{r}}
\contraction[1.0ex]{\{ }{\hat{p}}{^\dagger \hat{q}\,\hat{r}^\dagger  }{\hat{s}}
 \{ \hat{p}^\dagger \hat{q}\, \hat{r}^\dagger  \hat{s} \} 
+ x_{pqrs} 
\contraction[0.5ex]{\{ }{\hat{p}}{^\dagger }{\hat{q}} 
\contraction[0.5ex]{\{ \hat{p}^\dagger \hat{q}\, }{\hat{r}}{^\dagger  }{\hat{s}} 
\{ \hat{p}^\dagger \hat{q}\, \hat{r}^\dagger  \hat{s} \} \nonumber\\
&=&\sum_{p,q}  x_{pqqp} f_p^- f_q^+ + \sum_{p,r} x_{pprr} f_p^- f_r^-. \nonumber\\ 
\end{eqnarray}

However, more frequently, we need to evaluate the thermal average of a product of two or more normal-ordered products. We show that 
such a product is the sum of concatenated normal-ordered product plus all of its partial and full contractions excluding
internal contractions. An internal contraction is the one that involves at least one contraction within an original normal-ordered product. 
The following is an illustration of this rule:
\begin{eqnarray}
\{ \hat{p}^\dagger  \hat{q} \} \{ \hat{r}^\dagger \hat{s} \}  &=&  \{ \hat{p}^\dagger \hat{q} \,\hat{r}^\dagger  \hat{s} \}  
\contraction[0.5ex]{+ \{ }{\hat{p}}{^\dagger \hat{q}\,\hat{r}^\dagger  }{\hat{s}}
+ \{ \hat{p}^\dagger \hat{q} \, \hat{r}^\dagger \hat{s} \} 
\contraction[0.5ex]{+ \{ \hat{p}^\dagger }{\hat{q}}{\,}{\hat{r}}
+ \{ \hat{p}^\dagger \hat{q} \,\hat{r}^\dagger  \hat{s} \} 
\contraction[0.5ex]{+ \{ \hat{p}^\dagger }{\hat{q}}{\,}{\hat{r}}
\contraction[1.0ex]{+ \{ }{\hat{p}}{^\dagger \hat{q}\,\hat{r}^\dagger  }{\hat{s}}
+ \{ \hat{p}^\dagger \hat{q}\,\hat{r}^\dagger  \hat{s} \} . \nonumber\\
\end{eqnarray}
Internal contractions in this example are the ones that contract $\hat{p}^\dagger$ with $\hat{q}$ and/or 
$\hat{r}^\dagger$ with $\hat{s}$, which are therefore excluded from the right-hand side. 
This rule can be justified by comparing Eq.\ (\ref{pq,rs}) with
\begin{eqnarray}
\hat{p}^\dagger  \hat{q} \,\hat{r}^\dagger \hat{s} &=& \left( \hat{p}^\dagger  \hat{q} \right)\left( \hat{r}^\dagger \hat{s} \right) \nonumber \\
&=& \left( 
\contraction[0.5ex]{\{ \hat{p}^\dagger  \hat{q} \} + }{\hat{p}}{^\dagger  }{\hat{q}}
\{ \hat{p}^\dagger  \hat{q} \} + \hat{p}^\dagger  \hat{q} 
\right)
+ \left( 
\contraction[0.5ex]{\{ \hat{r}^\dagger \hat{s} \} +  }{\hat{r}}{^\dagger }{\hat{s}}
\{ \hat{r}^\dagger \hat{s} \} +  \hat{r}^\dagger \hat{s}
\right ) \nonumber \\
&=& \{ \hat{p}^\dagger  \hat{q} \} \{ \hat{r}^\dagger \hat{s} \}  
\contraction[0.5ex]{+  \{ \hat{p}^\dagger  \hat{q} \} \,  }{\hat{r}}{^\dagger }{\hat{s}}
+  \{ \hat{p}^\dagger  \hat{q} \} \,  \hat{r}^\dagger \hat{s}
\contraction[0.5ex]{+  }{\hat{p}}{^\dagger }{ \hat{q}} 
+  \hat{p}^\dagger  \hat{q}\, \{ \hat{r}^\dagger \hat{s} \}  
\contraction[0.5ex]{+  }{\hat{p}}{^\dagger }{ \hat{q}} 
\contraction[0.5ex]{+  \hat{p}^\dagger  \hat{q}\,  }{\hat{r}}{^\dagger }{\hat{s} }
+  \hat{p}^\dagger  \hat{q}\,  \hat{r}^\dagger \hat{s}. \nonumber\\ 
\end{eqnarray}
The last three terms are internal contractions, which cancel out the same in Eq.\ (\ref{pq,rs}) when equated
with the above.
Therefore, the thermal average of a product of normal-ordered products comes solely from
its full contractions excluding internal contractions, e.g.,
\begin{eqnarray}
\langle\langle I | x_{pq} y_{rs} \{\hat{p}^\dagger \hat{q}\}\{ \hat{r}^\dagger \hat{s}\}|I\rangle\rangle 
&=& 
\contraction[1.0ex]{\langle\langle I | x_{pq} y_{rs} \{}{\hat{p}}{^\dagger \hat{q}\}\{ \hat{r}^\dagger }{\hat{s}} 
\contraction[0.5ex]{\langle\langle I | x_{pq} y_{rs} \{\hat{p}^\dagger }{\hat{q}}{\}\{ }{\hat{r}}  
\langle\langle I | x_{pq} y_{rs} \{\hat{p}^\dagger \hat{q}\}\{ \hat{r}^\dagger \hat{s}\}|I\rangle\rangle  \nonumber \\
&=&  \sum_{p,q} x_{pq} y_{qp} f_p^- f_q^+. \label{example3}
\end{eqnarray}

\subsection{Hamiltonian and number operators\label{app:Hamiltonian}}

Let us express the number operator in the normal-ordered second quantization at finite temperature. 
\begin{eqnarray}
\hat{N} &\equiv& \sum_p \hat{p}^\dagger \hat{p} \nonumber\\
&=& \sum_p \left( \{ \hat{p}^\dagger \hat{p} \} 
\contraction[0.5ex]{+ }{\hat{p}}{^\dagger }{\hat{p}}
+ \hat{p}^\dagger \hat{p}
\right) \nonumber\\
&=& \sum_p \{ \hat{p}^\dagger \hat{p} \} + \sum_p f_p^- \nonumber\\
&=& \sum_p \{ \hat{p}^\dagger \hat{p} \}  + \bar{N}, \label{number}
\end{eqnarray}
where we used the zeroth-order electroneutrality condition [Eq.\ (\ref{Nbar})] in the last equality. Taking the thermal average of the above equation,
we immediately find $\langle \hat{N} \rangle = \bar{N}$. 

The Hamiltonian operator in second quantization is written as\cite{shavitt,szabo}
\begin{eqnarray}
\hat{H} &=& E_{\text{nuc.}} + \sum_{p,q} h_{pq} \hat{p}^\dagger \hat{q} + \frac{1}{4}\sum_{p,q,r,s} \langle pq || rs \rangle 
\hat{p}^\dagger \hat{q}^\dagger \hat{s}\, \hat{r} ,
\end{eqnarray}
where $E_{\text{nuc.}}$ is the nucleus-nucleus repulsion energy, $h_{pq}$ is the integral of the one-electron part of the Hamiltonian operator,
and $\langle pq || rs \rangle$ is the antisymmetrized
two-electron integral. This can be finite-temperature normal ordered as
\begin{eqnarray}
\hat{H} &=& E_{\text{nuc.}} + \sum_{p,q} h_{pq} \left( \{ \hat{p}^\dagger \hat{q} \} + 
\contraction[0.5ex]{}{\hat{p}}{^\dagger}{ \hat{q}}
\hat{p}^\dagger \hat{q}
 \right) \nonumber\\&& 
+ \frac{1}{4}\sum_{p,q,r,s} \langle pq || rs \rangle \left( 
\{ \hat{p}^\dagger \hat{q}^\dagger \hat{s}\, \hat{r} \} 
\contraction[0.5ex]{+ \{}{\hat{p}}{^\dagger \hat{q}^\dagger \hat{s}\, }{\hat{r}}
+ \{\hat{p}^\dagger \hat{q}^\dagger \hat{s}\, \hat{r} \}
\contraction[0.5ex]{+ \{}{\hat{p}}{^\dagger \hat{q}^\dagger }{\hat{s}}
+ \{\hat{p}^\dagger \hat{q}^\dagger \hat{s}\, \hat{r} \}
\right. \nonumber \\&& \left. 
\contraction[0.5ex]{+ \{\hat{p}^\dagger }{\hat{q}}{^\dagger \hat{s}\, }{\hat{r}}
+ \{\hat{p}^\dagger \hat{q}^\dagger \hat{s}\, \hat{r} \}
\contraction[0.5ex]{+ \{\hat{p}^\dagger }{\hat{q}}{^\dagger }{\hat{s}}
+ \{\hat{p}^\dagger \hat{q}^\dagger \hat{s}\, \hat{r} \}
\contraction[1.0ex]{+ \{}{\hat{p}}{^\dagger \hat{q}^\dagger \hat{s}\, }{\hat{r}}
\contraction[0.5ex]{+ \{\hat{p}^\dagger }{\hat{q}}{^\dagger }{\hat{s}}
+ \{\hat{p}^\dagger \hat{q}^\dagger \hat{s}\, \hat{r} \}
\contraction[1.0ex]{+ \{}{\hat{p}}{^\dagger \hat{q}^\dagger }{\hat{s}}
\contraction[0.5ex]{+ \{\hat{p}^\dagger }{\hat{q}}{^\dagger \hat{s}\, }{\hat{r}}
+ \{\hat{p}^\dagger \hat{q}^\dagger \hat{s}\, \hat{r} \}
\right) \nonumber\\
&=& E_{\text{HF}}(T) + \sum_{p,q} f_{pq}(T) \{ \hat{p}^\dagger \hat{q} \}  
+ \frac{1}{4}\sum_{p,q,r,s} \langle pq || rs \rangle 
\{ \hat{p}^\dagger \hat{q}^\dagger \hat{s} \,\hat{r} \},\nonumber\\ \label{Hamiltonian}
\end{eqnarray}
where $E_{\text{HF}}(T)$ and $f_{pq}(T)$ are the finite-temperature Hartree--Fock energy and Fock matrix element,\cite{Mermin} defined by
\begin{eqnarray}
E_{\text{HF}}(T) &=&  E_{\text{nuc.}} + \sum_{p} h_{pp} f_p^- + \frac{1}{2} \sum_{p,q} \langle pq || pq \rangle f_p^- f_q^-, \\
f_{pq}(T) &=& h_{pq} + \sum_r \langle pr || qr \rangle f_r^-.
\end{eqnarray}

Therefore, $E_{\text{HF}}(T)$ is identified as the following sum,
\begin{eqnarray}
E_{\text{HF}}(T) = \langle E^{(0)}_I\rangle + \langle E^{(1)}_I\rangle, \label{EHF}
\end{eqnarray}
 where the thermal averages of the zeroth- and first-order energy corrections\cite{Hirata1storder,Hirata2ndorder} are given by
\begin{eqnarray}
\langle E^{(0)}_I\rangle &=& E_{\text{nuc.}} + \sum_p \epsilon_p f_p^-,  \label{e0}  \\
\langle E^{(1)}_I\rangle &=& \sum_p F_{pp} f_p^- -\frac{1}{2} \sum_{p,q} \langle pq||pq \rangle f_p^- f_q^- \label{e1} 
\end{eqnarray}
with $F_{pq}$ being the difference between the finite-temperature Fock and diagonal zero-temperature Fock matrix elements,
\begin{eqnarray}
F_{pq} &=& f_{pq}(T) - \epsilon_p \delta_{pq} . \label{Fock}
\end{eqnarray}
Equation (\ref{EHF}) mirrors the well-known identity\cite{szabo} in the zero-temperature case, $E_{\text{HF}} = E^{(0)} + E^{(1)}$,
but it does not agree with either $U^{(0)} + U^{(1)}$ or $\Omega^{(0)} + \Omega^{(1)}$ at $T > 0$. 
They are instead related to one another by\cite{Hirata1storder,Hirata2ndorder} 
\begin{eqnarray}
U^{(0)} + U^{(1)} &=& E_{\text{nuc.}} + \sum_{p} \epsilon_p f_p^- + \sum_p F_{pp} f_p^-
\nonumber \\&&
 -\frac{1}{2} \sum_{p,q} \langle pq || pq \rangle f_p^- f_q^- 
  -\beta \sum_p (F_{pp} - \mu^{(1)} )\epsilon_p f_p^- f_p^+ \nonumber \\
&=& E_{\text{HF}}(T) +TS^{(1)}, \\
\Omega^{(0)} + \Omega^{(1)} &=& E_{\text{nuc.}} + \frac{1}{\beta} \sum_p \ln f_p^+ + 
\sum_p F_{pp} f_p^- 
\nonumber\\&& -\frac{1}{2} \sum_{p,q} \langle pq || pq \rangle f_p^- f_q^- - \mu^{(1)} \bar{N} \nonumber \\
&=& E_{\text{HF}}(T)  - (\mu^{(0)} + \mu^{(1)} ) \bar{N}-TS^{(0)}.
\end{eqnarray}

We employ the M\o ller--Plesset partitioning\cite{moller} of the Hamiltonian. 
The zeroth-order Hamiltonian and perturbation operators are then written in the normal-ordered form as
\begin{eqnarray}
\hat{H}_0 &\equiv& E_{\text{nuc.}} + \sum_p \epsilon_p \hat{p}^\dagger \hat{p} 
= \langle E^{(0)}_I\rangle +  \sum_p \epsilon_p \{ \hat{p}^\dagger \hat{p} \},  \label{H0} \\
\hat{V} &\equiv& 
\langle E^{(1)}_I\rangle + \sum_{p,q} F_{pq} \{ \hat{p}^\dagger \hat{q} \}
+ \frac{1}{4}\sum_{p,q,r,s} \langle pq || rs \rangle 
\{ \hat{p}^\dagger \hat{q}^\dagger \hat{s} \hat{r} \}. \nonumber\\ \label{V}
\end{eqnarray}

\subsection{Projector and resolvent operator\label{app:projector}}

In order to understand the second-quantization rule for the resolvent operator, let us first examine the effect of inserting 
the resolution-of-the-identity  in Eq.\ (\ref{example3}). Using Einstein's implied summation
of repeated spinorbital and state indexes, we can rewrite Eq.\ (\ref{example3}) as
\begin{eqnarray}
&& \langle \langle I | x_{pq} y_{rs} \{\hat{p}^\dagger \hat{q}\}\{ \hat{r}^\dagger \hat{s}\} | I \rangle\rangle \nonumber\\ 
&& = \langle \langle I | x_{pq}  \{\hat{p}^\dagger \hat{q}\}|Q \rangle \langle Q | \{ \hat{r}^\dagger \hat{s}\} y_{rs}| I \rangle\rangle \nonumber\\
&&= \langle \langle I |  x_{pq}  
\contraction[1.0ex]{\{}{\hat{p}}{^\dagger \hat{q}\}\{\hat{t}^\dagger }{\hat{u}}
\contraction[0.5ex]{\{\hat{p}^\dagger}{\hat{q}}{\}\{}{\hat{t}}
\contraction[0.3ex]{\{\hat{p}^\dagger \hat{q}\}\{\hat{t}^\dagger \hat{u}\}| I \rangle\langle I | \{\hat{u}^\dagger}{\hat{t}}{\}\{ }{\hat{r}}
\contraction[1.0ex]{\{\hat{p}^\dagger \hat{q}\}\{\hat{t}^\dagger \hat{u}\}|I \rangle\langle I | \{}{\hat{u}}{^\dagger \hat{t}\}\{ \hat{r}^\dagger }{\hat{s}}
\bcontraction[0.5ex]{\{\hat{p}^\dagger \hat{q}\}\{\hat{t}^\dagger }{\hat{u}}{\}| I \rangle\langle I| \{}{\hat{u}}
\bcontraction[1.0ex]{\{\hat{p}^\dagger \hat{q}\}\{}{\hat{t}}{^\dagger \hat{u}\}|I\rangle\langle I| \{\hat{u}^\dagger }{\hat{t}}
\{\hat{p}^\dagger \hat{q}\}\{\hat{t}^\dagger \hat{u}\}|I \rangle\langle I | \{\hat{u}^\dagger \hat{t}\}\{ \hat{r}^\dagger \hat{s}\}
y_{rs} |I \rangle \nonumber\\
&&= \sum_{p,q} x_{pq} y_{qp} f_p^- f_q^+,
\end{eqnarray} 
where $Q$ runs over all determinants, and hence $\sum |Q\rangle \langle Q| = 1$. In the second equality, we wrote $|Q\rangle = \{\hat{t}^\dagger \hat{u}\}| I \rangle$.
For the last equality to hold, we should regard a chain of contractions, $p$-$u$-$s$ or $q$-$t$-$r$, as a single contraction evaluated as $f_p^-$
or $f_q^+$, respectively, rather than view each as three consecutive contractions evaluated as $f_p^- f_p^+ f_p^-$ or $f_q^+ f_q^- f_q^+$, which is erroneous.

Then, the rule for evaluating thermal averages involving a resolvent operator is essentially the same. To explain the rule by a similar example, we may consider 
\begin{eqnarray}
&& \langle \langle I | x_{pq} y_{rs} \{\hat{p}^\dagger \hat{q}\}\hat{R}_I \{ \hat{r}^\dagger \hat{s}\} | I \rangle\rangle \nonumber\\ 
&& = \langle \langle I | x_{pq}  \{\hat{p}^\dagger \hat{q}\}|A \rangle (E_I^{(0)} - E_A^{(0)})^{-1} \langle A | \{ \hat{r}^\dagger \hat{s}\} y_{rs}| I \rangle\rangle \nonumber\\
&&= \langle \langle I |  x_{pq}  
\contraction[1.0ex]{\{}{\hat{p}}{^\dagger \hat{q}\}\{\hat{t}^\dagger }{\hat{u}}
\contraction[0.5ex]{\{\hat{p}^\dagger}{\hat{q}}{\}\{}{\hat{t}}
\contraction[0.3ex]{\{\hat{p}^\dagger \hat{q}\}\{\hat{t}^\dagger \hat{u}\}| I \rangle(\epsilon_u -\epsilon_t)^{-1}\langle I | \{\hat{u}^\dagger}{\hat{t}}{\}\{ }{\hat{r}}
\contraction[1.0ex]{\{\hat{p}^\dagger \hat{q}\}\{\hat{t}^\dagger \hat{u}\}|I \rangle(\epsilon_u -\epsilon_t)^{-1}\langle I | \{}{\hat{u}}{^\dagger \hat{t}\}\{ \hat{r}^\dagger }{\hat{s}}
\bcontraction[0.5ex]{\{\hat{p}^\dagger \hat{q}\}\{\hat{t}^\dagger }{\hat{u}}{\}| I \rangle(\epsilon_u -\epsilon_t)^{-1}\langle I| \{}{\hat{u}}
\bcontraction[1.0ex]{\{\hat{p}^\dagger \hat{q}\}\{}{\hat{t}}{^\dagger \hat{u}\}|I\rangle(\epsilon_u -\epsilon_t)^{-1}\langle I| \{\hat{u}^\dagger }{\hat{t}}
\{\hat{p}^\dagger \hat{q}\}\{\hat{t}^\dagger \hat{u}\}|I \rangle(\epsilon_u -\epsilon_t)^{-1}\langle I | \{\hat{u}^\dagger \hat{t}\}\{ \hat{r}^\dagger \hat{s}\}
y_{rs} |I \rangle \nonumber\\
&&= \sum_{p,q}^{\text{denom.} \neq 0} \frac{x_{pq} y_{qp} f_p^- f_q^+}{\epsilon_p - \epsilon_q},
\end{eqnarray} 
where $A$ stands for a determinant outside the degenerate subspace of $I$, which explains why
the summations over $p$ and $q$ must be restricted (``$\text{denom.}\neq 0$'') to nonzero denominators. 
In other words, $\hat{R}_I$ acts as an outer projector.

Let us define inner projector $\hat{P}_I = \sum | J \rangle \langle J|$, where $J$ runs over only those determinants that are degenerate with $I$. 
\begin{eqnarray}
&& \langle \langle I | x_{pq} y_{rs} \{\hat{p}^\dagger \hat{q}\}\hat{P}_I \{ \hat{r}^\dagger \hat{s}\} | I \rangle\rangle \nonumber\\ 
&& = \langle \langle I | x_{pq}  \{\hat{p}^\dagger \hat{q}\}|J \rangle \langle J | \{ \hat{r}^\dagger \hat{s}\} y_{rs}| I \rangle\rangle \nonumber\\
&&= \langle \langle I |  x_{pq}  
\contraction[1.0ex]{\{}{\hat{p}}{^\dagger \hat{q}\}\{\hat{t}^\dagger }{\hat{u}}
\contraction[0.5ex]{\{\hat{p}^\dagger}{\hat{q}}{\}\{}{\hat{t}}
\contraction[0.3ex]{\{\hat{p}^\dagger \hat{q}\}\{\hat{t}^\dagger \hat{u}\}| I \rangle\langle I | \{\hat{u}^\dagger}{\hat{t}}{\}\{ }{\hat{r}}
\contraction[1.0ex]{\{\hat{p}^\dagger \hat{q}\}\{\hat{t}^\dagger \hat{u}\}|I \rangle\langle I | \{}{\hat{u}}{^\dagger \hat{t}\}\{ \hat{r}^\dagger }{\hat{s}}
\bcontraction[0.5ex]{\{\hat{p}^\dagger \hat{q}\}\{\hat{t}^\dagger }{\hat{u}}{\}| I \rangle\langle I| \{}{\hat{u}}
\bcontraction[1.0ex]{\{\hat{p}^\dagger \hat{q}\}\{}{\hat{t}}{^\dagger \hat{u}\}|I\rangle\langle I| \{\hat{u}^\dagger }{\hat{t}}
\{\hat{p}^\dagger \hat{q}\}\{\hat{t}^\dagger \hat{u}\}|I \rangle\langle I | \{\hat{u}^\dagger \hat{t}\}\{ \hat{r}^\dagger \hat{s}\}
y_{rs} |I \rangle\rangle \nonumber\\
&&= \sum_{p,q}^{\text{denom.}=0} x_{pq} y_{qp} f_p^- f_q^+,
\end{eqnarray} 
where the summations over $p$ and $q$ are now restricted (``$\text{denom.}= 0$'') to the cases whose fictitious denominator ($\epsilon_p - \epsilon_q$) is zero.

%
\end{document}